\documentclass[a4paper,11pt]{article}
\usepackage{a4wide}
\usepackage{amsmath}
\usepackage{amsfonts}
\usepackage{amssymb}
\usepackage{bbm}
\usepackage{longtable}

\numberwithin{equation}{section} 

\newcommand{\vm}{\vec{m}}
\newcommand{\vn}{\vec{n}}

\begin{document}
\thispagestyle{empty}
\begin{flushright}
MPP-2008-160
\end{flushright}
\vspace{1cm}
\begin{center}
{\LARGE Matrix Factorizations, Massey Products and F--Terms for\\\medskip Two--Parameter Calabi--Yau Hypersurfaces} 
\end{center}
\vspace{8mm}
\begin{center}
{\large Johanna Knapp$^{\dagger}$\footnote{email: \tt{knapp@mppmu.mpg.de}}, Emanuel Scheidegger$^{\ast}$\footnote{email: {\tt emanuel.scheidegger@math.uni-augsburg.de}}}
\end{center}
\vspace{5mm}
\begin{center}
{\it $^{\dagger}$Max--Planck--Institut f\"ur Physik \\   
F\"ohringer Ring 6\\
D--80805 Munich\\ 
Germany}
\end{center}
\begin{center}
{\it $^{\ast}$Institut f\"ur Mathematik\\ 
Universit\"at Augsburg\\
D--86135 Augsburg\\ 
Germany
}
\end{center}
\vspace{15mm}
\begin{abstract}
\noindent We discuss B--type tensor product branes in mirrors of two--parameter Calabi--Yau hypersurfaces, using the language of matrix factorizations. We determine the open string moduli of the branes at the Gepner point. By turning on both bulk and boundary moduli we then deform the brane away from the Gepner point. Using the deformation theory of matrix factorizations we compute Massey products. These contain the information about higher order deformations and obstructions. The obstructions are encoded in the F--term equations, which we obtain from the Massey product algorithm. We show that the F--terms can be integrated to an effective superpotential. Our results provide an ingredient for open/closed mirror symmetry for these hypersurfaces.
\end{abstract}
\newpage
\setcounter{tocdepth}{1}
\tableofcontents
\setcounter{footnote}{0}
\section{Introduction}
Recently, there has been tremendous progress in the understanding of open string mirror symmetry for compact Calabi--Yau manifolds. In \cite{Walcher:2006rs,Morrison:2007bm} techniques have been introduced to compute open string disk instantons on the quintic through mirror symmetry. The papers \cite{Walcher:2007tp,Walcher:2007qp} discussed the calculation of higher genus open string BPS invariants by making use of an extension of the holomorphic anomaly equation. Walcher's approach to computing open BPS invariants has been shown to work well also for one--parameter hypersurfaces other than the quintic, as was demonstrated in \cite{Krefl:2008sj,Knapp:2008uw}. \\
A special feature of the open string mirror symmetry calculations of \cite{Walcher:2006rs,Morrison:2007bm} is that the quantities which are computed do not show explicit dependence of the open string moduli. In particular this entails that one does not have to find a mirror map for open moduli in order to calculate open string instanton numbers using mirror symmetry. In \cite{Walcher:2007tp} it has been argued that this situation is quite generic for the following reasons. Continuous open moduli may be quite rare at general points in the bulk moduli space. At special points with high symmetries, like the Gepner point, brane moduli are however very likely to be found. In Calabi--Yau threefolds such boundary deformations are generically obstructed. This is due to the fact that one can map, via Serre duality, the open string states corresponding to boundary deformations to open string states which encode the obstructions to these deformations. The closed string moduli enter the game by the observation that certain bulk deformations may be identified with obstructions to deformations on the boundary. Obstructions to deformations are encoded in the effective superpotential $\mathcal{W}_{eff}$, or, more precisely, in its critical locus. This is nothing but the set of solutions of F--term equations which determines the locus where supersymmetry is preserved. The upshot of the argument of \cite{Walcher:2007tp} is that the moduli appearing in $\mathcal{W}_{eff}$ are those which are obstructed, i.e. those which can take only discrete values, in particular in terms of bulk moduli. The generating function of open string disk instanton numbers is the domain wall tension, which is the difference between effective superpotentials, evaluated at solutions of F--term equations corresponding to two brane vacua. In all the known examples this quantity does not depend on continuous brane moduli. The brane moduli--independence of the domain wall tension does not imply that there are no open moduli present. They definitely appear in the effective superpotential and in the F--term equations which contain essential information about supersymmetric vacua. It is therefore interesting to have techniques to compute $\mathcal{W}_{eff}$ and F--terms which have the full dependence on brane and bulk parameters. An important first step in this direction has been made in \cite{Jockers:2008pe}, where $\mathcal{W}_{eff}$ has been calculated for one--parameter models by applying $N=1$ special geometry of \cite{Mayr:2001xk,Lerche:2001cw} to compact Calabi--Yau manifolds. Remarkably, this approach also provides flat coordinates and a mirror map for open moduli.\\\\ 
So far, all the examples for open string mirror symmetry on compact Calabi--Yau threefolds has only been done for hypersurfaces with a single bulk modulus. In this paper we will discuss one building block of the open mirror symmetry program in two--parameter Calabi--Yau hypersurfaces. Going to more complicated examples with more bulk moduli is interesting for several reasons. One--parameter hypersurfaces, and in particular the quintic, exhibit a large amount of symmetry which simplify the calculation a lot. However, in such special models the full mathematical structure of the problem may remain partially hidden. Even if one does not encounter new conceptual features in models with several parameters it is to be expected that new technical tools are needed to handle the increased complexity of the calculations. In closed string mirror symmetry this new ingredient was toric geometry. The techniques developed in \cite{Candelas:1993dm,Hosono:1993qy,Candelas:1994hw,Hosono:1994ax} proved to be the correct language to systematically address closed string mirror symmetry problems.\\
It is to be expected that toric geometry will also play a prominent role in open string mirror symmetry. In open string models, however, there is at least one other problem which has to be overcome before one can set the machinery of toric geometry to work which is needed to derive (inhomogeneous) Picard--Fuchs equations and compute the mirror map. The first step in the program is a suitable choice of D--brane. In this article we will discuss tensor product branes in two--parameter hypersurfaces and their open string moduli. We would like to answer the question which D--branes have moduli and whether deformations with these open moduli are obstructed. This gives us important information about the existence of brane vacua and possible domain walls separating these.\\
The context in which we approach this question will be B--type topological Landau--Ginzburg models. We consider two--parameter hypersurfaces which admit a Landau--Ginzburg description. We focus on those models which are tensor products of minimal models of type $A$. D--branes in these models can be described in terms of matrix factorizations of the Landau--Ginzburg potential. At the Gepner point certain matrix factorizations can be identified with boundary states in conformal field theory. We will focus on tensor product branes which correspond to Recknagel--Schomerus boundary states. At the Gepner point we compute open string moduli of a given brane. Due to the enhanced symmetry at the Gepner point one expects to have more control over D--branes than at other points in moduli space. By deforming the brane away from the Gepner point with both bulk and boundary moduli one gets constraints on the deformation parameters for the brane to remain a valid supersymmetric boundary condition. The constraints encode the obstructions to the deformations of a brane away from the Gepner point. These are precisely the F--terms which determine the brane vacua. For matrix factorizations there exists a an algorithm to compute deformed matrix factorizations and F--terms \cite{Siqveland1,siqvelandPHD}. In \cite{Knapp:2006rd} this method has been applied to $N=2$ minimal models and extended to include bulk deformations.\\
In this article we apply the deformation theory algorithm to tensor product branes in mirrors of two--parameter Calabi--Yau hypersurfaces. A similar discussion has already been done in \cite{Hori:2004ja}. The advantage of this approach is that it avoids several difficulties. Since we are directly computing the F--terms which encode the physical information about the brane vacua we do not necessarily have to do the calculation in flat coordinates and therefore can postpone the problem of finding those. Furthermore it is known that the effective superpotential in topological string calculations is only defined up to  (possibly non--linear) reparametrizations due to the underlying $A_{\infty}$--structure. Its critical locus, i.e. the F--terms, should however not depend on the choice of parametrization. This freedom is visible in the deformation theory algorithm but since the information encoded in the F--terms should be invariant we can make a particular choice without losing information.\\
The advantages of this approach are actually also its biggest drawbacks: The information about flat coordinates in moduli space and the mirror map, which is essential for mirror symmetry calculations, has to be found with different methods. Therefore it is not possible to extract disk instanton numbers from our results without further input. This will be discussed elsewhere.\\
Another inconvenience is that the calculation is technically challenging and rather cumbersome. This also has to do with the reparametrization freedom mentioned above. At every order in deformation theory one has to make certain choices. Although they do not influence the result there may be certain choices which simplify the calculations tremendously. Unfortunately, no criterion is known to pin down particularly simple parametrization. Another general problem of this approach is that in the presence of unobstructed moduli the algorithm never terminates, so that the deformation theory problem cannot be solved completely. \\
Despite these difficulties we can decide for most of the examples we consider which moduli are obstructed and compute the full F--terms by a brute force calculation. \\\\
This article is organized as follows: In section \ref{sec-general} we review the relevant details about matrix factorizations which are necessary for the discussion. In particular we describe the deformation theory algorithm. Furthermore we discuss which branes are not captured by our discussion. We go on to summarize the results of our calculations and discuss certain common properties of our examples. We also comment about interesting new features we have encountered. The subsequent five sections are the technical part of the paper where we discuss brane moduli and obstruction for representative examples of branes for each of the five two--parameter Calabi--Yau hypersurfaces. Section \ref{sec-conclusions} is devoted to concluding remarks. In the appendix we collect some tables which contain information about the moduli of tensor product branes in two--parameter Calabi--Yau hypersurfaces.\\\\
{\bf Acknowledgements:} We would like to thank the Singular Team for support. Part of this work was completed during the authors' visit at the Erwin Schr\"odinger Institute in Vienna at the workshop "Mathematical Challenges in String Phenomenology".
\section{Matrix Factorizations, Deformation Theory and F--terms}
\label{sec-general}
We discuss brane moduli of Recknagel--Schomerus branes on the mirror of two--parameter hypersurfaces in weighted $\mathbbm{CP}^4$, using the language of matrix factorizations. At the Gepner point the Landau Ginzburg theory is an orbifold of tensor product of minimal models of type $A_{d_i-2}$ with superpotential:
\begin{equation}
W=x_1^{d_1}+x_2^{d_2}+x_3^{d_3}+x_4^{d_4}+x_5^{d_5}\:/\Gamma_{GP}
\end{equation}
Since we are interested in the mirror, we have also taken into account the action of the Greene--Plesser orbifold group $\Gamma_{GP}$. We will focus on the five models which have Hodge number $h^{2,1}=2$. The characteristic data of these models can be found for instance in \cite{Hosono:1993qy}. We will give more details in the sections devoted to the specific models. We will consider the following type of matrix factorizations $Q^2=W\cdot\mathbbm{1}$:
\begin{equation}
\label{rs-mf}
Q=\sum_{i=1}^5x_i^{k_i}\eta_i+x_i^{d_i-k_i}\bar{\eta}_i
\end{equation}
The $\eta_i,\bar{\eta}_i$ are boundary fermions satisfying the Clifford algebra relations
\begin{equation}
\{\eta_i,\bar{\eta}_j\}=\delta_{ij}\qquad \{\eta_i,\eta_j\}=0.
\end{equation}
The matrix factorization (\ref{rs-mf}) is a tensor product of minimal model matrix factorizations, which can be identified with CFT boundary states:
\begin{equation}
Q^{(k)}=x^k\eta+x^{d-k}\bar{\eta}=\left(\begin{array}{cc}
0&x^k\\
x^{d-k}&0
\end{array}
\right)\qquad\Longleftrightarrow\qquad |L,S\rangle=|k-1,0\rangle
\end{equation}
The matrix factorization (\ref{rs-mf}) can then be identified with a Gepner model boundary state with label $L=|k_1-1,k_2-1,k_3-1,k_4-1,k_5-1\rangle$. We will use this convenient notation to label our matrix factorizations, even if it has been deformed away from the Gepner point\footnote{Since we only consider marginal deformations of a single brane and no tachyon condensation processes, these labels are also accurate for the deformed branes.}.\\
Given a matrix factorization $Q$ we can define a matrix $R$ of $U(1)$ $R$--charges via the condition that $Q$ has charge $1$. Given a $U(1)$ action $x_i\rightarrow\lambda^{\omega_i} x_i$, $R$ is chosen such that:
\begin{equation}
\label{rdef}
R\cdot Q(\lambda^{\omega_i} x_i)\cdot R^{-1}=\lambda Q(x_i)
\end{equation} 
If we have an orbifold action $g:\:x_i\rightarrow e^{2\pi i g^i/d}x_i$, where $d$ is the degree of the Landau--Ginzburg potential, the orbifold action can be extended to the boundary. A matrix factorization is orbifold invariant if we can find a matrix $\gamma$ such that:
\begin{equation}
\gamma\cdot Q(e^{2\pi i g^i/d}x_i)\cdot \gamma^{-1}=Q(x_i)
\end{equation}
\subsection{Brane Moduli}
In the context of matrix factorizations physical open string states are determined by the cohomology of $Q$. $Q$ acts on the open string states via a commutator or an anticommutator. In this way the cohomology elements come with a natural $\mathbbm{Z}_2$--grading.\\
Brane moduli $\Psi_i\in H^{odd}(Q)$ correspond to $\mathbbm{Z}_2$--odd, boundary preserving open string states with R--charge 1. They may be used to deform the matrix factorization (\ref{rs-mf}) away from the Gepner point. Open string states of (\ref{rs-mf}) are tensor products of minimal model open string states. The $\mathbbm{Z}_2$--odd, "fermionic", open string states of a type $A$ minimal model look as follows:
\begin{equation}
\label{mmfer}
\psi^{(k)}_l=\left(\begin{array}{cc}
0&x^{l}\\
-x^{d-2k+l}&0
\end{array}\right)
\qquad l=0,\ldots,k-1
\end{equation}
The $R$--charges of these fermions are $q_{\psi_l}=\frac{d-2k+2l}{d}$. Note that for our choice for $k$, we have $l\leq d-2k+l$, which means that the exponent of the lower left entry of the matrix is always greater or equal to the exponent of the upper right entry. \\
The bosonic open string states have a simpler structure:
\begin{equation}
\label{mmbos}
\phi^{(k)}_l=\left(\begin{array}{cc}
x^l&0\\
0&x^l
\end{array}\right)\qquad l=0,\ldots,k-1
\end{equation}
The $R$--charges are $q_{\phi_l}=\frac{2l}{d}$.\\
The R--charge $r$ of a general open string state $\Psi$ is computed as follows:
\begin{equation}
R\cdot \Psi(\lambda^{\omega_i} x_i)\cdot R^{-1}=\lambda^r \Psi(x_i),
\end{equation}
where $R$ was defined in (\ref{rdef}). An open string state is orbifold invariant if:
\begin{equation}
\gamma\cdot \Psi(e^{2\pi i g^i/d}x_i)\cdot \gamma^{-1}=\Psi(x_i)
\end{equation}
Since the open string states of $A$--type minimal models are uniquely determined by their $R$--charge $r$ and their $\mathbbm{Z}_2$--grading $a$, we can label open string states of (\ref{rs-mf}) by\footnote{All the tensor products are graded. See \cite{Ashok:2004zb,Hori:2004ja} for details about the explicit construction of tensor product matrix factorizations.} $r_1^{a_1}\otimes r_2^{a_2}\otimes r_3^{a_3}\otimes r_4^{a_4}\otimes r_5^{a_5}$.
\subsection{Obstructions}
Obstructions $\Phi_i\in H^{even}(Q)$ to deformations of branes are encoded in $\mathbbm{Z}_2$--even, charge $2$ boundary preserving operators. It is important to notice that these open string states have the correct $R$--charge and degree to be deformations of the Landau--Ginzburg superpotential. Obviously only a subset of these open string states, namely those proportional to the unit matrix, can also be interpreted as bulk deformations. It is these boundary preserving open string states which are directly responsible for the link between bulk and boundary deformations and therefore for the fact that boundary deformations can be obstructed by bulk deformations.\\
The deformation theory for Calabi--Yau manifolds in three complex dimensions is special because only then the open string states describing obstructions are Serre dual to the deformations, i.e. the brane moduli.
\begin{equation}
H^{even}(Q)\cong (H^{odd}(Q))^{\ast}
\end{equation} 
This isomorphism between deformations and obstructions is the reason why brane deformations are "in general" obstructed.\\
It is important to know what is the corresponding obstruction to each deformation. This is done by computing the Serre pairings $\langle\Psi_i\Phi_j\rangle$, where we denote the boundary deformations by $\Psi_i$ and the obstructions by $\Phi_i$. 
These amplitudes are easily computed by the Kapustin--Li residue formula \cite{Kapustin:2003ga}:
\begin{equation}
\label{kapustin}
\langle\Psi_i\Phi_j\rangle=\frac{1}{(2\pi i)^5}\oint d^5x\frac{\mathrm{STr}\left(\left(\partial Q\right)^{\wedge 5}\Psi_i\Phi_j\right)}{\partial_1W\cdots\partial_5W}
\end{equation}
Only for Serre dual pairs this integral is non--zero.
\subsection{Deformations and Higher Products}
\label{sec-massey}
We now give a description of the deformation theory algorithm which we use to calculate the F--terms. Our discussion of the algorithm follows \cite{Knapp:2007vc}. For mathematical background we refer to the papers \cite{Siqveland1,siqvelandPHD}.\\
We start with describing the original algorithm which only includes deformations with brane moduli. Consider a matrix factorization $Q$ with $Q^2=W\cdot\mathbbm{1}$ and calculate the open string spectrum:
\begin{equation}
\Psi_i\in H^{odd}(Q)\qquad \Phi_i\in H^{even}\qquad \mathrm{dim}H^{even}(Q)=\mathrm{dim}H^{odd}=N
\end{equation}
We now want to calculate the most general non--linear deformation of this matrix factorization, taking into account only deformations with $\mathbbm{z}_2$--odd states. We make the following ansatz:
\begin{equation}
\label{eq-qdefans}
Q_{def}=Q+\sum_{\vec{m}\in\bar{B}}\alpha_{\vec{m}}u^{\vec{m}}
\end{equation}
Here, $\vec{m}$ is a multi index: $u^{\vec{m}}=u_1^{m_1}u_2^{m_2}\ldots u_N^{m_N}$ and we define $|\vec{m}|=\sum_{i=1}^N m_i$. $\bar{B}$ describes the allowed set of vectors $\vm$.
$u_1,\ldots, u_N$ are deformation parameters associated to $\Psi_1,\ldots,\Psi_n$ and  $\alpha_{\vec{m}}$ are matrices to be determined recursively in $|\vm|$. At the order $|\vec{m}|=1$ (linear deformations) they are defined to be the odd cohomology elements:
\begin{equation}
\alpha_{(1,0,\ldots,0)}=\Psi_1\quad \alpha_{(0,1,\ldots,0)}=\Psi_2\quad\ldots\quad \alpha_{(0,\ldots,0,1)}=\Psi_{N} 
\end{equation}
Now we impose the matrix factorization condition on $Q_{def}$:
\begin{eqnarray}
\label{eq-qdeffact1}
Q_{def}^2&\stackrel{!}{=}&W\cdot\mathbbm{1}+\sum_{j=1}^N{\hat{f}}_j(u)\Phi_j\\
\label{eq-qdeffact2}
&\sim& Q^2+\sum_{\vec{m}}[Q,\alpha_{\vec{m}}]u^{\vec{m}}+\sum_{\vec{m}_1+\vec{m}_2=\vec{m}}\underbrace{\alpha_{\vec{m}_1}\cdot\alpha_{\vec{m}_2}}_{y(\vec{m})}u^{\vec{m}}
\end{eqnarray}
Note that imposing $Q_{def}^2=W\cdot\mathbbm{1}$ does not work and we must employ the more general condition in (\ref{eq-qdeffact1}). Obviously, the matrix factorization condition only holds if we demand that ${\hat{f}}_i(u)=0$. The relations ${\hat{f}}_i(u)$ span the same ideal as the vanishing relations $f_i(u)$ of the power series ring of deformations $\mathbbm{C}[[u]]/(f_i(u))$. At the same time these relations determine the critical locus of the effective superpotential.\\
In the second line of the above equation we naively inserted the ansatz (\ref{eq-qdefans}). $y(\vm)$ is called 'matric Massey product' \cite{Siqveland1,siqvelandPHD}. Equation (\ref{eq-qdeffact2}) is actually only correct up to order $|\vm|=2$. At higher orders the definition of the Massey products gets modified due to the presence of the $f_i(u)$, as we will show below. The 'method of computing formal moduli' of \cite{Siqveland1,siqvelandPHD} provides an algorithm to calculate the $f_i(u)$ and the $\alpha_{\vm}$ explicitly at all orders in $\vm$.\\
Let us first look at the lowest orders, where (\ref{eq-qdeffact2}) is correct. At linear order $|\vm|=1$ in the deformation parameters, the second term in (\ref{eq-qdeffact2}) is zero, since the $\alpha_{\vm}$ are the fermionic cohomology elements and the second term becomes the physical state condition. The first Massey product $y(\vm)$ appears at order $|\vm|=2$. We can calculate this product explicitly, since all the $\alpha_{\vm}$ at order $|\vm|=1$ are known. $y(\vm)$ can take the following values:
\begin{itemize}
\item $y(\vm)\notin H^{even}(Q)$. In this case we can find an $\alpha_{m}$ with $|\vm|=2$ such that
\begin{equation}
[Q,\alpha_{\vec{m}}]\equiv-\beta_{\vm}=y(\vec{m}).
\end{equation}
Thus, the second and the third term in (\ref{eq-qdeffact2}) cancel at order $|\vm|=2$ and we produced new $\alpha_{\vm}$'s and thus can calculate Massey products at higher order. 
\item $y(\vm)\in H^{even}(Q)$, i.e. $y(\vm)=c\:\Phi_k$, where $c$ is some number. Clearly, this cannot be cancelled by a term $[Q,\alpha_{\vec{m}}]$ since the $\Phi_i$ are by definition not $Q$--exact. Thus, we have encountered an obstruction. The obstructions are encoded in the polynomial $f_k(u)$ associated to $\Phi_k$ in the following way:
\begin{equation}
f_k=c u^{\vm}
\end{equation}
\end{itemize}
We can now continue this algorithm to higher orders in $|\vm|$. There, however, some subtleties arise due to the presence of the $f_i(u)$. They impose relations among the $u^{\vm}$, which have to be incorporated into the algorithm. One has to introduce various bases for allowed vectors $\vm$. Furthermore the definition of the higher order Massey products has to be modified as compared to the naive definition of (\ref{eq-qdeffact2}). The deformation theory construction of \cite{Siqveland1,siqvelandPHD} yields the following results:\\
For a vector $\vn\in B'_{i+1}$, $i>0$ ($B'_{i+1}$ to be defined momentarily) the Massey product $y(\vn)$ is given by:
\begin{equation}
\label{eq-massey}
y(\vn)=\sum_{|\vm|\leq i+1}\sum_{\stackrel{\vm_1+\vm_2=\vm}{\vm_i\in \bar{B}_{i}}}\beta'_{\vm,\vn}\alpha_{\vm_1}\cdot\alpha_{\vm_2}
\end{equation}
The coefficients $\beta'_{\vm,\vn}$ can be determined from the unique relation
\begin{equation}
\label{eq-masseycoeff1}
u^{\vn}=\sum_{\vm\in\bar{B}'_{i+1}}\beta'_{\vn,\vm}u^{\vm}+\sum_{j=1}^N\beta'_{\vn,j}f^i_j
\end{equation}
for each $\vn\in\mathbbm{N}^N$ with $|\vn|\leq i+1$. If the Massey product is $y(\vn)=c\:\Phi_k$ then we get a contribution to the $k$--th polynomial $f_k(u)$:
\begin{equation}
\label{eq-masseyobs}
f^{i+1}_k=f_k^{i}+\sum_{\vn\in B'_{i+1}}c\: u^{\vn},
\end{equation}
where the superscript gives the order in $u$.\\
The $\alpha_{\vm}$ are defined as follows. For each vector $\vm$ in a basis $B_{i+1}$ we can find a matrix $\alpha_{\vm}$  such that\footnote{Since many of the $\beta_{\vm}$ appear several times in a given example, we will suppress the $\vm$--labels of the $\beta_{\vm}$ in the following sections.}:
\begin{equation}
\label{eq-masseydef}
[Q,\alpha_{\vm}]=-\beta_{\vm}=-\sum_{l=0}^{(i+1)-2}\sum_{\vn\in B'_{2+l}}\beta_{\vn,\vm}y(\vn),
\end{equation}
where the coefficients $\beta_{\vn,\vm}$ are given by the unique relation
\begin{equation}
\label{eq-masseycoeff2}
u^{\vn}=\sum_{\vm\in\bar{B}_{i+1}}\beta_{\vn,\vm}u^{\vm}.
\end{equation}
The various bases $B$, $\bar{B}$, $B'$, $\bar{B}'$ are defined recursively. One starts by setting $\bar{B}_1=\{\vn\in\mathbbm{N}^{N}|\:|\vn|\leq1\}$ and $B_1=\{\vn\in\mathbbm{N}^N|\:|\vn|=1\}$. For $i\geq 1$, $B'_{i+1}$ is then defined as a basis for $m^{i+1}/(m^{i+2}+m^{i+1}\cap m(f^i_1,\ldots,f^i_N))$, where $m=(u_1,\ldots,u_N)$ defines the maximal ideal of $\mathbbm{C}[[u]]/(f_i(u))$. In most cases, the elements $\{u^{\vn}\}_{\vn\in B'_{i+1}}$ can be chosen such that $u^{\vn}=u_k\cdot u^{\vm}$ for some $\vm\in \bar{B}_i$ and $u_k\in\{u_1,\ldots ,u_N\}$. One defines $\bar{B}'_{i+1}=\bar{B}_i\cup B'_{i+1}$. Finally, $B_{i+1}$ is a basis for $(m^{i+1}+(f_1^i,\ldots,f_N^i))/(m^{i+2}+(f_1^{i+1},\ldots,f_N^{i+1}))$ such that $B_{i+1}\subseteq B'_{i+1}$. We set $\bar{B}_{i+1}=\bar{B}_i\cup B_{i+1}$.\\\\
With these definitions it is now possible to calculate the critical locus $f_i(u)$ of the effective superpotential along with the deformed matrix factorization $Q_{def}$. The algorithm, which we will refer to as the 'Massey product algorithm', looks as follows \cite{Siqveland1}:
\begin{itemize}
\item Choose a matrix factorization $Q$ and calculate the open string spectrum, where $\Psi_j\in H^{odd}(Q)$ and $\Phi_{j}\in H^{even}(Q)$, where $j=1,\ldots,N$.
\item Set $\alpha_{\vec{e}_j}=\Psi_j$, where $\vec{e}_j$ are the canonical basis vectors of $\mathbbm{R}^N$. Furthermore associate a deformation parameter $u_k$ to every $\Psi_k$.
\item For each $i\geq 0$ perform the following steps:
\begin{itemize}
\item Calculate the bases $B'_{i+1}$ and $\bar{B}'_{i+1}$.
\item Determine the coefficients $\beta'_{\vm,\vn}$ from the relations (\ref{eq-masseycoeff1}).
\item Calculate the Massey products $y(\vn)$ defined in (\ref{eq-massey}).
\item Determine $f_j^{i+1}$ using (\ref{eq-masseyobs}).
\item Choose bases $B_{i+1}$ and $\bar{B}_{i+1}$.
\item Calculate the coefficients $\beta_{\vm,\vn}$ from the relations (\ref{eq-masseycoeff2}).
\item Choose suitable $\alpha_{\vm}$ according to (\ref{eq-masseydef}).
\end{itemize}
\item If the algorithm terminates at a given order, integrate (homogeneous linear combinations of) the $f_i$ in order to obtain $\mathcal{W}_{eff}$.
\item Calculate the deformed matrix factorization:
\begin{equation}
\label{eq-masseyqdef}
Q_{def}=Q+\sum_{\vm\in B}\alpha_{\vm}u^{\vm}, \qquad B=\bigcup_i B_i
\end{equation}
\end{itemize}
Let us supplement a few comments. The choice of $\alpha_{\vm}$ is ambiguous. Taking different $\alpha_{\vm}$ also results in different $f_i$. The effective superpotentials obtained from these different choices are related via field redefinitions of the $u_i$, where field redefinition means in this case that every $u_k$ can be replaced by a power series in terms of the $u_i$. This freedom reflects the reparametrization freedom one has in the underlying $A_{\infty}$--structure.\\
For matrix factorizations in Landau--Ginzburg orbifolds we cannot expect that the algorithm terminates due to the presence of marginal deformations which may or may not be obstructed at higher orders. For Calabi--Yau threefolds the algorithm may terminate since obstructions are expected to be present. However, this is not guaranteed. Some moduli may still be exactly marginal, i.e. unobstructed to all orders in deformation theory. In that case one may at least argue that the F--terms do not get new contributions from a given order on. Then integration to an effective superpotential is still possible.\\
Furthermore note that some of the $f_i(u)$ may remain zero throughout the calculation. The power series ring of deformations is then defined as $\mathbbm{C}[[u_1,\ldots,u_N]]/(f_1,\ldots,f_r)$, where $r\leq N$.
\subsubsection{Bulk deformations}
\label{sec-bulkmassey}
So far, we have only discussed deformations of a brane with open moduli. One can however also deform the brane with bulk parameters. In the B--model setup this corresponds to complex structure deformations. The task is to include bulk deformations of the Landau--Ginzburg superpotential at the Gepner point into the deformation theory algorithm. This has already been done for minimal model examples in \cite{Knapp:2006rd}\footnote{In this case the bulk deformations were added to boundary deformed matrix factorizations. This was possible because for minimal models the deformation theory algorithm always terminates after a finite number of steps. In our case, the algorithm may not terminate. This forces us to look at bulk and boundary deformations at the same time.}. For our two--parameter models we would like to deform the superpotential $W(x_i)$ into:
\begin{equation}
W(x_i,\varphi_1,\varphi_2)=W(x_i)+\varphi_1\phi_1+\varphi_2\phi_2,
\end{equation}
where $\varphi_1,\varphi_2$ are the two bulk moduli and $\phi_1,\phi_2$ are elements of the bulk chiral ring.\\
How to proceed crucially depends on the properties of $\phi_1,\phi_2$. If the bulk deformation is also an element of the boundary chiral ring, i.e. if the bulk deformation multiplied with the unit matrix is a charge $2$ bosonic open string state (this is precisely an obstruction!), the associated bulk modulus enters the F--term related to this open string state. This adds linear terms in the bulk moduli to the F--terms. If it is possible to integrate these terms to an effective superpotential these contributions give terms $u\varphi$, where $u$ is a boundary modulus. This encodes the values of the bulk--boundary two--point disk amplitudes. \\
If, however, the bulk deformation is $Q$--exact, the Massey product algorithm implies that this must be related to a deformation. This deformation, call it $\alpha_{(1,0,\ldots,0)}$, is a linear deformation of the matrix factorization where the deformation parameter is the bulk modulus. This deformation has the property that $\{Q,\alpha_{(1,0,\ldots,0)}\}=\phi\mathbbm{1}$, where $\phi$ is the $Q$--exact bulk deformation. So, to first order, we have:
\begin{equation}
Q_{def}=Q+\varphi\alpha_{(1,0,\ldots,0)}+\ldots
\end{equation} 
Squaring this, we get:
\begin{equation}
Q_{def}^2=W(x_i)+\varphi\phi+\ldots,
\end{equation}
which is nothing but the bulk deformed Landau--Ginzburg potential $W(x_i,\varphi)$ plus higher order terms.\\
To summarize, depending on whether the bulk deformation is also an open string state or not we either have to modify the F--terms or introduce bulk deformations of the matrix factorizations. From then on, we can formally apply the Massey product algorithm as discussed above. We simply extend the vectors $\vm$ to include also the bulk deformations. We have chosen the convention that we add the new entries at the beginning of the vector.\\
Note that this discussion is not a rigorous mathematical derivation of the algorithm like it was given in \cite{Siqveland1,siqvelandPHD} for the original setup. It rather relies on the observation that the structure does not change by adding bulk deformations. As we will demonstrate in the following sections, the extended algorithm works well. 
\subsection{How to check the Results}
Given such a complex algorithm to compute deformations, it is already a strong indication of consistency if one has obtained a deformed matrix factorization which squares to the (deformed) Landau--Ginzburg superpotential modulo F--terms, and the obstructions are constraints which can be integrated to give an effective superpotential $\mathcal{W}_{eff}$. Still, it is desirable to have an independent consistency check for the results. This is provided by the interpretation of $\mathcal{W}_{eff}$ as the generating function of disk amplitudes. Differentiation of this effective superpotential should give back the F--terms. The disk amplitudes entering $\mathcal{W}_{eff}$ which do not contain integrated insertions of bulk-- or boundary--operators can be computed by the residue formula of Kapustin and Li \cite{Kapustin:2003ga}. Two kinds of amplitudes can be calculated. The first is the bulk--boundary two--point function:
\begin{equation}
\langle\phi_i\Psi_j\rangle=\frac{1}{(2\pi i)^5}\oint d^5x\frac{\phi_i\mathrm{STr}\left(\left(\partial Q\right)^{\wedge 5}\Psi_j\right)}{\partial_1W\cdots\partial_5W}
\end{equation}
Furthermore, we can calculate the disk amplitude with three boundary insertions:
\begin{equation}
\langle\Psi_i\Psi_j\Psi_k\rangle=\frac{1}{(2\pi i)^5}\oint d^5x\frac{\mathrm{STr}\left(\left(\partial Q\right)^{\wedge 5}\Psi_i\Psi_j\Psi_k\right)}{\partial_1W\cdots\partial_5W}
\end{equation}
Comparing to the F--terms/$\mathcal{W}_{eff}$ obtained from the Massey product calculation, this gives a non--trivial check for the lowest orders in the deformation parameters. Having computed the correlators there are a few steps to be done in order to obtain the effective superpotential. We will now briefly describe these steps, following \cite{Herbst:2004jp}. \\
A disk amplitude with an arbitrary number of bulk and boundary\footnote{In contrast to the rest of the paper $\Psi$ can be $\mathbbm{Z}_2$--even or $\mathbbm{Z}_2$--odd here.} insertions is defined as follows:
\begin{eqnarray}
\label{eq-bdef}
B_{a_0\ldots a_m;i_1\ldots i_n}&:=&(-1)^{\tilde{a}_1+\ldots+\tilde{a}_{m-1}}\left\langle\Psi_{a_0}\Psi_{a_1}P\int\Psi^{(1)}_{a_2}\ldots\int\Psi^{(1)}_{a_{m-1}}\Psi_{a_m}\int\phi^{(2)}_{i_1}\ldots \int\phi^{(2)}_{i_n} \right\rangle \nonumber \\
&=&-\left\langle\phi_{i_1}\Psi_{a_0}P\int\Psi^{(1)}_{a_1}\ldots\int\Psi^{(1)}_{a_m}\int\phi_{i_2}^{(2)}\ldots\int\phi_{i_n}^{(2)}\right\rangle,
\end{eqnarray}
where 
\begin{equation}
\int\phi_i^{(2)}\equiv\int\phi_i^{(1,1)}=\int_{D_2}[G,[\bar{G},\phi_i]]dz\:d\bar{z}
\end{equation}
are the bulk descendants, with $D_2$ the disk and $G$ the twisted fermionic current, and
\begin{equation}
\int\Psi_a^{(1)}=\int_{\tau_l}^{\tau_r}[G,\Psi_a]d\tau
\end{equation}
are the boundary descendants. The integral runs, from a suitably chosen position $\tau_l$ to the left of the operator to a position $\tau_r$ to its right, along the boundary of the disk. The boundary integrals in (\ref{eq-bdef}) have to be path ordered, and $P$ denotes the path ordering operator. We have also introduced a suspended grade $\tilde{a}$ of the boundary fields $\Psi_a$:
\begin{equation}
\tilde{a}:=|\Psi_a|+1,
\end{equation}
where $|\Psi_a|$ is the $\mathbbm{Z}_2$--degree of the boundary field. Since the amplitudes are completely symmetric with respect to the bulk insertions we can introduce generating functions for the bulk perturbations which satisfy the following property:
\begin{equation}
B_{a_0\ldots a_m;i_1\ldots i_n}=\partial_{i_1}\ldots\partial_{i_n}\mathcal{F}_{a_0\ldots a_m}(\varphi)|_{\varphi=0}\qquad{\textrm{where}}\quad \partial_{i_k}=\frac{\partial}{\partial\varphi_{i_k}}
\end{equation}
To calculate $\mathcal{W}_{eff}$, we associate parameters $s_a=(u_a,v_a)$ to every open string insertion. The commuting $u_a$ are the parameters of the fermionic open string states, the $v_a$ are anticommuting and go with the bosons. The effective superpotential is then the sum of the symmetrized amplitudes with parameters $s_a$:
\begin{equation}
\mathcal{W}_{eff}(s;\varphi)=\sum_{m\geq 1}\frac{1}{m}s_{a_m}\ldots s_{a_1}\mathcal{A}_{a_1\ldots a_m}(\varphi),
\end{equation}
where 
\begin{equation}
\mathcal{A}_{a_0\ldots a_m}:=(m-1)!\mathcal{F}_{(a_0,\ldots,a_m)}:=\frac{1}{m}\sum_{\sigma\in S_m}\eta(\sigma;a_0,\ldots,a_m)\mathcal{F}_{a_{\sigma(0)}\ldots a_{\sigma(m)}},
\end{equation}
$\sigma$ is a permutation and $\eta$ is the sign one obtains from permuting the variables $s_a$. \\
To summarize, using the Kapustin--Li formula we can compute the lowest contributions to the effective superpotential and therefore get an independent, non--trivial check for deformation theory calculation.
\subsection{Branes which are not discussed in the examples}
In the following sections we will discuss in detail the deformations of matrix factorizations corresponding to Recknagel--Schomerus boundary states. We will start at the Gepner point and deform a single brane with bulk-- and boundary moduli. If the deformation theory algorithm terminates after a finite number of steps we will end up with a matrix factorization of the bulk--deformed Landau--Ginzburg superpotential. We will now mention some other constructions which will not be discussed in the remainder of the paper.
\subsubsection{Short Orbit Branes}
In \cite{Krefl:2008sj} the branes corresponding to Recknagel--Schomerus boundary states with maximal $L$--label have been identified to be mirror to the Lagrangians defined by the real hypersurface equations. For hypersurfaces with even degree these maximal branes are reducible, i.e they can be decomposed into "short orbit branes"\cite{Fuchs:2000gv,Brunner:2004zd}. In the matrix factorization language this means that one can define projectors $P_{\pm}$ and a pair of matrix factorizations:
\begin{equation}
Q_{\pm}=P_{\pm}QP_{\pm}
\end{equation} 
In \cite{Krefl:2008sj} these branes have been shown to exhibit BPS domain walls, and instanton numbers have been computed. Since the domain wall separates two brane vacua, this should be visible in the solutions of the F--term equations. So it is a natural question to ask whether this can be seen in a deformation theory calculation. There we immediately run into a problem. The Massey product algorithm works very well for deformations of a single matrix factorization. For the short orbit branes, however, it is suggestive that one has to consider $Q_{+}$ and $Q_{-}$ together since the pair already exists at the Gepner point. A generalization of the deformation theory algorithm to systems with multiple matrix factorizations has not been worked out and does not seem to be straight forward. Let us however report a suggestive observation. For one--parameter models one can define precisely one, and for two--parameter models at least one, set of fermionic charge $1$ boundary changing operators $\{\Psi_{+-},\Psi_{-+}\}$ from $Q_+$ to $Q_-$ and back such that the symmetric product
\begin{equation}
\label{twobrane-massey}
\Psi_{+-}\cdot\Psi_{-+}+\Psi_{-+}\cdot\Psi_{+-}
\end{equation}
gives a bulk deformation. It is tempting to conclude that the two brane vacua separated by a domain wall come from an F--term which is produced by deforming the combined $Q_+,Q_-$--system with the boundary changing open string states.
\subsubsection{"Incomplete" Deformations}
In all the examples we will discuss in the following sections, we will always deform the matrix factorization with all allowed open and closed string deformations, in order to get the full F--terms. It is an interesting question to ask what happens if we only turn on some of the (brane and bulk) moduli. We will now focus on an interesting special case which occurs quite frequently and has also played a role in \cite{Knapp:2008uw}. Consider a matrix factorization $Q$ of a Landau--Ginzburg superpotential at the Gepner point. Assume that there is an open string state $\Psi$ with the property:
\begin{equation}
\Psi^2=\phi\cdot\mathbbm{1},
\end{equation}
where $\phi$ is a bulk deformation. We can thus define a deformed matrix factorization,
\begin{equation}
Q_{def}^{lin}=Q+u\Psi,
\end{equation}
which has the following factorization property:
\begin{equation}
(Q_{def}^{lin})^2=W+u^2\phi
\end{equation}
Obviously, this is a matrix factorization of the bulk deformed Landau--Ginzburg superpotential $W+\varphi\phi$ if and only if the following constraint is satisfied:
\begin{equation}
\label{fakefterm}
u^2-\varphi=0
\end{equation}
Here, $\varphi$ is the bulk modulus. This immediately leads to a pair of matrix factorizations of the deformed Landau--Ginzburg potential:
\begin{equation}
\label{fakeqpm}
Q_{\pm}=Q\pm\sqrt{\varphi}\Psi
\end{equation}
In \cite{Knapp:2008uw} domain wall tensions have been computed for such branes. Note however that the constraint (\ref{fakefterm}) is (part of) an F--term if and only if the bulk deformation $\phi$ is also a physical open string state. If not, one would have to introduce a bulk deformation of the brane for $\phi$, and usually one finds that in this case the brane modulus is unobstructed and the F--terms are zero. On the other hand, (\ref{fakeqpm}) is a well-defined matrix factorization of the deformed Landau--Ginzburg potential which does not seem to care whether the bulk deformation is $Q$--exact or not. It comes from a deformation with a boundary modulus and by the "fake" F--term (\ref{fakefterm}) the deformed matrix factorization naturally comes as a pair $Q_{\pm}$. Thus, one has a very natural setup for BPS domain walls and in \cite{Knapp:2008uw} a non--zero domain wall tension has been computed for a such brane where the bulk deformation was indeed $Q$--exact. It is an obvious question to ask whether such domain walls related to partial deformations of a brane are qualitatively different to those which can be related to F--terms.\\
One could integrate (\ref{fakefterm}) to a cubic effective superpotential. However, if this equation is not associated to an obstruction the result does not match with the correlators one can compute with the Kapustin--Li formula. This implies that, in order to get the correct effective superpotential, one has to take into account the combined bulk--boundary deformation of the brane. Nevertheless, the equation (\ref{fakefterm}) seems to define two distinguished points in moduli space. \\
In our discussion of two--parameter hypersurfaces we will not highlight this type of linearly deformed matrix factorization. It exists whenever an open string state squares to one of the bulk deformations. As can be deduced from the examples, this happens quite often.
\subsubsection{Permutation Branes}
It is well--known that the the Recknagel--Schomerus boundary states do not always generate the full lattice of RR--charges. The branes which do the job are the (generalized) permutation branes \cite{Recknagel:2002qq}. They also have a convenient description in terms of matrix factorizations \cite{Brunner:2005fv,Enger:2005jk,Fredenhagen:2005an,Caviezel:2005th}. It would be interesting to discuss moduli and deformation theory of these branes. Since these matrix factorizations have a more complicated polynomial structure, calculations of open string states are technically more challenging, in particular if one intends to deal with large classes of examples and therefore has to rely on efficient computer code. We will postpone the discussion of permutation branes to future work.
\subsection{Summary of the Results}
In the following five sections we will discuss in great detail deformations and F-terms of tensor product branes on the mirrors of two--parameter hypersurfaces in weighted $\mathbbm{CP}^4$. Since this is quite technical we summarize the relevant steps here:
\begin{itemize}
\item For every model, go through the list of tensor product matrix factorizations corresponding to Recknagel--Schomerus boundary states at the Gepner point and compute the brane moduli.
\item Classify the branes according to the number and structure\footnote{By structure we mean their decomposition in terms of minimal model open string states (\ref{mmfer}), (\ref{mmbos}).} of the brane moduli. 
\item For each class, pick a specific tensor product brane (we choose the one with maximal $L$--label inside the respective class) and compute the higher order deformations of F--terms using the Massey product algorithm.
\item Integrate the F--terms to the effective superpotential.
\item Check the consistency of the results by computing two-- and three--point functions using the Kapustin--Li residue formula.
\end{itemize}
\subsubsection{Common Features}
Our main tool of calculation will be the the Massey product algorithm. This is a quite complicated procedure and we have decided to display the calculations in great detail in order to expose its strengths and drawbacks. Depending on whether moduli are obstructed or not the calculation will proceed in different ways. There are two extreme cases of what can happen.
\begin{itemize}
\item The algorithm terminates after a finite number of steps. This can happen for obstructed and unobstructed moduli. In order to get higher order deformations there must be Massey products which yield something $Q$--exact. If all higher products are zero or obstructed from some order on, the algorithm will terminate at some point because the products are no longer defined. If we have for example deformations $\alpha_{\vm}$ up to $|\vm|=3$ there are no more higher products to compute at order $|\vm|=7$ because there simply are no $\alpha$'s left we could multiply. For unobstructed moduli this typically happens when many higher products are zero. If the moduli are obstructed then this happens because contributions to the F--terms do not give new deformations and remove certain elements in the basis $B'$ which encode the information about the allowed Massey products.
\item The brane modulus is unobstructed and the algorithm never terminates. This happens whenever recurring patterns appear in the algorithm. By recurring patterns we mean that Massey products keep producing the same $Q$--exact expressions which lead to the same deformations at ever higher orders. A nice example for this case is the brane $L=(5,2,2,1,0)$ on the mirror of the degree $12$ hypersurface in $\mathbbm{P}(12234)$. This example is discussed in section \ref{sec-12234twomod}. 
\end{itemize}
Note that it hardly ever happens that these extreme cases appear in an isolated fashion for branes with several open moduli. Usually the deformation theory problem will be a combination thereof. In particular it is notoriously difficult to identify recurring structures and to decide whether the algorithm terminates or not. This is why, in some complicated cases, we only managed to make precise statements up to a certain order in deformation theory.
\subsubsection{Issues and new aspects}
We now comment on some interesting issues which arise in the deformation theory calculations.\\
For some branes we observe a new phenomenon which has not been encountered in minimal model examples. In our examples, the branes $L=(6,2,2,2,0)$ (see section \ref{sec-12227fourmod}) and $L=(4,2,2,2,0)$ (see section \ref{sec-12227threemod}) on the mirror of the hypersurface $\mathbbm{P}(12227)[14]$ and the brane $L=(8,8,8,0,0)$ (see section \ref{sec-11169fourmod}) on the mirror of  $\mathbbm{P}(12227)[14]$  have obstructed moduli but the deformed matrix factorization does not square to (\ref{eq-qdeffact1}), but rather to:
\begin{equation}
\label{bulk-strange}
Q_{def}^2=W+\sum_{i=1}^N\bar{f}_i(u_i,\varphi_i)\Phi_i+\tilde{f}_i(u_i,\varphi_i)\lambda_i
\end{equation}
There are two issues. Firstly, the prefactors $\bar{f}_i(u_i,\varphi_i)$ of the $\Phi_i$ which determine the F--terms may not be easily separated. Given an open string state $\Phi_k$ it may happen that the moduli dependent prefactors of the matrix entries may differ up to the F--terms of the other $\Phi_i$. In principle, this is not inconsistent but the factorization $f_i\Phi_i$ only works up to F--terms. The second new issue concerns the third summand on the righthand side of (\ref{bulk-strange}). The $\lambda_i$ are not in the $Q$--cohomology (i.e. they are neither $Q$--closed nor $Q$--exact!). However, the moduli--dependent prefactors of the $\lambda_i$ are (combinations of) the F--terms associated to the obstructions. Therefore, also these additional terms are not inconsistent and do not contain extra information, at least in all examples where we have found them. Note that this phenomenon is not related to the extension of the algorithm to bulk moduli since it also occurs when these are turned off.\\
A further interesting novel feature has occurred for the branes $L=(6,2,2,2,0)$ (see section \ref{sec-12227fourmod}) and $L=(4,2,2,2,0)$ (see section \ref{sec-12227threemod}) on the mirror of $\mathbbm{P}(12227)[14]$. For these branes the F--terms can only be integrated for a particular choice of higher order deformations $\alpha_{\vm}$. This is unexpected since different choices of $\alpha_{\vm}$ should not produce qualitatively different results. In particular the physical information in the F--terms should not change. Note however, that the deformation theory only sees the obstructions, which are the F--terms, but does not "know" that these constraints are actually the critical locus of $\mathcal{W}_{eff}$. The existence of an effective superpotential does not enter into the algorithm or the mathematical structure behind it. From that point of view it is actually quite remarkable that the constraints one gets are really integrable and it may well be that this property does not persist for arbitrary choices of deformations. Still, it is a very interesting question to find out why some branes exhibit this problem while others do not. The fact that the problematic branes also have the unusual factorization property (\ref{bulk-strange}) implies that the new phenomena are not independent. \\
A big inconvenience of the Massey product algorithm is that the choice of higher deformations $\alpha_{\vm}$ is not unique. No distinguished basis of deformations is known. In our calculation we loosely stuck to the rule that the deformation matrices should have as few entries as possible and that, if possible, all monomial entries should be equal. It could be that a different choice of deformations would exhibit recurring structures more clearly or may even cause the algorithm to terminate earlier than with another choice. However, in most cases there are usually only a few possible choices for a new deformation which lead to the same results. The physical information which is contained in the F--terms should of course not depend on the choice of higher order deformations.
\section{The model $\mathbbm{P}(11222)[8]/\mathbbm{Z}_8\times (\mathbbm{Z}_4)^2$}
The Landau--Ginzburg superpotential (at the Gepner point) associated to this degree $8$ hypersurface is:
\begin{equation}
W=x_1^8+x_2^8+x_3^4+x_4^4+x_5^4
\end{equation}
To get the mirror we impose the following $\mathbbm{Z}_8\times (\mathbbm{Z}_4)^2$ orbifold action:
\begin{eqnarray}
g_1:&&(1,7,0,0,0)\nonumber\\
g_2:&&(2,0,6,0,0)\nonumber\\
g_3:&&(2,0,0,6,0),
\end{eqnarray}
where $g_j:\:x_i\rightarrow e^{2\pi i g_j^i/d}x_i$.\\
For later convenience we also give the two bulk deformations:
\begin{eqnarray}
\phi_1&=&x_1^4x_2^4\nonumber\\
\phi_2&=&x_1x_2x_3x_4x_5 
\end{eqnarray}
In Table \ref{tab11222mod} we list the branes and moduli of this model.
\subsection{Discussion of Moduli}
Let us now discuss in more detail the moduli of the "maximal" brane with label $L=(3,3,1,1,1)$. As we can see from the tables, this brane has the maximum number of moduli. Furthermore, we observe that, as we decrease the entries of the $L$--label, the number of moduli changes but their structure does not. For that reason, if we have discussed the moduli of this maximal brane, we have discussed all the others as well. The only thing that can change are the entries of the fermionic minimal model components of the brane. Note that this change is mild in the sense that the Massey product of an open string state with itself (not with others, however!) is always the same for a modulus with a definite structure.\\
So let us discuss the structure of the boundary moduli for the $(3,3,1,1,1)$ brane.  We denote the open string state labelled by $\frac{1}{2}^1\otimes\frac{1}{2}^1\otimes 0^1\otimes 0^1\otimes 0^1$ with $\Psi_1$. It looks as follows:
\begin{equation}
\label{11222mod1}
\Psi_1=\left(\begin{array}{cc}
0&x_1^2\\
-x_1^2&0
\end{array}\right)\otimes
\left(\begin{array}{cc}
0&x_2^2\\
-x_2^2&0
\end{array}\right)\otimes
\left(\begin{array}{cc}
0&1\\
-1&0
\end{array}\right)\otimes
\left(\begin{array}{cc}
0&1\\
-1&0
\end{array}\right)\otimes
\left(\begin{array}{cc}
0&1\\
-1&0
\end{array}\right)
\end{equation}
This open modulus also appears on the other branes which have moduli. The only difference for the other branes is that the first two matrices can also have the form\footnote{Of course, the $R$--charges remain the same, as indicated in the label.} $\left(\begin{array}{cc}0&x\\-x^3&0\end{array}\right)$ or $\left(\begin{array}{cc}0&1\\-x^4&0\end{array}\right)$, depending on the values $k_1,k_2$ in the matrix factorization. The second open string state which has label $\frac{1}{2}^0\otimes\frac{1}{2}^0\otimes 0^1\otimes 0^1\otimes 0^1$ has a similar structure:
\begin{equation}
\label{11222mod2}
\Psi_2=\left(\begin{array}{cc}
x_1^2&0\\
0&x_1^2
\end{array}\right)\otimes
\left(\begin{array}{cc}
x_2^2&0\\
0&x_1^2
\end{array}\right)\otimes
\left(\begin{array}{cc}
0&1\\
-1&0
\end{array}\right)\otimes
\left(\begin{array}{cc}
0&1\\
-1&0
\end{array}\right)\otimes
\left(\begin{array}{cc}
0&1\\
-1&0
\end{array}\right)
\end{equation}
This open string state is the same for every brane with the label $\frac{1}{2}^0\otimes\frac{1}{2}^0\otimes 0^1\otimes 0^1\otimes 0^1$.
\subsection{Obstructions}
Let us now give the explicit expressions for the charge $0$, $\mathbbm{Z}_2$--even open string states which encode the obstructions to the deformations with (\ref{11222mod1}) and (\ref{11222mod2}). Our first obstruction state $\Phi_1$ is nothing but the bulk deformation $\phi_2$:
\begin{equation}
\label{11222ob1}
\Phi_1=\left(\begin{array}{cc}
x_1&0\\
0&x_1
\end{array}\right)\otimes
\left(\begin{array}{cc}
x_2&0\\
0&x_2
\end{array}\right)\otimes
\left(\begin{array}{cc}
x_3&0\\
0&x_3
\end{array}\right)\otimes
\left(\begin{array}{cc}
x_4&0\\
0&x_4
\end{array}\right)\otimes
\left(\begin{array}{cc}
x_5&0\\
0&x_5
\end{array}\right)
\end{equation}
It has the structure $\frac{1}{4}^0\otimes\frac{1}{4}^0\otimes\frac{1}{2}^0\otimes\frac{1}{2}^0\otimes\frac{1}{2}^0$. This state is Serre dual to (\ref{11222mod1}).\\
The obstruction which is Serre dual to (\ref{11222mod2}) has the structure  $\frac{1}{4}^1\otimes\frac{1}{4}^1\otimes\frac{1}{2}^0\otimes\frac{1}{2}^0\otimes\frac{1}{2}^0$:
\begin{equation}
\label{11222ob2}
\Phi_2=\left(\begin{array}{cc}
0&x_1\\
-x_1&0
\end{array}\right)\otimes
\left(\begin{array}{cc}
0&x_2\\
-x_2&0
\end{array}\right)\otimes
\left(\begin{array}{cc}
x_3&0\\
0&x_3
\end{array}\right)\otimes
\left(\begin{array}{cc}
x_4&0\\
0&x_4
\end{array}\right)\otimes
\left(\begin{array}{cc}
x_5&0\\
0&x_5
\end{array}\right)
\end{equation}
\subsection{Massey Products and F--terms}
In this section we compute Massey products and F--terms for the branes in our model. We have to distinguish two cases. There are three branes which have both moduli (\ref{11222mod1}) and (\ref{11222mod2}) and three which have only the modulus (\ref{11222mod1}). In order to capture the relevant information it will be enough to discuss only one example of each class. We choose the brane with the maximal $L$--label for each case.
\subsubsection{Two Brane Moduli}
The maximal brane with two moduli has label $L=(3,3,1,1,1)$. The brane moduli are those given in (\ref{11222mod1}) and (\ref{11222mod2}). We observe that the bulk modulus $\phi_2=x_1x_2x_3x_4x_5$ is always in the boundary cohomology and contributes to the $F$--term associated with the obstruction (\ref{11222ob1}). The bulk modulus $\phi_1=x_1^4x_2^4$ is $Q$--exact and we have to take care of this by adding a deformation $\alpha_{(1,0,0)}$ such that $\{Q,\alpha_{(1,0,0)}\}=\phi_1$. One easily checks that\footnote{This choice is not unique.}:
\begin{equation}
\alpha_{(1,0,0)}=x_2^4\bar{\eta}_1
\end{equation}
Now we are ready to compute the Massey products to second order in deformation theory. The following ones are non--zero:
\begin{eqnarray}
y_{(1,0,1)}&=&\{\alpha_{(1,0,0)},\Psi_1\}=\beta_1(x_1,x_2)\nonumber\\
y_{(0,2,0)}&=&\Psi_1\cdot\Psi_1=-x_1^4x_2^4\mathbbm{1}\nonumber\\
y_{(0,1,1)}&=&\{\Psi_1,\Psi_2\}=\beta_2(x_1,x_2)\nonumber\\
y_{(0,0,2)}&=&\Psi_2\cdot\Psi_2=x_1^4x_2^4\mathbbm{1}
\end{eqnarray}
Here we wrote $\beta_i(x_i)$ for non--diagonal $Q$--exact states, indicating their variable dependence in parentheses. We note that all these Massey products are $Q$--exact and have to be canceled by deformations at order $3$. We do not need to do this explicitly since all the exact states we get as well as the higher deformations and brane moduli do not contain the variables $x_3,x_4,x_5$, whereas they appear in both obstructions. Therefore, at any order, no Massey product can ever be proportional to an obstruction. Therefore we only get deformations but no obstructions and thus no contribution to the F--terms. The only F--term we have is:
\begin{equation}
f_1:\qquad \varphi_2=0,
\end{equation}
where $\varphi_1$ is the bulk modulus associated to $\phi_2$. This tells us that the bulk deformation $\phi_2$ is not allowed in the presence of our brane. The F--terms can be integrated to the following effective superpotential:
\begin{equation}
\mathcal{W}_{eff}=u_1\varphi_2
\end{equation}
Furthermore we conclude that the brane $L=(3,3,1,1,1)$ has two unobstructed boundary moduli. \\
The above structure arguments were enough to determine the full F--terms. What we do not know from this reasoning is whether we need a finite or an infinite number of deformations to obtain a matrix factorization of the deformed Landau--Ginzburg superpotential. So, let us do some more steps in deformation theory. At order two, we get four new deformations, two of which are very simple:
\begin{eqnarray}
\alpha_{(0,2,0)}=-\alpha_{(0,0,2)}&=&\alpha_{(1,0,0)}\nonumber\\
\alpha_{(1,1,0)}&=&-x_1^2x_2^2(\eta_5-\bar{\eta}_5)(\eta_4-\bar{\eta}_4)(\eta_3-\bar{\eta}_3)\bar{\eta}_2\eta_2 \nonumber\\
\alpha_{(0,1,1)}&=&-2x_2^4(\eta_2-\bar{\eta}_2)\bar{\eta}_1\eta_1
\end{eqnarray} 
These deformations are not unique. For our particular choice, we get the following Massey products at order $3$:
\begin{eqnarray}
y_{(1,2,0)}&=&\{\alpha_{(1,0,0)},\alpha_{(0,2,0)}\}+\{\alpha_{(1,1,0)},\Psi_1\}=\frac{1}{2}\beta_2(x_1,x_2)\nonumber\\
y_{(1,1,1)}&=&\{\alpha_{(1,1,0)},\Psi_2\}+\{\alpha_{(1,0,0)},\alpha_{(0,1,1)}\}=\beta_3(x_1,x_2)\nonumber\\
y_{(0,3,0)}&=&\{\alpha_{(0,2,0)},\Psi_1\}=\beta_2(x_1,x_2)\nonumber\\
y_{(0,2,1)}&=&\{\alpha_{(0,2,0)},\Psi_2\}+\{\alpha_{(0,1,1)},\Psi_1\}=\beta_4(x_1,x_2)\nonumber\\
y_{(0,1,2)}&=&\{\alpha_{(0,1,1)},\Psi_2\}+\{\alpha_{(0,0,2)},\Psi_1\}=-\beta_1(x_1,x_2)
\end{eqnarray}
All the other possible Massey products at this order are $0$. Three of the five new deformations are easy to find:
\begin{eqnarray}
\alpha_{(1,2,0)}&=&\frac{1}{2}\alpha_{(0,1,1)}\nonumber\\
\alpha_{(0,3,0)}=-\alpha_{(0,1,2)}&=&\alpha_{(1,1,0)}\nonumber\\
\alpha_{(1,1,1)}&=&-2x_2^4\bar{\eta}_1\eta_2\bar{\eta}_2\nonumber\\
\alpha_{(0,2,1)}&=&-2x_1^2x_2^2(\eta_5-\bar{\eta}_5)(\eta_4-\bar{\eta}_4)(\eta_3-\bar{\eta}_3)\eta_2(\eta_1-\bar{\eta}_1)
\end{eqnarray}
With this specific choice there are the following nonzero Massey products at order $4$:
\begin{eqnarray}
y_{(2,2,0)}&=&\alpha_{(1,1,0)}\cdot\alpha_{(1,1,0)}+\{\alpha_{(1,0,0)},\alpha_{(1,2,0)}\}=\frac{1}{2}\beta_3(x_1,x_2)\nonumber\\
y_{(1,3,0)}&=&\{\alpha_{(1,1,0)},\alpha_{(0,2,0)}\}+\{\alpha_{(1,2,0)},\Psi_1\}+\{\alpha_{(1,0,0)},\alpha_{(0,3,0)}\}=\frac{1}{2}\beta_4\nonumber\\
y_{(0,4,0)}&=&\alpha_{(0,2,0)}\cdot\alpha_{(0,2,0)}+\{\alpha_{(0,3,0)},\Psi_1\}=\frac{1}{2}\beta_2(x_1,x_2)\nonumber\\
y_{(0,3,1)}&=&\{\alpha_{(0,2,0)},\alpha_{(0,1,1)}\}+\{\alpha_{(0,3,0)},\Psi_2\}+\{\alpha_{(0,2,1)},\Psi_1\}=\beta_5(x_1,x_2)\nonumber\\
y_{(0,2,2)}&=&\{\alpha_{(0,2,0)},\alpha_{(0,0,2)}\}+\alpha_{(0,1,1)}\cdot\alpha_{(0,1,1)}+\{\alpha_{(0,2,1)},\Psi_2\}+\{\alpha_{(0,1,2)},\Psi_1\}=\beta_6(x_1,x_2)\nonumber\\
y_{(0,1,3)}&=&\{\alpha_{(0,1,1)},\alpha_{(0,0,2)}\}+\{\alpha_{(0,1,2)},\Psi_1\}=-\beta_3(x_1,x_2)
\end{eqnarray}
Four of these six new deformations are identical to deformations at lower order:
\begin{eqnarray}
\alpha_{(2,2,0)}=-\frac{1}{2}\alpha_{(0,1,3)}&=&\frac{1}{2}\alpha_{(1,1,1)}\nonumber\\
\alpha_{(1,3,0)}&=&\frac{1}{2}\alpha_{(0,2,1)}\nonumber\\
\alpha_{(0,4,0)}&=&\frac{1}{2}\alpha_{(0,1,1)}\nonumber\\
\alpha_{(0,3,1)}&=&-2x_2^4\eta_2\bar{\eta}_1\bar{\eta}_2\nonumber\\
\alpha_{(0,2,2)}&=&3x_2^4\eta_2\bar{\eta}_1\eta_1+x_2^4\bar{\eta}_2\bar{\eta}_1\eta_1
\end{eqnarray}
Before we give up, let us list the non--zero Massey products at order five in deformation theory:
\begin{eqnarray}
y_{(1,4,0)}&=&\{\alpha_{(1,1,0)},\alpha_{(0,3,0)}\}+\{\alpha_{(0,2,0)},\alpha_{(1,2,0)}\}+\{\alpha_{(1,3,0)},\Psi_1\}+\{\alpha_{(0,4,0)},\alpha_{(1,0,0)}\}=\beta_7(x_1,x_2)\nonumber\\
y_{(1,3,1)}&=&\{\alpha_{(1,1,0)},\alpha_{(0,2,1)}\}+\{\alpha_{(0,2,0)},\alpha_{(1,1,1)}\}+\{\alpha_{(0,1,1)},\alpha_{(1,2,0)}\}+\{\alpha_{(1,3,0)},\Psi_2\}\nonumber\\
&&+\{\alpha_{(0,3,1)},\alpha_{(1,0,0)}\}=\beta_8(x_1,x_2)\nonumber\\
y_{(0,5,0)}&=&\{\alpha_{(0,2,0)},\alpha_{(0,3,0)}\}+\{\alpha_{(0,4,0)},\Psi_1\}=\frac{1}{2}\beta_4(x_1,x_2)\nonumber\\
y_{(0,4,1)}&=&\{\alpha_{(0,2,0)},\alpha_{(0,2,1)}\}+\{\alpha_{(0,1,1)},\alpha_{(0,3,0)}\}+\{\alpha_{(0,3,1)},\Psi_1\}+\{\alpha_{(0,4,0)},\Psi_2\}=-2\beta_1(x_1,x_2)\nonumber\\
y_{(0,3,2)}&=&\{\alpha_{(0,2,0)},\alpha_{(0,1,2)}\}+\{\alpha_{(0,1,1)},\alpha_{(0,2,1)}\}+\{\alpha_{(0,0,2)},\alpha_{(0,3,0)}\}+\{\alpha_{(0,2,2)},\Psi_1\}\nonumber\\
&&+\{\alpha_{(0,3,1)},\Psi_2\}=-\frac{3}{2}\beta_4(x_1,x_2)
\end{eqnarray}
We notice a recurring pattern: certain deformations (or linear combinations thereof) are produced at every order. This suggests that the algorithm keeps going on forever. This is however no proof since these expressions are not produced in the same way at every order but rather come out of increasingly complicated combinations of deformations. Furthermore it is not excluded that there exists a choice of deformations for which the algorithm terminates after a finite number of steps.
\subsubsection*{Correlators}
We can test our results by computing three--point amplitudes on the disk and bulk-to-boundary two--point functions, using the residue formula of Kapustin and Li. For this brane, there is only one non--zero correlator:
\begin{equation}
\langle\Psi_1\phi_2\rangle=1
\end{equation}
This correlator is consistent with the single F--term and the effective superpotential we have.
\subsubsection{One Brane Modulus}
Let us now discuss the tensor product brane $L=(3,1,1,1,1)$, which has only one brane modulus $\Psi_1$ which has the same charge decomposition as $(\ref{11222mod1})$. Furthermore, we still have the boundary deformation which produces the bulk deformation $\phi_1$, which is the same as in the example with two brane moduli but will now be labelled as $\alpha_{(1,0)}$. There are only two non--zero Massey products at order $2$:
\begin{eqnarray}
y_{(1,1)}&=&\{\alpha_{(1,0)},\Psi_1\}=\beta_1(x_1,x_2)\nonumber\\
y_{(0,2)}&=&\Psi_1\cdot\Psi_1=-x_1^4x_2^4\mathbbm{1}=\beta_2(x_1,x_2)
\end{eqnarray}
Again, all possible deformations of $Q$ only depend on $x_1,x_2$ and therefore their higher products can never contribute to the obstructions. As in the two--moduli case, the only F--term is $\varphi_2=0$, which renders one bulk deformation inconsistent. The effective superpotential is $\mathcal{W}_{eff}=u_1\varphi_2$. The single boundary modulus of this brane is unobstructed. This example, or rather the brane $L=(1,1,1,1,1)$ has already been discussed in \cite{Hori:2004ja}. \\
Let us furthermore calculate the deformations up to order five in order to collect evidence that the deformation theory algorithm does not terminate. At order two we have gained two more deformations, $\alpha_{(1,1)}$ and $\alpha_{(0,2)}$:
\begin{eqnarray}
\alpha_{(0,2)}&=&\alpha_{(1,0)}\nonumber\\
\alpha_{(1,1)}&=&-x_1^2x_2^2(\eta_5-\bar{\eta}_5)(\eta_4-\bar{\eta}_4)(\eta_3-\bar{\eta}_3)\bar{\eta}_2\eta_2 
\end{eqnarray}
At order three, the following Massey products are non--zero:
\begin{eqnarray}
y_{(1,2)}&=&\{\alpha_{(0,2)},\alpha_{(1,0)}\}+\{\alpha_{(1,1)},\Psi_1\}=\beta_3(x_1,x_2)\nonumber\\
y_{(0,3)}&=&\{\alpha_{(0,2)},\Psi_1\}=\beta_1(x_1,x_2)
\end{eqnarray}
We make the following choice for the $\alpha's$:
\begin{eqnarray}
\alpha_{(0,3)}&=&\alpha_{(1,1)}\nonumber\\
\alpha_{(1,2)}&=&x_2^4\eta_1\eta_2\bar{\eta}_1+x_2^6\eta_1\bar{\eta}_1\bar{\eta}_2
\end{eqnarray}
 At order four the following Massey products are non--zero:
\begin{eqnarray}
y_{(2,2)}&=&\{\alpha_{(1,2)},\alpha_{(1,0)}\}+\alpha_{(1,1)}\cdot\alpha_{(1,1)}=\beta_4(x_1,x_2)\nonumber\\
y_{(1,3)}&=&\{\alpha_{(1,1)},\alpha_{(0,2)}\}+\{\alpha_{(1,2)},\Psi_1\}+\{\alpha_{(0,3)},\alpha_{(1,0)}\}=\beta_5(x_1,x_2)\nonumber\\
y_{(0,4)}&=&\alpha_{(0,2)}\cdot\alpha_{(0,2)}+\{\alpha_{(0,3)},\Psi_1\}=\beta_3(x_1,x_2)
\end{eqnarray}
There are three new deformations:
\begin{eqnarray}
\alpha_{(0,4)}&=&\alpha_{(1,2)}\nonumber\\
\alpha_{(1,3)}&=&x_1^2(\eta_5-\bar{\eta}_5)(\eta_4-\bar{\eta}_4)(\eta_3-\bar{\eta}_3)\eta_2(\eta_1-\bar{\eta}_1) \nonumber\\
\alpha_{(2,2)}&=&-x_2^4\bar{\eta}_1\bar{\eta}_2\eta_2
\end{eqnarray}
With our particular choice of deformations, we find the following non--zero Massey products at order five:
\begin{eqnarray}
y_{(2,3)}&=&\{\alpha_{(1,1)},\alpha_{(1,2)}\}+\{\alpha_{(2,2)},\Psi_1\}+\{\alpha_{(1,3)},\alpha_{(1,0)}\}=-\beta_1(x_1,x_2)\nonumber\\
y_{(1,4)}&=&\{\alpha_{(1,1)},\alpha_{(0,3)}\}+\{\alpha_{(0,2)},\alpha_{(1,2)}\}+\{\alpha_{(1,3)},\Psi_1\}+\{\alpha_{(0,4)},\alpha_{(1,0)}\}=\beta_4(x_1,x_2)\nonumber\\
y_{(0,5)}&=&\{\alpha_{(0,2)},\alpha_{(0,3)}\}+\{\alpha_{(0,4)},\Psi_1\}=\beta_5(x_1,x_2)
\end{eqnarray}
Just like in the two--moduli case, deformations from lower orders seem to reappear at higher orders which suggests that the procedure never stops.
\subsubsection*{Correlators}
The Kapustin--Li formula only yields one non--vanishing correlator:
\begin{equation}
\langle\Psi_1\phi_2\rangle=1
\end{equation}
This is again consistent with the F--term $\varphi_2=0$ and $\mathcal{W}_{eff}$.
\section{The model $\mathbbm{P}(11226)[12]/\mathbbm{Z}_{12}\times (\mathbbm{Z}_6)^2$}
Here we have the following superpotential:
\begin{equation}
W=x_1^{12}+x_2^{12}+x_3^6+x_4^6+x_5^2
\end{equation}
There is a $\mathbbm{Z}_{12}\times (\mathbbm{Z}_6)^2$ orbifold action which has the following generators:
\begin{eqnarray}
g_1:&&(1,11,0,0,0)\nonumber\\
g_2:&&(2,0,10,0,0)\nonumber\\
g_3:&&(2,0,0,10,0)
\end{eqnarray}
The two bulk moduli are:
\begin{eqnarray}
\phi_1&=&x_1^6x_2^6\nonumber\\
\phi_2&=&x_1x_2x_3x_4x_5
\end{eqnarray}
Via the equations of motion of $x_5$ there is an alternative way to write the modulus $\phi_2$. We can also choose $\phi_2=x_1^2x_2^2x_3^2x_4^2$.\\
We list the tensor product branes which have moduli in table \ref{tab11226mod}.
\subsection{Discussion of Moduli}
The branes in this model can have at most four moduli. We will give them explicitly for the maximal branes labeled by $L=(5,5,2,2,0)$. The modulus with labels $\frac{1}{6}^1\otimes\frac{1}{6}^1\otimes\frac{1}{3}^0\otimes\frac{1}{3}^0\otimes 0^1$ looks as follows:
\begin{equation}
\label{11226mod1}
\Psi_1=\left(\begin{array}{cc}
0&x_1\\
-x_1&0
\end{array}\right)\otimes
\left(\begin{array}{cc}
0&x_2\\
-x_2&0
\end{array}\right)\otimes
\left(\begin{array}{cc}
x_3&0\\
0&x_3
\end{array}\right)\otimes
\left(\begin{array}{cc}
x_4&0\\
0&x_4
\end{array}\right)\otimes
\left(\begin{array}{cc}
0&1\\
-1&0
\end{array}\right)
\end{equation}
The modulus with labels $\frac{1}{6}^0\otimes\frac{1}{6}^0\otimes\frac{1}{3}^0\otimes\frac{1}{3}^0\otimes 0^1$ has the following form:
\begin{equation}
\label{11226mod2}
\Psi_2=\left(\begin{array}{cc}
x_1&0\\
0&x_1
\end{array}\right)\otimes
\left(\begin{array}{cc}
x_2&0\\
0&x_2
\end{array}\right)\otimes
\left(\begin{array}{cc}
x_3&0\\
0&x_3
\end{array}\right)\otimes
\left(\begin{array}{cc}
x_4&0\\
0&x_4
\end{array}\right)\otimes
\left(\begin{array}{cc}
0&1\\
-1&0
\end{array}\right)
\end{equation}
The third modulus with labels $\frac{1}{2}^1\otimes\frac{1}{2}^1\otimes 0^1\otimes 0^1\otimes 0^1$ looks like this:
\begin{equation}
\label{11226mod3}
\Psi_3=\left(\begin{array}{cc}
0&x_1^3\\
-x_1^3&0
\end{array}\right)\otimes
\left(\begin{array}{cc}
0&x_2^3\\
-x_2^3&0
\end{array}\right)\otimes
\left(\begin{array}{cc}
0&1\\
-1&0
\end{array}\right)\otimes
\left(\begin{array}{cc}
0&1\\
-1&0
\end{array}\right)\otimes
\left(\begin{array}{cc}
0&1\\
-1&0
\end{array}\right)
\end{equation}
The fourth modulus with labels  $\frac{1}{2}^0\otimes\frac{1}{2}^0\otimes 0^1\otimes 0^1\otimes 0^1$ is represented by the following $32\times 32$ matrix:
\begin{equation}
\label{11226mod4}
\Psi_4=\left(\begin{array}{cc}
x_1^3&0\\
0&x_1^3
\end{array}\right)\otimes
\left(\begin{array}{cc}
x_2^3&0\\
0&x_2^3
\end{array}\right)\otimes
\left(\begin{array}{cc}
0&1\\
-1&0
\end{array}\right)\otimes
\left(\begin{array}{cc}
0&1\\
-1&0
\end{array}\right)\otimes
\left(\begin{array}{cc}
0&1\\
-1&0
\end{array}\right)
\end{equation}
\subsection{Obstructions}
For the $L=(5,5,2,2,0)$ brane there are four bosonic open string states which are the obstructions to the deformations of (\ref{11226mod1})--(\ref{11226mod4}).\\
The obstruction Serre dual to (\ref{11226mod1}) has structure $\frac{2}{3}^0\otimes\frac{2}{3}^0\otimes\frac{1}{3}^1\otimes\frac{1}{3}^1\otimes0^0$:
\begin{equation}
\label{11226ob1}
\Phi_1=\left(\begin{array}{cc}
x_1^4&0\\
0&x_1^4
\end{array}\right)\otimes
\left(\begin{array}{cc}
x_2^4&0\\
0&x_2^4
\end{array}\right)\otimes
\left(\begin{array}{cc}
0&x_3\\
-x_3&0
\end{array}\right)\otimes
\left(\begin{array}{cc}
0&x_4\\
-x_4&0
\end{array}\right)\otimes
\left(\begin{array}{cc}
1&0\\
0&1
\end{array}\right)
\end{equation}
The obstruction Serre dual to (\ref{11226mod2}) has structure $\frac{2}{3}^1\otimes\frac{2}{3}^1\otimes\frac{1}{3}^1\otimes\frac{1}{3}^1\otimes0^0$:
\begin{equation}
\label{11226ob2}
\Phi_2=\left(\begin{array}{cc}
0&x_1^4\\
-x_1^4&0
\end{array}\right)\otimes
\left(\begin{array}{cc}
0&x_2^4\\
-x_2^4&0
\end{array}\right)\otimes
\left(\begin{array}{cc}
0&x_3\\
-x_3&0
\end{array}\right)\otimes
\left(\begin{array}{cc}
0&x_4\\
-x_4&0
\end{array}\right)\otimes
\left(\begin{array}{cc}
1&0\\
0&1
\end{array}\right)
\end{equation}
The obstruction corresponding to (\ref{11226mod3}) has structure $\frac{1}{3}^0\otimes\frac{1}{3}^0\otimes\frac{2}{3}^0\otimes\frac{2}{3}^0\otimes0^0$:
\begin{equation}
\label{11226ob3}
\Phi_3=\left(\begin{array}{cc}
x_1^2&0\\
0&x_1^2
\end{array}\right)\otimes
\left(\begin{array}{cc}
x_2^2&0\\
0&x_2^2
\end{array}\right)\otimes
\left(\begin{array}{cc}
x_3^2&0\\
0&x_3^2
\end{array}\right)\otimes
\left(\begin{array}{cc}
x_4^2&0\\
0&x_4^2
\end{array}\right)\otimes
\left(\begin{array}{cc}
1&0\\
0&1
\end{array}\right)
\end{equation}
This is the bulk deformation $x_1^2x_2^2x_3^2x_4^2$. Finally we have a charge $2$ boson $\frac{1}{3}^1\otimes\frac{1}{3}^1\otimes\frac{2}{3}^0\otimes\frac{2}{3}^0\otimes0^0$ which is Serre dual to (\ref{11226mod4}):
\begin{equation}
\label{11226ob4}
\Phi_4=\left(\begin{array}{cc}
0&x_1^2\\
-x_1^2&0
\end{array}\right)\otimes
\left(\begin{array}{cc}
0&x_2^2\\
-x_2^2&0
\end{array}\right)\otimes
\left(\begin{array}{cc}
x_3^2&0\\
0&x_3^2
\end{array}\right)\otimes
\left(\begin{array}{cc}
x_4^2&0\\
0&x_4^2
\end{array}\right)\otimes
\left(\begin{array}{cc}
1&0\\
0&1
\end{array}\right)
\end{equation}
\subsection{Massey Products and F--terms}
Now we discuss deformations and obstructions of the branes on this hypersurface. There are five classes of branes.
\subsubsection{Four Brane Moduli}
The brane with maximal $L$--label $(5,5,2,2,0)$ has four moduli which are given explicitly in (\ref{11226mod1})--(\ref{11226mod4}). Furthermore all tensor product branes with labels $L_1,L_2\geq 4$ and $L_3,L_4=2$ have four moduli. The bulk deformation $\phi_1=x_1^6x_2^6$ is $Q$--exact for every brane, whereas the other bulk modulus, written as $\phi_2=x_1^2x_2^2x_3^2x_4^2$, coincides with the boundary modulus (\ref{11226ob3}). In order to take care of the $Q$--exact bulk deformation we define:
\begin{equation}
\alpha_{(1,0,0,0,0)}=x_2^6\bar{\eta}_1
\end{equation}
There are several non--vanishing Massey products at order $2$:
\begin{eqnarray}
y_{(1,1,0,0,0)}&=&\{\alpha_{(1,0,0,0,0)},\Psi_1\}=\beta_1(x_1,x_2,x_3,x_4)\nonumber\\
y_{(1,0,0,1,0)}&=&\{\alpha_{(1,0,0,0,0)},\Psi_3\}=\beta_2(x_1,x_2)\nonumber\\
y_{(0,2,0,0,0)}&=&\Psi_1\cdot\Psi_1=x_1^2x_2^2x_3^2x_4^2\:\mathbbm{1}=\Phi_3 \nonumber\\
y_{(0,1,1,0,0)}&=&\{\Psi_1,\Psi_2\}=-2\Phi_4\nonumber\\
y_{(0,1,0,1,0)}&=&\{\Psi_1,\Psi_3\}=2\Phi_1\nonumber\\
y_{(0,1,0,0,1)}&=&\{\Psi_1,\Psi_4\}=2\Phi_2\nonumber\\
y_{(0,0,2,0,0)}&=&\Psi_2\cdot\Psi_2=-x_1^2x_2^2x_3^2x_4^2\:\mathbbm{1}=-\Phi_3 \nonumber\\
y_{(0,0,1,1,0)}&=&\{\Psi_2,\Psi_3\}=-2\Phi_2\nonumber\\
y_{(0,0,1,0,1)}&=&\{\Psi_2,\Psi_4\}=2\Phi_1\nonumber\\
y_{(0,0,0,2,0)}&=&-x_1^6x_2^6\mathbbm{1}=\beta_3(x_1,x_2)\nonumber\\
y_{(0,0,0,1,1)}&=&\beta_4(x_1,x_2)\nonumber\\
y_{(0,0,0,0,2)}&=&x_1^6x_2^6\mathbbm{1}=-\beta_3(x_1,x_2)
\end{eqnarray}
All the Massey products which are $Q$--exact states lead to second order deformations of the matrix factorization:
\begin{eqnarray}
\alpha_{(0,0,0,2,0)}=-\alpha_{(0,0,0,0,2)}&=&\alpha_{(1,0,0,0,0)}\nonumber\\
\alpha_{(1,1,0,0,0)}&=&x_1x_2x_3x_4(\eta_5-\bar{\eta}_5)\bar{\eta}_2\eta_2 \nonumber\\
\alpha_{(1,0,0,1,0)}&=&x_1^3x_2^3(\eta_5-\bar{\eta}_5)(\eta_4-\bar{\eta}_4)(\eta_3-\bar{\eta}_3)\bar{\eta}_2\eta_2 \nonumber\\
\alpha_{(0,0,0,1,1)}&=&-2x_2^6(\eta_2-\bar{\eta}_2)\bar{\eta}_1\eta_1
\end{eqnarray}
Since this model is quite tricky, let us list the F--terms at order two:
\begin{eqnarray}
\label{11226-order2f}
f_1^{(2)}:&\quad&u_1u_3+u_2u_4=0\nonumber\\
f_2^{(2)}:&\quad&u_1u_4-u_2u_3=0\nonumber\\
f_3^{(2)}:&\quad&u_1^2-u_2^2+\varphi_2=0\nonumber\\
f_4^{(2)}:&\quad&u_1u_2=0
\end{eqnarray}
The choice of basis $B'_3$ for the Massey products at order three is slightly problematic since the equations above are not quite independent when they are extended to order three. If we, for instance, multiply the first equation with $u_3$ and the second equation with $u_4$ we find the following:
\begin{equation}
\left.\begin{array}{c}
u_2u_3u_4=-u_1u_3^2\\
u_2u_3u_4=u_1u_4^2
\end{array}\right\}\Rightarrow u_1u_4^2=-u_1u_3^2
\end{equation}
Such relations reduce the dimension of the basis and are easily overlooked. It is helpful to use the computer algebra program Singular \cite{GPS05} to compute a basis of monomials of order three in a quotient ring defined by the ideal defined by (\ref{11226-order2f}) multiplied by $\{u_1,u_2,u_3,u_4\}$.\\
At order $3$ there are five non--zero Massey products which do not get corrected by the obstructions at order two:
\begin{eqnarray}
y_{(1,0,0,2,0)}&=&\{\alpha_{(1,0,0,1,0)},\Psi_3\}+\{\alpha_{(0,0,0,2,0)},\alpha_{(1,0,0,0,0)}\}=\frac{1}{2}\beta_4(x_1,x_2)\nonumber\\
y_{(1,0,0,1,1)}&=&\{\alpha_{(1,0,0,1,0)},\Psi_4\}+\{\alpha_{(0,0,0,1,1)},\alpha_{(1,0,0,0,0)}\}=\beta_5(x_1,x_2)\nonumber\\
y_{(0,0,0,3,0)}&=&\{\alpha_{(0,0,0,2,0)},\Psi_3\}=\beta_2(x_1,x_2)\nonumber\\
y_{(0,0,0,2,1)}&=&\{\alpha_{(0,0,0,2,0)},\Psi_4\}+\{\alpha_{(0,0,0,1,1)},\Psi_3\}=\beta_6(x_1,x_2)\nonumber\\
y_{(0,0,0,1,2)}&=&\{\alpha_{(0,0,0,1,1)},\Psi_4\}+\{\alpha_{(0,0,0,0,2)},\Psi_3\}=-\beta_2(x_1,x_1)
\end{eqnarray}
Furthermore there are the following products which get contributions from the obstructions:
\begin{eqnarray}
y_{(0,0,1,0,2)}&=&\{\alpha_{(0,0,0,0,2)},\Psi_2\}-\{\alpha_{(0,0,0,1,1)},\Psi_1\}=\beta_7(x_1,x_2,x_3,x_4)\nonumber\\
y_{(0,0,1,1,1)}&=&\{\alpha_{(0,0,0,1,1)},\Psi_2\}-\{\alpha_{(0,0,0,2,0)},\Psi_1\}+\{\alpha_{(0,0,0,0,2)},\Psi_1\}=\frac{1}{2}\beta_1(x_1,x_2,x_3,x_4)\nonumber\\
y_{(1,0,1,0,1)}&=&-\{\alpha_{(1,1,0,0,0)},\Psi_3\}-\{\alpha_{(1,0,0,1,0)},\Psi_1\}=2\Phi_2\nonumber\\
y_{(1,0,2,0,0)}&=&\{\alpha_{(1,1,0,0,0)},\Psi_1\}=-\Phi_4
\end{eqnarray}
Four of the six new deformations coincide with deformations at lower order:
\begin{eqnarray}
\alpha_{(0,0,0,3,0)}=-\alpha_{(0,0,0,1,2)}&=&-\alpha_{(1,0,0,1,0)}\nonumber\\
\alpha_{(1,0,0,2,0)}&=&\frac{1}{2}\alpha_{(0,0,0,1,1)}\nonumber\\
\alpha_{(0,0,1,1,1)}&=&-2\alpha_{(1,1,0,0,0)}\nonumber\\
\alpha_{(1,0,0,1,1)}&=&-2x_2^6\bar{\eta}_1((\mathbbm{1}+\eta_2\bar{\eta}_2)-2\eta_1) \nonumber\\
\alpha_{(0,0,0,2,1)}&=&-2x_1^3x_2^3(\eta_5-\bar{\eta}_5)(\eta_4-\bar{\eta}_4)(\eta_3-\bar{\eta}_3)\eta_2(\eta_1-\bar{\eta}_1)\nonumber\\
\alpha_{(0,0,1,0,2)}&=&-2x_1x_2x_3x_4(\eta_5-\bar{\eta}_5)\eta_2(\eta_1-\bar{\eta}_1)
\end{eqnarray}
At order four, the following Massey products contribute:
\begin{eqnarray}
y_{(2,0,0,2,0)}&=&\{\alpha_{(1,0,0,2,0)},\alpha_{(1,0,0,0,0)}\}+\alpha_{(1,0,0,1,0)}\cdot\alpha_{(1,0,0,1,0)}=\frac{1}{2}\beta_5(x_1,x_2)\nonumber\\
y_{(1,0,0,3,0)}&=&\{\alpha_{(1,0,0,1,0)},\alpha_{(0,0,0,2,0)}\}+\{\alpha_{(1,0,0,2,0)},\Psi_3\}+\{\alpha_{(0,0,0,3,0)},\alpha_{(1,0,0,0,0)}\}=\frac{1}{2}\beta_6(x_1,x_2)\nonumber\\
y_{(0,0,0,4,0)}&=&\{\alpha_{(0,0,0,3,0)},\Psi_3\}+\alpha_{(0,0,0,2,0)}\cdot\alpha_{(0,0,0,2,0)}=\frac{1}{2}\beta_4(x_1,x_2)\nonumber\\
y_{(0,0,0,3,1)}&=&\{\alpha_{(0,0,0,2,0)},\alpha_{(0,0,0,1,1)}\}+\{\alpha_{(0,0,0,3,0)},\Psi_4\}+\{\alpha_{(0,0,0,2,1)},\Psi_3\}=\beta_8(x_1,x_2)\nonumber\\
y_{(0,0,0,2,2)}&=&\{\alpha_{(0,0,0,2,0)},\alpha_{(0,0,0,0,2)}\}+\alpha_{(0,0,0,1,1)}\cdot\alpha_{(0,0,0,1,1)}+\{\alpha_{(0,0,0,2,1)},\Psi_4\}\nonumber\\
&&+\{\alpha_{(0,0,0,1,2)},\Psi_3\}=\beta_9(x_1,x_2)\nonumber\\
y_{(0,0,0,1,3)}&=&\{\alpha_{(0,0,0,1,1)},\alpha_{(0,0,0,0,2)}\}+\{\alpha_{(0,0,0,1,2)},\Psi_4\}=-\beta_5(x_1,x_2)
\end{eqnarray}
There are further non-zero products which involve the relations:
\begin{eqnarray}
y_{(1,0,1,0,2)}&=&\{\alpha_{(0,0,1,0,2)},\alpha_{(1,0,0,0,0)}\}-\{\alpha_{(0,0,0,0,2)},\Psi_2\}+\{\alpha_{(1,1,0,0,0)},\alpha_{(0,0,0,1,1)}\}\nonumber\\ &&+\{\alpha_{(1,0,0,1,1)},\Psi_1\}=9\beta_1(x_1,x_2,x_3,x_4)\nonumber\\
y_{(1,0,1,1,1)}&=&\{\alpha_{(0,0,1,1,1)},\alpha_{(1,0,0,0,0)}\}-\{\alpha_{(1,1,0,0,0)},\alpha_{(0,0,0,2,0)}\}-\{\alpha_{(1,0,0,2,0)},\Psi_1\}\nonumber\\
&&-\{\alpha_{(0,0,0,1,1)},\Psi_1\}=\frac{1}{2}\beta_7(x_1,x_2,x_3,x_4)\nonumber\\
y_{(0,0,1,0,3)}&=&-\{\alpha_{(0,0,0,2,1)},\Psi_2\}+\{\alpha_{(0,0,1,0,2)},\Psi_4\}-\{\alpha_{(0,0,1,1,1)},\Psi_3\}-\{\alpha_{(0,0,0,1,2)},\Psi_1\}\nonumber\\
&&+\{\alpha_{(0,0,0,3,0)},\Psi_1\}=-4\Phi_2\nonumber\\
y_{(0,0,1,1,2)}&=&\{\alpha_{(0,0,0,1,2)},\Psi_2\}-\{\alpha_{(0,0,0,3,0)},\Psi_2\}+\{\alpha_{(0,0,1,0,2)},\Psi_3\}+\{\alpha_{(0,0,1,1,1)},\Psi\}\nonumber\\
&&-\{\alpha_{(0,0,0,2,1)},\Psi_1\}=4\Phi_1
\end{eqnarray}
Due to increasing complexity, we do not continue with the iteration and note the following F--terms at order four:
\begin{eqnarray}
f_1^{(4)}:&\quad&u_1u_3+u_2u_4+4u_2u_3u_4^2=0\nonumber\\
f_2^{(4)}:&\quad&u_1u_4-u_2u_3-4u_2u_4^3+2\varphi_1u_2u_4=0\nonumber\\
f_3^{(4)}:&\quad&u_1^2-u_2^2+\varphi_2=0\nonumber\\
f_4^{(4)}:&\quad&-2u_1u_2-\varphi_1u_2^2=0
\end{eqnarray}
We can make some statements about the higher order deformations. We notice that some exact states from lower orders reappear at higher orders. This suggests that the algorithm does not terminate. Furthermore we note that in particular exact states which contain the variables $\{x_1,x_2,x_3,x_4\}$ are among these recurring deformations which implies that obstructions may occur at very high orders. Therefore the chances to get the full F--terms from deformation theory are not very good.
\subsubsection*{Correlators}
The two-- and three--point correlators are consistent with the F--terms at order two.
\begin{eqnarray}
\langle\Psi_1\Psi_1\Psi_3\rangle&=&-1\nonumber\\
\langle\Psi_2\Psi_2\Psi_3\rangle&=&\phantom{-}1\nonumber\\
\langle\Psi_1\Psi_2\Psi_4\rangle=\langle\Psi_1\Psi_4\Psi_2\rangle&=&-1\nonumber\\
\langle\Psi_1\phi_2\rangle&=&-1
\end{eqnarray}
\subsubsection{Three Brane Moduli}
Tensor product branes with $3\leq L_1\leq 5$, $L_2=3$ and $L_3,L_4=2$ have three brane moduli. For concreteness, we will discuss the brane $L=(5,3,2,2,0)$. The modulus with the same charge distribution as (\ref{11226mod1}) is no longer present. That is why the remaining three moduli will be called $\Psi_2,\Psi_3,\Psi_4$ here. The bulk modulus $\phi_1$ is $Q$--exact, and the associated boundary deformation is:
\begin{equation}
\alpha_{(1,0,0,0)}=x_2^6\bar{\eta}_1
\end{equation}
The non--zero Massey products at order two are the following:
\begin{eqnarray}
y_{(1,0,1,0)}&=&\{\alpha_{(1,0,0,0)},\Psi_3\}=\beta_1(x_1,x_2)\nonumber\\
y_{(0,2,0,0)}&=&\Psi_2\cdot\Psi_2=-x_1^2x_2^2x_3^2x_4^2\mathbbm{1}=-\Phi_3\nonumber\\
y_{(0,1,1,0)}&=&\{\Psi_2,\Psi_3\}=-2\Phi_2\nonumber\\
y_{(0,1,0,1)}&=&\{\Psi_2,\Psi_4\}=\beta_2(x_1,x_2,x_3,x_4)\nonumber\\
y_{(0,0,2,0)}&=&\Psi_3\cdot\Psi_3=-x_1^6x_2^6\mathbbm{1}=\beta_3(x_1,x_2)\nonumber\\
y_{(0,0,1,1)}&=&\{\Psi_3,\Psi_4\}=\beta_4(x_1,x_2)\nonumber\\
y_{(0,0,0,2)}&=&\Psi_4\cdot\Psi_4=x_1^6x_2^6\mathbbm{1}=-\beta_3(x_1,x_2)
\end{eqnarray}
There are five new deformations:
\begin{eqnarray}
\alpha_{(0,0,2,0)}=-\alpha_{(0,0,0,2)}&=&\alpha_{(1,0,0,0)}\nonumber\\
\alpha_{(1,0,1,0)}&=&-x_1^3x_2^3(\eta_5-\bar{\eta}_5)(\eta_4-\bar{\eta}_4)(\eta_3-\bar{\eta}_3)\eta_2\bar{\eta}_2 \nonumber\\
\alpha_{(0,1,0,1)}&=&2x_1^4x_3x_4(\eta_4-\bar{\eta}_4)(\eta_3-\bar{\eta}_3)\bar{\eta}_2 \nonumber\\ 
\alpha_{(0,0,1,1)}&=&2x_2^4\eta_1\eta_2\bar{\eta}_1+2x_2^8\eta_1\bar{\eta}_1\bar{\eta}_2
\end{eqnarray}
Given our particular choice of deformations, there are five non--zero Massey products:
\begin{eqnarray}
y_{(1,0,2,0)}&=&\{\alpha_{(1,0,1,0)},\Psi_3\}+\{\alpha_{(0,0,2,0)},\alpha_{(1,0,0,0)}\} =-\frac{1}{2}\beta_4(x_1,x_2)\nonumber\\
y_{(1,0,1,1)}&=&\{\alpha_{(1,0,1,0)},\Psi_4\}+\{\alpha_{(0,0,1,1)},\alpha_{(1,0,0,0)}\}=\beta_5(x_1,x_2)\nonumber\\
y_{(0,0,3,0)}&=&\{\alpha_{(0,0,2,0)},\Psi_3\}=\beta_1(x_1,x_2)\nonumber\\
y_{(0,0,2,1)}&=&\{\alpha_{(0,0,2,0)},\Psi_4\}+\{\alpha_{(0,0,1,1)},\Psi_3\}=\beta_6(x_1,x_2)\nonumber\\
y_{(0,0,1,2)}&=&\{\alpha_{(0,0,0,2)},\Psi_3\}+\{\alpha_{(0,0,1,1)},\Psi_4\}=-\beta_1(x_1,x_2)
\end{eqnarray} 
At this order we get five new deformations and no contribution to the F--terms. Three of the deformations we have encountered previously:
\begin{eqnarray}
\alpha_{(0,0,3,0)}=-\alpha_{(0,0,1,2)}&=&\alpha_{(1,0,1,0)}\nonumber\\
\alpha_{(1,0,1,2)}&=&-\frac{1}{2}\alpha_{(0,0,1,1)}\nonumber\\
\alpha_{(1,0,1,1)}&=&2x_2^6\bar{\eta}_1\eta_2\bar{\eta}_2\nonumber\\
\alpha_{(0,0,2,1)}&=&2x_1^3x_2^5(\eta_5-\bar{\eta}_5)(\eta_4-\bar{\eta}_4)(\eta_3-\bar{\eta}_3)\bar{\eta}_2(\eta_1-\bar{\eta}_1)
\end{eqnarray}
At order four, there are seven new deformations:
\begin{eqnarray}
y_{(2,0,2,0)}&=&\alpha_{(1,0,1,0)}\cdot\alpha_{(1,0,1,0)}+\{\alpha_{(1,0,2,0)},\alpha_{(1,0,0,0)}\}=-\frac{1}{2}\beta_5(x_1,x_2)\nonumber\\
y_{(1,0,3,0)}&=&\{\alpha_{(1,0,1,0)},\alpha_{(0,0,2,0)}\}+\{\alpha_{(1,0,2,0)},\Psi_3\}+\{\alpha_{(0,0,3,0)},\alpha_{(1,0,0,0)}\}=-\frac{1}{2}\beta_6(x_1,x_2)\nonumber\\
y_{(1,0,2,1)}&=&\{\alpha_{(1,0,1,0)},\alpha_{(0,0,1,1)}\}+\{\alpha_{(1,0,2,0)},\Psi_4\}+\{\alpha_{(1,0,1,1)},\Psi_3\}\nonumber\\
&&+\{\alpha_{(0,0,2,1)},\alpha_{(1,0,0,0)}\}=2\beta_1(x_1,x_2)\nonumber\\
y_{(0,0,4,0)}&=&\alpha_{(0,0,2,0)}\cdot\alpha_{(0,0,2,0)}+\{\alpha_{(0,0,3,0)},\Psi_3\}=-\frac{1}{2}\beta_4(x_1,x_2)\nonumber\\
y_{(0,0,3,1)}&=&\{\alpha_{(0,0,1,1)},\alpha_{(0,0,2,0)}\}+\{\alpha_{(0,0,3,0)},\Psi_4\}+\{\alpha_{(0,0,2,1)},\Psi_3\}=\beta_7(x_1,x_2)\nonumber\\
y_{(0,0,2,2)}&=&\{\alpha_{(0,0,2,0)},\alpha_{(0,0,0,2)}\}+\alpha_{(0,0,1,1)}\cdot\alpha_{(0,0,1,1)}+\{\alpha_{(0,0,2,1)},\Psi_4\}\nonumber\\
&&+\{\alpha_{(0,0,1,2)},\Psi_3\}=\beta_8(x_1,x_2)\nonumber\\
y_{(0,0,1,3)}&=&\{\alpha_{(0,0,1,1)},\alpha_{(0,0,0,2)}\}+\{\alpha_{(0,0,1,2)},\Psi_4\}=-\beta_5(x_1,x_2)
\end{eqnarray}
Having seven deformations at order seven four huge combinatorics at order five. Since recurring deformations are already visible at this order there is not much hope that the algorithm will terminate at order five. Therefore we will content ourselves with arguing that there are no further obstructions at order five. Massey products at order five are either products of order four deformations with the brane moduli or products of order two and order three deformations. Since the order four deformations only contain $x_1,x_2$ an obstruction can only be produced by multiplication with $\Psi_2$ but such products are not allowed due to the obstructions. From the other possible combination, only products containing $\alpha_{(0,1,0,1)}$ may lead to an obstruction. However, also these products are forbidden due to the F--terms at lower order.\\
Thus, (up to order five) we have found the following F--terms:
\begin{eqnarray}
f^{(5)}_2:&\quad&u_2u_3=0\nonumber\\
f^{(5)}_3:&\quad&-u_2^2+\varphi_2=0\nonumber\\
f^{(5)}_4:&\quad&0
\end{eqnarray}
There are two solutions to these equations:
\begin{eqnarray}
&&u_2=0,\varphi_2=0\nonumber\\
&&u_3=0,u_2=\pm\sqrt{\varphi_2}
\end{eqnarray}
The first solution tells us that, if $u_2=0$, the remaining boundary deformation is unobstructed and the bulk deformation $x_1^2x_2^2x_3^2x_4^2$ is not allowed. The second solution shows the existence of a BPS domain wall.\\
The F--terms can be integrated to the following effective superpotential:
\begin{equation}
\mathcal{W}_{eff}=u_2^2u_3-\varphi_2u_3
\end{equation}
\subsubsection*{Correlators}
For the given brane there are two non--zero correlators which can be computed with the residue formula:
\begin{eqnarray}
\langle\Psi_2\Psi_2\Psi_3\rangle&=&-1\nonumber\\
\langle\Psi_3\phi_2\rangle&=&\phantom{-}1
\end{eqnarray}
This is consistent with the F--terms and $\mathcal{W}_{eff}$.
\subsubsection{Two Brane Moduli -- Case A}
The first two--moduli case we discuss is a brane where the moduli (\ref{11226mod1}) and (\ref{11226mod2}) are present. This happens for $4\leq L_1,L_2\leq5$ if at least one of $L_3$ and $L_4$ is $1$. The maximal possible label for such a brane is $L=(5,5,2,1,0)$. In this case both bulk deformations are $Q$--exact. The associated first order deformations of the matrix factorization are:
\begin{eqnarray}
\alpha_{(1,0,0,0)}&=&x_2^6\bar{\eta}_1\nonumber\\
\alpha_{(0,1,0,0)}&=&x_1^2x_2^2x_3^2\bar{\eta}_4
\end{eqnarray}
There are four non--vanishing Massey products at order $2$:
\begin{eqnarray}
y_{(1,0,1,0)}&=&\{\alpha_{(1,0,0,0)},\Psi_1\}=\beta_1(x_1,x_2,x_3.x_4)\nonumber\\
y_{(0,0,2,0)}&=&\Psi_1\cdot\Psi_1=x_1^2x_2^2x_3^2x_4^2\mathbbm{1}=\beta_2(x_1,x_2,x_3,x_4)\nonumber\\
y_{(0,0,1,1)}&=&\{\Psi_1,\Psi_2\}=\beta_3(x_1,x_2,x_3,x_4)\nonumber\\
y_{(0,0,0,2)}&=&\Psi_2\cdot\Psi_2=-x_1^2x_2^2x_3^2x_4^2=-\beta_2(x_1,x_2,x_3,x_4)
\end{eqnarray}
The order two deformations are:
\begin{eqnarray}
\alpha_{(0,0,0,2)}=-\alpha_{(0,0,2,0)}&=&\alpha_{(1,0,0,0)}\nonumber\\
\alpha_{(1,0,1,0)}&=&x_1x_2x_3x_4(\eta_5-\bar{\eta}_5)\eta_2\bar{\eta}_2 \nonumber \nonumber\\
\alpha_{(0,0,1,1)}&=&2x_1^2x_2^2x_3^2\bar{\eta}_4(\eta_2-\bar{\eta}_2)(\eta_1-\bar{\eta}_1)
\end{eqnarray}
At order three, only two Massey products are non--zero:
\begin{eqnarray}
y_{(1,0,2,0)}&=&\{\alpha_{(1,0,1,0)},\Psi_1\}+\{\alpha_{(0,0,2,0)},\alpha_{(1,0,0,0)}\}=-\frac{1}{2}\beta_3(x_1,x_2,x_3,x_4)\nonumber\\
y_{(1,0,1,1)}&=&\{\alpha_{(1,0,1,0)},\Psi_2\}+\{\alpha_{(0,0,1,1)},\alpha_{(1,0,0,0)}\}=\beta_4(x_1,x_2,x_2,x_4)
\end{eqnarray}
The new deformations are:
\begin{eqnarray}
\alpha_{(1,0,2,0)}&=&-\frac{1}{2}\alpha_{(0,0,1,1)}\nonumber\\
\alpha_{(1,0,1,1)}&=&-2x_1^2x_2^2x_3^3\bar{\eta}_4\eta_2\bar{\eta}_2
\end{eqnarray}
With this choice of deformations there is only one non--zero Massey product at order four:
\begin{equation}
y_{(2,0,2,0)}=\alpha_{(1,0,1,0)}\cdot\alpha_{(1,0,1,0)}+\{\alpha_{(1,0,2,0)},\alpha_{(1,0,0,0)}\}=-\frac{1}{2}\beta_4(x_1,x_2,x_3,x_4)
\end{equation}
We have encountered this deformation before:
\begin{equation}
\alpha_{(2,0,2,0)}=-\frac{1}{2}\alpha_{(1,0,2,0)}
\end{equation}
Since there is only a small number of deformations, it is easy to go to higher order. In fact it is no problem to compute the Massey products up to order eight. It turns out that all these higher products are zero. Since the last deformations has been found at order four, there are no more products to define at order nine. Therefore the algorithm terminates at order eight. The complete deformed matrix factorization is:
\begin{eqnarray}
Q_{def}&=&Q+\varphi_1\alpha_{(1,0,0,0)}+\varphi_2\alpha_{(0,1,0,0)}+u_1\Psi_1+u_2\Psi_2\nonumber\\ 
&&+\varphi_1 u_1\alpha_{(1,0,1,0)}+u_1^2\alpha_{(0,0,2,0)}+u_1u_2\alpha_{(0,0,1,1)}+u_2^2\alpha_{(0,0,0,2)}\nonumber\\
&&+\varphi_1u_1^2\alpha_{(1,0,2,0)}+\varphi_1u_1u_2\alpha_{(1,0,1,1)}+\varphi_1^2u_1^2\alpha_{(2,0,2,0)}
\end{eqnarray}
Upon squaring $Q_{def}$ all the brane moduli dependence miraculously cancels and we are left with the bulk deformed Landau--Ginzburg superpotential. All the F--terms are zero.
\subsubsection*{Correlators}
All of the correlators with unintegrated insertions are zero. This is consistent with the Massey product calculation.
\subsubsection{Two Brane Moduli -- Case B}
There is a second class of branes which have two moduli. They have $2\leq L_1\leq 5$ and $L_2=L_3=L_4=2$. We discuss the brane $L=(5,2,2,2,0)$. The two moduli are (\ref{11226mod2}) and (\ref{11226mod3}). Only the bulk deformation $x_1^6x_2^6$ is $Q$--exact, and we define:
\begin{equation}
\alpha_{(1,0,0)}=x_2^6\bar{\eta}_1
\end{equation}
The non--zero Massey products are:
\begin{eqnarray}
y_{(1,0,1)}&=&\{\alpha_{(1,0,0)},\Psi_3\}=\beta_1(x_1,x_2)\nonumber\\
y_{(0,2,0)}&=&\Psi_2\cdot\Psi_2=-x_1^2x_2^2x_3^2x_4^2\mathbbm{1}=-\Phi_3\nonumber\\
y_{(0,1,1)}&=&\{\Psi_2,\Psi_3\}=-2\Phi_2\nonumber\\
y_{(0,0,2)}&=&-x_1^6x_2^6\mathbbm{1}=\beta_2(x_1,x_2)
\end{eqnarray}
There are two deformations:
\begin{eqnarray}
\alpha_{(0,0,2)}&=&\alpha_{(1,0,0)}\nonumber\\
\alpha_{(1,0,1)}&=&x_1^3x_2^3(\eta_5-\bar{\eta}_5)(\eta_4-\bar{\eta}_4)(\eta_3-\bar{\eta}_3)\eta_2\bar{\eta}_2
\end{eqnarray}
At order three there are two non--vanishing Massey products:
\begin{eqnarray}
y_{(1,0,2)}&=&\{\alpha_{(1,0,1)},\Psi_3\}+\{\alpha_{(0,0,2)},\alpha_{(1,0,0)}\}=\beta_3(x_1,x_2)\nonumber\\
y_{(0,0,3)}&=&\{\alpha_{(0,0,2)},\Psi_3\}=\beta_1(x_1,x_2)
\end{eqnarray}
For the deformations we find:
\begin{eqnarray}
\alpha_{(0,0,3)}&=&\alpha_{(1,0,1)}\nonumber\\
\alpha_{(1,0,2)}&=&-x_1^6(\eta_1-\bar{\eta}_1)\eta_2\bar{\eta}_2
\end{eqnarray}
Going to order four, the following products are non--zero:
\begin{eqnarray}
y_{(1,0,3)}&=&\{\alpha_{(1,0,1)},\alpha_{(0,0,2)}\}+\{\alpha_{(1,0,2)},\Psi_3\}+\{\alpha_{(0,0,3)},\alpha_{(1,0,0)}\}=\beta_4(x_1,x_2)\nonumber\\
y_{(0,0,4)}&=&\{\alpha_{(0,0,3)},\Psi_3\}=\beta_3(x_1,x_2)
\end{eqnarray}
The corresponding deformations are:
\begin{eqnarray}
\alpha_{(0,0,4)}&=&\alpha_{(0,2,1)}\nonumber\\
\alpha_{(1,0,3)}&=&(\eta_5-\bar{\eta}_5)(\eta_4-\bar{\eta}_4)(\eta_3-\bar{\eta}_3)\eta_1(-x_1^3\eta_2+x_1^3x_2^6\bar{\eta}_2)
\end{eqnarray}
At order five, three Massey products lead to new deformations:
\begin{eqnarray}
y_{(0,0,5)}&=&\{\alpha_{(0,0,2)},\alpha_{(0,0,3)}\}+\{\alpha_{(0,0,4)},\Psi_3\}=\beta_4(x_1,x_2)\nonumber\\
y_{(1,0,4)}&=&\{\alpha_{(1,0,2)},\alpha_{(0,0,2)}\}+\{\alpha_{(1,0,1)},\alpha_{(0,0,3)}\}+\{\alpha_{(1,0,3)},\Psi_3\}\nonumber\\ &&+\{\alpha_{(0,0,4)},\alpha_{(1,0,0)}\}=x_1^6x_2^6\mathbbm{1}=-\beta_2(x_1,x_2)\nonumber\\
y_{(2,0,3)}&=&\{\alpha_{(1,0,1)},\alpha_{(1,0,2)}\}+\{\alpha_{(1,0,3)},\alpha_{(1,0,0)}\}=-\beta_1(x_1,x_2)
\end{eqnarray}
We notice recurring patterns of deformations at every order. This suggests that the deformation theory algorithm does not terminate. Although we cannot prove this rigorously, we observe that there will be no further F--terms. The only way to get an obstruction at higher order would be to compute a Massey product of a deformation with $\Psi_2$ (which is the only one that contains the variable $x_4$) but all these products are forbidden due to the deformations.\\
Therefore, the F--terms are:
\begin{eqnarray}
f_2:&\quad&u_2u_3=0\nonumber\\
f_3:&\quad&u_2^2-\varphi_2=0
\end{eqnarray}
The two non--trivial solutions are:
\begin{eqnarray}
&& u_2=0,\varphi_2=0\nonumber\\
&& u_3=0,u_2=\pm \sqrt{\varphi_2}
\end{eqnarray}
The F--terms are easily integrated:
\begin{equation}
\mathcal{W}_{eff}=u_2^2u_3-u_3\varphi_2
\end{equation}
Note that this is the same effective superpotential as we had in the three--moduli example.
\subsubsection*{Correlators}
All the correlators we can get from the residue formula turn out to be consistent with $\mathcal{W}_{eff}$:
\begin{eqnarray}
\langle\Psi_2\Psi_2\Psi_3\rangle&=&-1\nonumber\\
\langle\psi_3\phi_2\rangle&=&\phantom{-}1
\end{eqnarray}
\subsubsection{One Brane Modulus}
Finally, all remaining branes with $L_i\geq1$ have one modulus (\ref{11226mod2}). Both bulk deformations are $Q$--exact. The maximal brane with this property has label $L=(5,3,2,1,0)$. The first order bulk deformations are:
\begin{eqnarray}
\alpha_{(1,0,0)}&=&x_2^6\bar{\eta}_1\nonumber\\
\alpha_{(0,1,0)}&=&x_1^2x_2^2x_3^2\bar{\eta}_4
\end{eqnarray}
The only non--zero Massey product is:
\begin{equation}
y_{(0,0,2)}=\Psi_2\cdot\Psi_2=-x_1^2x_2^2x_3^2x_4^2\mathbbm{1}
\end{equation}
To cancel this exact term we define $\alpha_{(0,0,2)}=-\alpha_{(0,1,0)}$. Since this anticommutes with everybody, the algorithm breaks at order three. There are no F--terms and the full deformation is:
\begin{equation}
Q_{def}=Q+\varphi_1\alpha_{(1,0,0)}+\varphi_2\alpha_{(0,1,0)}+u_2\Psi_2+u_2^2\alpha_{(0,0,2)}
\end{equation}
It is easy to check that this squares to the bulk--deformed Landau--Ginzburg superpotential. \\
We conclude that this brane has an unobstructed boundary modulus.
\subsubsection*{Correlators}
All the correlators we can compute are zero. This is consistent with the fact that the boundary modulus of this brane is unobstructed.
\section{The model $\mathbbm{P}(12234)[12]/(\mathbbm{Z}_6)^2$}
The Landau--Ginzburg superpotential associated to this degree $12$ hypersurface is:
\begin{equation}
W=x_1^{12}+x_2^6+x_3^6+x_4^4+x_5^3
\end{equation}
We impose the following $(\mathbbm{Z}_6)^2$ orbifold action:
\begin{eqnarray}
g_1:&&(2,10,0,0,0)\nonumber\\
g_2:&&(2,0,10,0,0)
\end{eqnarray}
The two bulk deformations are:
\begin{eqnarray}
\phi_1&=&x_1^6x_4^2\nonumber\\
\phi_2&=&x_1x_2x_3x_4x_5
\end{eqnarray}
This model has only few branes which have moduli. We list them in table \ref{tab12234mod}.
\subsection{Discussion of Moduli}
We give the explicit expressions for the brane moduli of the tensor product brane with the highest $L$--label $L=(5,2,2,1,0)$. The charge $1$ fermion with $R$--charges $\frac{1}{2}^1\otimes0^1\otimes 0^1\otimes\frac{1}{2}^0\otimes 0^0$ is:
\begin{equation}
\label{12234mod1}
\Psi_1=\left(\begin{array}{cc}
0&x_1^3\\
-x_1^3&0
\end{array}\right)\otimes
\left(\begin{array}{cc}
0&1\\
-1&0
\end{array}\right)\otimes
\left(\begin{array}{cc}
0&1\\
-1&0
\end{array}\right)\otimes
\left(\begin{array}{cc}
x_4&0\\
0&x_4
\end{array}\right)\otimes
\left(\begin{array}{cc}
1&0\\
0&1
\end{array}\right)
\end{equation}
The second modulus with structure $\frac{1}{2}^0\otimes0^1\otimes 0^1\otimes\frac{1}{2}^1\otimes 0^0$ is:
\begin{equation}
\label{12234mod2}
\Psi_2=\left(\begin{array}{cc}
x_1^3&0\\
0&x_1^3
\end{array}\right)\otimes
\left(\begin{array}{cc}
0&1\\
-1&0
\end{array}\right)\otimes
\left(\begin{array}{cc}
0&1\\
-1&0
\end{array}\right)\otimes
\left(\begin{array}{cc}
0&x_4\\
-x_4&0
\end{array}\right)\otimes
\left(\begin{array}{cc}
1&0\\
0&1
\end{array}\right)
\end{equation}
\subsection{Obstructions}
The the open string states describing the obstructions to the above deformations look as follows. The obstruction which is Serre dual to (\ref{12234mod1}) has charges $\frac{1}{3}^0\otimes\frac{2}{3}^0\otimes\frac{2}{3}^0\otimes 0^1\otimes \frac{1}{3}^1$:
\begin{equation}
\label{12234ob1}
\Phi_1=\left(\begin{array}{cc}
x_1^2&0\\
0&x_1^2
\end{array}\right)\otimes
\left(\begin{array}{cc}
x_2^2&0\\
0&x_2^2
\end{array}\right)\otimes
\left(\begin{array}{cc}
x_3^2&0\\
0&x_3^2
\end{array}\right)\otimes
\left(\begin{array}{cc}
0&1\\
-1&0
\end{array}\right)\otimes
\left(\begin{array}{cc}
0&1\\
-x_5&0
\end{array}\right)
\end{equation}
The Serre dual of (\ref{12234mod2}) is
\begin{equation}
\label{12234ob2}
\Phi_2=\left(\begin{array}{cc}
0&x_1^2\\
-x_1^2&0
\end{array}\right)\otimes
\left(\begin{array}{cc}
x_2^2&0\\
0&x_2^2
\end{array}\right)\otimes
\left(\begin{array}{cc}
x_3^2&0\\
0&x_3^2
\end{array}\right)\otimes
\left(\begin{array}{cc}
1&0\\
0&1
\end{array}\right)\otimes
\left(\begin{array}{cc}
0&1\\
-x_5&0
\end{array}\right)
\end{equation}
This has structure $\frac{1}{3}^1\otimes\frac{2}{3}^0\otimes\frac{2}{3}^0\otimes 0^0\otimes \frac{1}{3}^1$.
\subsection{Massey Products and F--terms}
In this model we have to distinguish between branes with two moduli and one modulus. We present examples in the following subsections. One special feature of this model is that both bulk deformations are always $Q$--exact for all tensor product branes. For our particular examples we can choose the following first order deformations of the matrix factorization to produce these:
\begin{eqnarray}
\alpha_{(1,0,0,(0))}&=&x_4^2\bar{\eta}_1\nonumber\\
\alpha_{(0,1,0,(0))}&=&x_1x_2x_3x_4\bar{\eta}_5
\end{eqnarray}
The index vector of the $\alpha$'s has three or four entries, depending on whether we have one or two brane moduli.
\subsubsection{Two Moduli}
\label{sec-12234twomod}
The maximal brane with two moduli has labels $L=(5,2,2,1,0)$. The marginal open string states are (\ref{12234mod1}) and (\ref{12234mod2}). At order two in deformation theory there are the following non--zero Massey products:
\begin{eqnarray}
y_{(1,0,1,0)}&=&\{\alpha_{(1,0,0,0)},\Psi_1\}=\beta_1(x_1,x_4)\nonumber\\
y_{(0,0,2,0)}&=&\Psi_1\cdot\Psi_1=x_1^6x_4^2\mathbbm{1}=\beta_2(x_1,x_4)\nonumber\\
y_{(0,0,0,2)}&=&\Psi_2\cdot\Psi_2=x_1^6x_4^2\mathbbm{1}=\beta_2(x_1,x_4)
\end{eqnarray}
All these products are $Q$--exact and only contain the variables $x_1,x_4$. In order to cancel these terms we we have to find higher order deformations of the matrix factorization, and these will also only contain $x_1,x_4$. Also, the open string states only contain only these two variables. The obstructions (\ref{12234ob1}) and (\ref{12234ob2}), however, contain the variables $x_2,x_3,x_5$. These can never be obtained by computing Massey products at any order. \\
Thus, we conclude that for this class of branes, both boundary deformations are unobstructed. There are no F--terms. \\
This could be the end of the story, but what is also interesting is to find out whether there are any signs that the deformation theory algorithm terminates at a finite order. Therefore, we go on and compute the deformations up to order five. The deformations at order two can be chosen as follows:
\begin{eqnarray}
\alpha_{(0,0,0,2)}=\alpha_{(0,0,2,0)}&=&-\alpha_{(1,0,0,0)}\nonumber\\
\alpha_{(1,0,1,0)}&=&x_1^3x_4\eta_4(\eta_3-\bar{\eta}_3)(\eta_2-\bar{\eta}_2)
\end{eqnarray} 
At order three, there are three non--zero Massey products:
\begin{eqnarray}
y_{(0,0,3,0)}&=&\{\alpha_{(0,0,2,0)},\Psi_1\}=-\beta_1(x_1,x_4)\nonumber\\
y_{(0,0,1,2)}&=&\{\alpha_{(0,0,0,2)},\Psi_1\}=-\beta_1(x_1,x_4)\nonumber\\
y_{(1,0,1,1)}&=&\{\alpha_{(1,0,1,0)},\Psi_2\}=x_1^6x_4^2\mathbbm{1}=\beta_2(x_1,x_4)
\end{eqnarray} 
The deformations are easily computed:
\begin{eqnarray}
\alpha_{(0,0,3,0)}=\alpha_{(0,0,1,2)}&=&-\alpha_{(1,0,1,0)}\nonumber\\
\alpha_{(1,0,1,0)}&=&-\alpha_{(1,0,0,0)}
\end{eqnarray}
At order four, there are again three Massey products which lead to new deformations:
\begin{eqnarray}
y_{(1,0,2,1)}&=&\{\alpha_{(1,0,1,1)},\Psi_1\}=-\beta_1(x_1,x_4)\nonumber\\
y_{(0,0,3,1)}&=&\{\alpha_{(0,0,3,0)},\Psi_2\}=x_1^6x_4^2\mathbbm{1}=\beta_2(x_1,x_4)\nonumber\\
y_{(0,0,1,3)}&=&\{\alpha_{(0,0,1,2)},\Psi_2\}=-x_1^6x_4^2\mathbbm{1}=-\beta_2(x_1,x_4)
\end{eqnarray}
The new deformations are:
\begin{eqnarray}
\alpha_{(1,0,2,1)}&=&-\alpha_{(1,0,1,0)}\nonumber\\
\alpha_{(0,0,1,3)}=-\alpha_{(0,0,3,1)}&=&\alpha_{(1,0,0,0)}
\end{eqnarray}
At order five we find the following:
\begin{eqnarray}
y_{(1,0,2,2)}&=&\{\alpha_{(1,0,1,0)},\alpha_{(0,0,1,2)}\}+\{\alpha_{(1,0,2,1)},\Psi_2\}=x_1^6x_4^2\mathbbm{1}=\beta_2(x_1,x_4)\nonumber\\
y_{(0,0,4,1)}&=&\{\alpha_{(0,0,3,1)},\Psi_1\}=-\beta_1(x_1,x_4)\nonumber\\
y_{(0,0,2,3)}&=&\{\alpha_{(0,0,1,3)},\Psi_1\}=\beta_1(x_1,x_4)
\end{eqnarray}
This is a particularly nice and simple example of a recurring pattern where it is obvious that the deformation theory algorithm does not terminate. There are only two kinds of higher order deformations which anticommute among each other. Thus, the only contribution to higher Massey products can come products with first order deformations. We find:
\begin{equation}
\{\alpha_{(1,0,0,0)},\Psi_1\}=\beta_1(x_1,x_4)\Rightarrow\alpha_{(1,0,1,0)}\longrightarrow \{\alpha_{(1,0,1,0)},\Psi_2\}=\beta_2(x_1,x_4)\Rightarrow\alpha_{(1,0,0,0)}
\end{equation} 
This structure repeats in a two--periodic way and stops the algorithm from terminating.
\subsubsection*{Correlators}
In agreement with the deformation theory, all the correlators which can be computed by the Kapustin--Li formula are $0$.
\subsubsection{One Modulus -- Case A}
As an example for a brane with one modulus we discuss the brane with labels $L=(5,2,2,0,0)$. For the class of branes represented by this model, only the open string state (\ref{12234mod2}) is left over. There is only one non--zero Massey product at order $2$:
\begin{equation}
y_{(0,0,2)}=\Psi_2\cdot\Psi_2=x_1^6x_4^2\mathbbm{1}=\beta_1(x_1,x_4)
\end{equation}
This can be canceled by deforming the brane with $-\alpha_{(1,0,0)}$ at second order in deformation theory. There are no further Massey products at higher order. The deformed $Q$--operator is:
\begin{equation}
Q_{def}=Q+\varphi_1\alpha_{(1,0,0)}+\varphi_2\alpha_{(0,1,0)}+u_2\Psi_2-u_2^2\alpha_{(1,0,0)}
\end{equation}
This deformed matrix factorization squares precisely to the deformed Landau--Ginzburg superpotential. There are no F-terms, so the boundary deformation is unobstructed. 
\subsubsection*{Correlators}
As expected, all the correlators which can be computed by the Kapustin--Li formula are $0$.
\subsubsection{One Modulus -- Case B}
The final class of branes in this model is represented by the brane with label $L=(2,2,2,1,0)$. It has one modulus (\ref{12234mod1}). There are two non--zero Massey products at order $2$:
\begin{eqnarray}
y_{(1,0,1)}&=&\{\alpha_{(1,0,0)},\Psi_1\}=\beta_1(x_1,x_4)\nonumber\\
y_{(0,0,2)}&=&\Psi_1\cdot\Psi_1=x_1^6x_4^2\mathbbm{1}=\beta_2(x_1,x_4)
\end{eqnarray}
By the same argument as in the two-moduli case, the obstruction can never be reached by Massey products at any order in deformation theory. Therefore, the brane modulus is unobstructed.\\
Let us proceed to higher orders in deformation theory in order to find out whether the number of deformations is finite or infinite. There are two deformations at order two:
\begin{eqnarray}
\alpha_{(0,0,2)}&=&-\alpha_{(1,0,0)}\nonumber\\
\alpha_{(1,0,1)}&=&x_1^3x_4\eta_4(\eta_3-\bar{\eta}_3)(\eta_2-\bar{\eta}_2)
\end{eqnarray}
At order three, there is only one non--zero Massey product:
\begin{eqnarray}
y_{(0,0,3)}&=&\{\alpha_{(0,0,2)},\Psi_1\}=-\beta_1(x_1,x_4)
\end{eqnarray}
The corresponding deformation has already been computed: $\alpha_{(0,0,3)}=-\alpha_{(1,0,1)}$. Going to higher orders in deformation theory, we find that all further Massey products are zero. Having no more deformations at our disposition, the algorithm terminates at order seven. The deformed matrix factorization is:
\begin{eqnarray}
Q_{def}&=&Q+\varphi_1\alpha_{(1,0,0)}+\varphi_2\alpha_{(0,1,0)}+u_1\Psi_1\nonumber\\
&&+\varphi_1u_1\alpha_{(1,0,1)}+u_2^2\alpha_{(0,0,2)}+u_3^2\alpha_{(0,0,3)}
\end{eqnarray}
This squares to the bulk deformed Landau--Ginzburg superpotential.
\subsubsection*{Correlators}
All the correlators which can be computed by the Kapustin--Li formula are $0$, which confirms that the modulus is unobstructed.
\section{The model $\mathbbm{P}(12227)[14]/(\mathbbm{Z}_7)^2$}
The Landau--Ginzburg superpotential associated to this degree $14$ hypersurface is:
\begin{equation}
W=x_1^{14}+x_2^7+x_3^7+x_4^7+x_5^2
\end{equation}
We impose the following $(\mathbbm{Z}_7)^2$ orbifold action:
\begin{eqnarray}
g_1:&&(2,12,0,0,0)\nonumber\\
g_2:&&(2,0,12,0,0)
\end{eqnarray}
The two bulk deformations are:
\begin{eqnarray}
\phi_1&=&x_1^7x_5\nonumber\\
\phi_2&=&x_1x_2x_3x_4x_5
\end{eqnarray}
In this model we can again write $\phi_2=x_1^2x_2^2x_3^2x_4^2$ via the equations of motion of $x_5$. Furthermore, we can also rewrite $\phi_1=x_1^8x_2x_3x_4$. We list the tensor product branes with moduli in tables \ref{tab12227mod}.
\subsection{Discussion of Moduli}
The maximal brane $L=(6,2,2,2,0)$ has four moduli. The open string state with label $\frac{1}{7}^0\otimes\frac{2}{7}^0\otimes\frac{2}{7}^0\otimes \frac{2}{7}^0\otimes 0^1$ looks as follows:
\begin{equation}
\label{12227mod1}
\Psi_1=\left(\begin{array}{cc}
x_1&0\\
0&x_1
\end{array}\right)\otimes
\left(\begin{array}{cc}
x_2&0\\
0&x_2
\end{array}\right)\otimes
\left(\begin{array}{cc}
x_3&0\\
0&x_3
\end{array}\right)\otimes
\left(\begin{array}{cc}
x_4&0\\
0&x_4
\end{array}\right)\otimes
\left(\begin{array}{cc}
0&1\\
-1&0
\end{array}\right)
\end{equation}
The open string state $\frac{1}{7}^1\otimes\frac{2}{7}^0\otimes\frac{2}{7}^0\otimes \frac{2}{7}^0\otimes 0^0$ has the following explicit form:
\begin{equation}
\label{12227mod2}
\Psi_2=\left(\begin{array}{cc}
0&x_1\\
-x_1&0
\end{array}\right)\otimes
\left(\begin{array}{cc}
x_2&0\\
0&x_2
\end{array}\right)\otimes
\left(\begin{array}{cc}
x_3&0\\
0&x_3
\end{array}\right)\otimes
\left(\begin{array}{cc}
x_4&0\\
0&x_4
\end{array}\right)\otimes
\left(\begin{array}{cc}
1&0\\
0&1
\end{array}\right)
\end{equation}
The open string state $\Psi_3$ has the structure $\frac{4}{7}^1\otimes\frac{1}{7}^1\otimes\frac{1}{7}^1\otimes\frac{1}{7}^1\otimes 0^1$:
\begin{equation}
\label{12227mod3}
\Psi_3=\left(\begin{array}{cc}
0&x_1^4\\
-x_1^4&0
\end{array}\right)\otimes
\left(\begin{array}{cc}
0&1\\
-x_2&0
\end{array}\right)\otimes
\left(\begin{array}{cc}
0&1\\
-x_3&0
\end{array}\right)\otimes
\left(\begin{array}{cc}
0&1\\
-x_4&0
\end{array}\right)\otimes
\left(\begin{array}{cc}
0&1\\
-1&0
\end{array}\right)
\end{equation}
Finally, we have a state with label $\frac{4}{7}^0\otimes\frac{1}{7}^1\otimes\frac{1}{7}^1\otimes\frac{1}{7}^1\otimes 0^0$:
\begin{equation}
\label{12227mod4}
\Psi_4=\left(\begin{array}{cc}
x_1^4&0\\
0&x_1^4
\end{array}\right)\otimes
\left(\begin{array}{cc}
0&1\\
-x_2&0
\end{array}\right)\otimes
\left(\begin{array}{cc}
0&1\\
-x_3&0
\end{array}\right)\otimes
\left(\begin{array}{cc}
0&1\\
-x_4&0
\end{array}\right)\otimes
\left(\begin{array}{cc}
1&0\\
0&1
\end{array}\right)
\end{equation}
\subsection{Obstructions}
The obstruction associated with (\ref{12227mod1}) has structure $\frac{5}{7}^1\otimes\frac{3}{7}^1\otimes\frac{3}{7}^1\otimes\frac{3}{7}^1\otimes 0^0$:
\begin{equation}
\label{12227ob1}
\Phi_1=\left(\begin{array}{cc}
0&x_1^5\\
-x_1^5&0
\end{array}\right)\otimes
\left(\begin{array}{cc}
0&x_2\\
-x_2^2&0
\end{array}\right)\otimes
\left(\begin{array}{cc}
0&x_3\\
-x_3^2&0
\end{array}\right)\otimes
\left(\begin{array}{cc}
0&x_4\\
-x_4^2&0
\end{array}\right)\otimes
\left(\begin{array}{cc}
1&0\\
0&1
\end{array}\right)
\end{equation}
The Serre dual boson to (\ref{12227mod2}) with label  $\frac{5}{7}^0\otimes\frac{3}{7}^1\otimes\frac{3}{7}^1\otimes\frac{3}{7}^1\otimes 0^1$ explicitly looks as follows:
\begin{equation}
\label{12227ob2}
\Phi_2=\left(\begin{array}{cc}
x_1^5&0\\
0&x_1^5
\end{array}\right)\otimes
\left(\begin{array}{cc}
0&x_2\\
-x_2^2&0
\end{array}\right)\otimes
\left(\begin{array}{cc}
0&x_3\\
-x_3^2&0
\end{array}\right)\otimes
\left(\begin{array}{cc}
0&x_4\\
-x_4^2&0
\end{array}\right)\otimes
\left(\begin{array}{cc}
0&1\\
-1&0
\end{array}\right)
\end{equation}
The obstruction corresponding to (\ref{12227mod3}) has structure $\frac{2}{7}^0\otimes\frac{4}{7}^0\otimes\frac{4}{7}^0\otimes\frac{4}{7}^0\otimes 0^0$:
\begin{equation}
\label{12227ob3}
\Phi_3=\left(\begin{array}{cc}
x_1^2&0\\
0&x_1^2
\end{array}\right)\otimes
\left(\begin{array}{cc}
x_2^2&0\\
0&x_2^2
\end{array}\right)\otimes
\left(\begin{array}{cc}
x_3^2&0\\
0&x_3^2
\end{array}\right)\otimes
\left(\begin{array}{cc}
x_4^2&0\\
0&x_4^2
\end{array}\right)\otimes
\left(\begin{array}{cc}
1&0\\
0&1
\end{array}\right)
\end{equation}
This is proportional to the bulk deformation $x_1^2x_2^2x_3^2x_4^2$.\\
Finally, we have the boson $\frac{2}{7}^1\otimes\frac{4}{7}^0\otimes\frac{4}{7}^0\otimes\frac{4}{7}^0\otimes 0^1$ which is associated with (\ref{12227mod4}):
\begin{equation}
\label{12227ob4}
\Phi_4=\left(\begin{array}{cc}
0&x_1^2\\
-x_1^2&0
\end{array}\right)\otimes
\left(\begin{array}{cc}
x_2^2&0\\
0&x_2^2
\end{array}\right)\otimes
\left(\begin{array}{cc}
x_3^2&0\\
0&x_3^2
\end{array}\right)\otimes
\left(\begin{array}{cc}
x_4^2&0\\
0&x_4^2
\end{array}\right)\otimes
\left(\begin{array}{cc}
0&1\\
-1&0
\end{array}\right)
\end{equation}
\subsection{Massey Products and F--terms}
\subsubsection{Four Moduli}
\label{sec-12227fourmod}
We start by discussing the brane with labels $L=(6,2,2,2,0)$. This brane has four moduli, which are listed in (\ref{12227mod1})--(\ref{12227mod4}). The bulk deformation $\phi_1=x_1^8x_2x_3x_3$ is $Q$--exact. It can be produced in the matrix factorization by the following deformation:
\begin{equation}
\alpha_{(1,0,0,0,0)}=x_1x_2x_3x_4\bar{\eta}_1
\end{equation}
The first set of non--zero Massey products is:
\begin{eqnarray}
y_{(1,0,1,0,0)}&=&\{\alpha_{(1,0,0,0,0)},\Psi_2\}=-x_1^2x_2^2x_3^2x_4^2\mathbbm{1}=-\Phi_3\nonumber\\
y_{(1,0,0,1,0)}&=&\{\alpha_{(1,0,0,0,0)},\Psi_3\}=-\Phi_2\nonumber\\
y_{(0,2,0,0,0)}&=&\Psi_1\cdot\Psi_1=-x_1^2x_2^2x_3^2x_4^2\mathbbm{1}=-\Phi_3\nonumber\\
y_{(0,1,0,1,0)}&=&\{\Psi_1,\Psi_3\}=-2\Phi_1\nonumber\\
y_{(0,0,2,0,0)}&=&\Psi_2\cdot\Psi_2=-\Phi_3\nonumber\\
y_{(0,0,1,1,0)}&=&\{\Psi_2,\Psi_3\}=-2\Phi_2\nonumber\\
y_{(0,0,0,2,0)}&=&\Psi_3\cdot\Psi_3=-x_1^8x_2x_3x_4\mathbbm{1}=\beta_1(x_1,x_2,x_3,x_4)\nonumber\\
y_{(0,0,0,1,1)}&=&\{\Psi_3,\Psi_4\}=\beta_2(x_1,x_2,x_3,x_4)\nonumber\\
y_{(0,0,0,0,2)}&=&\Psi_4\cdot\Psi_4=x_1^8x_2x_3x_4\mathbbm{1}=-\beta_1(x_1,x_2,x_3,x_4)
\end{eqnarray}
We get three new deformations and, for reasons explained below, we choose the most general parametrization:
\begin{eqnarray}
\alpha_{(0,0,0,1,1)}&=&x_1x_2x_3x_4(\eta_5-\bar{\eta}_5)((2-k_1)\mathbbm{1}-2\eta_1\bar{\eta}_1)\nonumber\\
\alpha_{(0,0,0,0,2)}&=&x_1x_2x_3x_4((-1+k_2)\eta_1-k_2\bar{\eta}_1)\nonumber\\
\alpha_{(0,0,0,2,0)}&=&x_1x_2x_3x_4((1-k_3)\eta_1+k_3\bar{\eta}_1),
\end{eqnarray}
for arbitrary $\{k_1,k_2,k_3\}$.\\
At order $3$, there are six non--zero Massey products two of which get extra contributions due to the obstructions:
\begin{eqnarray}
y_{(0,0,1,0,2)}&=&\{\alpha_{(0,0,0,0,2)},\Psi_2\}=(2k_2-1)\Phi_3\nonumber\\
y_{(0,0,0,3,0)}&=&\{\alpha_{(0,0,0,2,0)},\Psi_3\}=(1-2k_3)\Phi_2\nonumber\\
y_{(0,0,0,2,1)}&=&\{\alpha_{(0,0,0,2,0)},\Psi_4\}+\{\alpha_{(0,0,0,1,1)},\Psi_3\}=-2(k_1-1)\Phi_1\nonumber\\
y_{(0,0,0,1,2)}&=&\{\alpha_{(0,0,0,1,1)},\Psi_4\}+\{\alpha_{(0,0,0,0,2)},\Psi_3\}=(2k_2-1)\Phi_2\nonumber\\
y_{(1,0,0,0,2)}&=&\{\alpha_{(0,0,0,0,2)},\alpha_{(1,0,0,0,0)}\}=(k_2-1)\Phi_3\nonumber\\
y_{(1,0,0,2,0)}&=&\{\alpha_{(0,0,0,2,0)},\alpha_{(1,0,0,0,0)}\}-\frac{1}{2}\{\alpha_{(0,0,0,2,0)},\Psi_2\}=\frac{1}{2}\Phi_3\nonumber\\
y_{(1,0,0,1,1)}&=&\{\alpha_{(0,0,0,1,1)},\alpha_{(1,0,0,0,0)}\}-\frac{1}{2}\{\alpha_{(0,0,0,1,1)},\Psi_2\}=\Phi_4
\end{eqnarray}
Finally, there are five more products at order four:
\begin{eqnarray}
y_{(0,0,0,0,4)}&=&\alpha_{(0,0,0,0,2)}\cdot\alpha_{(0,0,0,0,2)}=k_2(k_2-1)\Phi_3\nonumber\\
y_{(0,0,0,4,0)}&=&\alpha_{(0,0,0,2,0)}\cdot\alpha_{(0,0,0,2,0)}+\frac{1-2k_3}{2}\{\alpha_{(0,0,0,2,0)},\Psi_2\}=\frac{1-2k_3+2k_3^2}{2}\Phi_3\nonumber\\
y_{(0,0,0,3,1)}&=&\{\alpha_{(0,0,0,2,0)},\alpha_{(0,0,0,1,1)}\}-(k_1-1)\{\alpha_{(0,0,0,2,0)},\Psi_1\}+\frac{1-2k_3}{2}\{\alpha_{(0,0,0,1,1)},\Psi_2\}=\Phi_4\nonumber\\
y_{(0,0,0,1,3)}&=&\{\alpha_{(0,0,0,1,1)},\alpha_{(0,0,0,0,2)}\}+\frac{2k_2-1}{2}\{\alpha_{(0,0,0,1,1)},\Psi_2\}=-\Phi_4\nonumber\\
y_{(0,0,0,2,2)}&=&\alpha_{(0,0,0,1,1)}\cdot\alpha_{(0,0,0,1,1)}+\{\alpha_{(0,0,0,0,2)},\alpha_{(0,0,0,2,0)}\}-(k_1-1)\{\alpha_{(0,0,0,1,1)},\Psi_1\}\nonumber\\
&&+\frac{2k_2-1}{2}\{\alpha_{(0,0,0,2,0)},\Psi_2\}=-\frac{1+4k_1-2k_1^2}{2}\Phi_3
\end{eqnarray}
With that, all higher order directions are obstructed and the algorithm terminates. Collecting the contributions to the F--terms, we find:
\begin{eqnarray}
f_1:&\quad& -2u_1u_3-2(k_1-1)u_3^2u_4=0\nonumber\\
f_2:&\quad&-\varphi_1u_3-2u_2u_3+(1-2k_3)u_3^3+(2k_2-1)u_3u_4^2=0 \nonumber\\
f_3:&\quad&\varphi_2-\varphi_1u_2-u_1^2-u_2^2+(k_2-1)\varphi_1u_4^2+(2k_2-1)u_2u_4^2+\frac{1}{2}\varphi_1u_3^2+\frac{1-2k_3+2k_3^2}{2}u_3^4\nonumber\\
&\quad&-\frac{1+4k_1-2k_1^2}{2}u_3^2u_4^2-k_2(k_2-1)u_4^4=0 \nonumber\\
f_4:&\quad&\varphi_1u_3u_4+u_3^3u_4-u_3u_4^3=0
\end{eqnarray}
Surprisingly, we find that these equations are only integrable to an effective superpotential if we choose a particular parametrization of the deformations, namely:
\begin{equation}
k_1=1\qquad k_2=k_3=\frac{1}{2}
\end{equation}
With that, we get the following $\mathcal{W}_{eff}$:
\begin{eqnarray}
\mathcal{W}_{eff}&=&\varphi_2u_3-u_1^2u_3-\varphi_1u_2u_3-u_2^2u_3+\frac{1}{6}\varphi_1u_3^3+\frac{1}{20}u_3^5-\frac{1}{2}\varphi_1u_3u_4^2-\frac{1}{2}u_3^3u_4^2+\frac{1}{4}u_3u_4^4\nonumber\\
\end{eqnarray}
The deformed matrix factorization is:
\begin{eqnarray}
Q_{def}&=&Q+\varphi_1\alpha_{(1,0,0,0,0)}+u_1\Psi_1+u_2\Psi_2+u_3\Psi_3+u_4\Psi_4\nonumber\\
&&+u_3u_4\alpha_{(0,0,0,1,1)}+u_3^2\alpha_{(0,0,0,2,0)}+u_4^2\alpha_{(0,0,0,0,2)}
\end{eqnarray}
Squaring this and inserting the particular values for the $k_i$, one does not find precisely the F--terms listed above but rather:
\begin{eqnarray}
\bar{f}_1:&\quad& -2u_1u_3=0\nonumber\\
\bar{f}_2:&\quad&-\varphi_1u_3-2u_2u_3=0 \nonumber\\
\bar{f}_3:&\quad&\varphi_2-\varphi_1u_2-u_1^2-u_2^2\pm 2u_1u_3u_4+\frac{1}{2}\varphi_1u_3^2-\frac{1}{2}\varphi_1u_4^2-\frac{3}{2}u_3^4u_4^2+\frac{1}{4}u_3^4+\frac{1}{4}u_4^4=0\nonumber\\
\bar{f}_4:&\quad&2u_2u_3u_4-u_3^3u_4+u_3u_4^3=0\quad \mathrm{or}\nonumber\\
&\quad&2\varphi_1u_3u_4+2u_2u_3u_4+u_3^3u_4-u_3u_4^3=0
\end{eqnarray}
This is one of the examples where there are different, yet consistent, constraints in front of the entries of the obstructions $\Phi_i$. This phenomenon may be related to the observation that not every allowed choice of higher deformations leads to integrable F--terms.
\subsubsection*{Correlators}
Computing disk amplitudes with the Kapustin--Li residue formula, we find the following non--zero ones:
\begin{eqnarray}
\langle\Psi_1\Psi_1\Psi_3\rangle&=&-1\nonumber\\
\langle\Psi_2\Psi_2\Psi_3 \rangle&=&-1\nonumber\\
\langle\Psi_1\Psi_2\Psi_4\rangle=-\langle\Psi_1\Psi_4\Psi_2\rangle&=&\phantom{-}1\nonumber\\
\langle\Psi_3\phi_2\rangle&=&\phantom{-}1
\end{eqnarray}
Since the contribution of the the correlators $\langle\Psi_1\Psi_2\Psi_4\rangle$ and $\langle\Psi_1\Psi_4\Psi_2\rangle$ to the effective superpotential cancel, this result is in agreement with the F--terms.
\subsubsection{Three Moduli}
\label{sec-12227threemod}
There is just one tensor product brane which has three open moduli. It has label $L=(4,2,2,2,0)$. The moduli have the same charges as (\ref{12227mod1}), (\ref{12227mod3}) and (\ref{12227mod4}). For this brane, the bulk deformation $\phi_1$ is $Q$--exact. We define:
\begin{equation}
\alpha_{(1,0,0,0)}=x_1^5x_2x_3x_4\bar{\eta}_1
\end{equation}
At order two in deformation theory the following Massey products are non--zero:
\begin{eqnarray}
y_{(1,0,1,0)}&=&\{\alpha_{(1,0,0,0)},\Psi_3\}=\beta_1(x_1,x_2,x_3,x_4)\nonumber\\
y_{(0,2,0,0)}&=&\Psi_1\cdot\Psi_1=-x_1^2x_2^2x_3^2x_4^2\mathbbm{1}=-\Phi_3\nonumber\\
y_{(0,1,1,0)}&=&\{\Psi_1,\Psi_3\}=-2\Phi_1\nonumber\\
y_{(0,0,2,0)}&=&\Psi_3\cdot\Psi_3=-x_1^8x_2x_3x_4\mathbbm{1}=\beta_2(x_1,x_2,x_3,x_4)\nonumber\\
y_{(0,0,1,1)}&=&\{\Psi_3,\Psi_4\}=\beta_3(x_1,x_2,x_3,x_4)\nonumber\\
y_{(0,0,0,2)}&=&\Psi_4\cdot\Psi_4=x_1^8x_2x_3x_4\mathbbm{1}=-\beta_2(x_1,x_2,x_3,x_4)
\end{eqnarray}
The four exact Massey products lead to new deformations for which we will again choose the most general possible deformation: 
\begin{eqnarray}
\alpha_{(0,0,2,0)}=-\alpha_{(0,0,0,2)}&=&\alpha_{(1,0,0,0)}\nonumber\\
\alpha_{(1,0,1,0)}&=&x_2x_3x_4(\eta_5-\bar{\eta}_5)[-\eta_2\eta_3\eta_4\bar{\eta}_1-x_4\eta_2\eta_3\bar{\eta}_1\bar{\eta}_4+x_3\eta_2\eta_4\bar{\eta}_1\bar{\eta}_3-x_3\eta_3\eta_4\bar{\eta}_1\bar{\eta}_2\nonumber\\
&&-x_3x_4\eta_2\bar{\eta}_1\bar{\eta}_3\bar{\eta}_4+x_2x_4\eta_3\bar{\eta}_1\bar{\eta}_2\bar{\eta}_4-x_2x_3\eta_4\bar{\eta}_1\bar{\eta}_2\bar{\eta}_3-x_2x_3x_4\bar{\eta}_1\bar{\eta}_2\bar{\eta}_3\bar{\eta}_4]\nonumber\\
\alpha_{(0,0,1,1)}&=&x_1x_2x_3x_4(\eta_5-\bar{\eta}_5)((2-k_1)\mathbbm{1}-2\eta_1\bar{\eta}_1)
\end{eqnarray}
At order $3$, there are four non--zero Massey products:
\begin{eqnarray}
y_{(1,0,2,0)}&=&\{\alpha_{(1,0,1,0)},\Psi_3\}=-x_1^2x_2^2x_3^2x_4^2\mathbbm{1}=-\Phi_3\nonumber\\
y_{(0,0,3,0)}&=&\{\alpha_{(0,0,2,0)},\Psi_3\}=\beta_1(x_1,x_2,x_3,x_4)\nonumber\\
y_{(0,0,2,1)}&=&\{\alpha_{(0,0,1,1)},\Psi_3\}=-2(k_1-1)\Phi_1\nonumber\\
y_{(0,0,1,2)}&=&\{\alpha_{(0,0,1,1)},\Psi_4\}+\{\alpha_{(0,0,0,2)},\Psi_3\}=-\beta_1(x_1,x_2,x_3,x_4)
\end{eqnarray}
Computing the third order deformations $\alpha_{(0,0,3,0)}$ and $\alpha_{(0,0,1,2)}$ is simple, since they coincide with a second order deformation:
\begin{eqnarray}
\alpha_{(0,0,3,0)}=-\alpha_{(0,0,1,2)}=\alpha_{(1,0,1,0)}
\end{eqnarray}
At order $4$ there are two non--zero Massey products:
\begin{eqnarray}
y_{(0,0,4,0)}&=&\alpha_{(0,0,2,0)}\cdot\alpha_{(0,0,2,0)}+\{\alpha_{(0,0,3,0)},\Psi_3\}=-x_1^2x_2^2x_3^2x_4^2\mathbbm{1}=-\Phi_3 \nonumber\\
y_{(0,0,2,2)}&=&\alpha_{(0,0,1,1)}\cdot\alpha_{(0,0,1,1)}+\{\alpha_{(0,0,0,2)},\alpha_{(0,0,2,0)}\}+\{\alpha_{(0,0,1,2)},\Psi_3\}\nonumber\\
&&-\{\alpha_{(0,0,1,1)},\Psi_1\}=(k_1-1)^2\Phi_3
\end{eqnarray}
We do not get additional deformations and one can show that all higher order Massey products are $0$. Collecting the Massey products which give obstructions, we find the following F--terms:
\begin{eqnarray}
f_1:&\quad&-2u_1u_3-2(k_1-1)u_3^2u_4=0\nonumber\\
f_3:&\quad&\varphi_2-u_1^2-\varphi_1u_3^2-u_3^4+(k_1-1)^2u_3^2u_4^2=0\nonumber\\
f_4:&\quad&0
\end{eqnarray}
Like in the the four--moduli case these equations integrate to an effective superpotential only if $k_1=1$. In that case we get:
\begin{equation}
\mathcal{W}_{eff}=\varphi_2u_3-u_1^2u_3-\frac{1}{3}\varphi_1u_3^3-\frac{1}{5}u_3^5
\end{equation}
The deformed matrix factorization is:
\begin{eqnarray}
Q_{def}&=&Q+\varphi_1\alpha_{(1,0,0,0)}+u_1\Psi_1+u_3\Psi_3+u_4\Psi_4+u_3^2\alpha_{(0,0,2,0)}+u_4^2\alpha_{(0,0,0,2)}\nonumber\\
&&+\varphi_1u_3\alpha_{(1,0,1,0)}+u_3u_4\alpha_{(0,0,1,1)}+u_3^3\alpha_{(0,0,3,0)}+u_3u_4^2\alpha_{(0,0,1,2)}
\end{eqnarray}
This squares to the deformed Landau--Ginzburg superpotential modulo F--terms. Furthermore, there is an additional new feature. There are further terms in $Q_{def}^2$ which are proportional to a matrix which is not in the $Q$--cohomology. Normally, this indicates that one has made a mistake in the deformation theory calculation. 
One finds however that this term comes with a prefactor $u_1u_3$ which is nothing but the F--term associated to $\Phi_1$. Therefore, this additional contribution is ugly but consistent.\\
The F--terms one gets from $Q_{def}^2$ with $k_1=1$ are:
\begin{eqnarray}
\bar{f}_1:&\quad&-2u_1u_3=0\nonumber\\
\bar{f}_3:&\quad&\varphi_2-u_1^2-\varphi_1u_3^2-u_3^4\pm2u_1u_3u_4=0\nonumber\\
\bar{f}_4:&\quad&0
\end{eqnarray}
Here, again, the monomial entries in $\Phi_3$ come with different but consistent constraints.
\subsubsection*{Correlators}
All the two-- and three--point correlators have values compatible with the F--terms:
\begin{eqnarray}
\langle\Psi_1\Psi_1\Psi_3\rangle&=&-1\nonumber\\
\langle\Psi_3\phi_2\rangle&=&\phantom{-}1
\end{eqnarray}
\subsubsection{Two Moduli -- Case A}
One class of tensor product branes with two moduli has open string states (\ref{12227mod1}) and (\ref{12227mod2}). We will discuss the example $L=(6,2,2,1,0)$. For this brane, and for all the others in this class, both bulk deformations are $Q$--exact. These are produced by the following first order deformations of the matrix factorization:
\begin{eqnarray}
\alpha_{(1,0,0,0)}&=&x_1x_2x_3x_4\bar{\eta}_1\nonumber\\
\alpha_{(0,1,0,0)}&=&x_1^2x_2^2x_3^2\bar{\eta}_4
\end{eqnarray}
There are two non--zero Massey products at order $2$:
\begin{eqnarray}
y_{(1,0,0,1)}&=&\{\alpha_{(1,0,0,0)},\Psi_2\}=-x_1^2x_2^2x_3^2x_4^2\mathbbm{1}=\beta_1(x_1,x_2,x_3,x_4)\nonumber\\
y_{(0,0,2,0)}&=&\Psi_1\cdot\Psi_1=-x_1^2x_2^2x_3^2x_4^2\mathbbm{1}=\beta_1(x_1,x_2,x_3,x_4)\nonumber\\
y_{(0,0,0,2)}&=&\Psi_2\cdot\Psi_2=-x_1^2x_2^2x_3^2x_4^2\mathbbm{1}=\beta_1(x_1,x_2,x_3,x_4)
\end{eqnarray}
Both of these expression are $Q$--exact and we find $\alpha_{(1,0,0,1)}=\alpha_{(0,0,2,0)}=\alpha_{(0,0,0,2)}=\alpha_{(0,1,0,0)}$. Since $\alpha_{(0,1,0,0)}$ anticommutes with everybody else, there are no higher order Massey products, and we are done. All the F--terms are $0$, and the moduli are unobstructed. The deformed matrix factorization is:
\begin{eqnarray}
Q_{def}&=&Q+\varphi_1\alpha_{(1,0,0,0)}+\varphi_2\alpha_{(0,1,0,0)}+u_1\Psi_1+u_2\Psi_2+u_2^2\alpha_{(0,0,2,0)}+\varphi_1u_2\alpha_{(1,0,0,1)}\nonumber\\
&&+u_1^2\alpha_{(0,0,2,0)}+u_2^2\alpha_{(0,0,0,2)}
\end{eqnarray}
\subsubsection*{Correlators}
All the correlators which can be easily computed are zero for this brane. This is in agreement with the Massey products.
\subsubsection{Two Moduli -- Case B}
The second type of brane with two moduli is represented by the brane with label $L=(3,2,2,2,0)$. In this model, the bulk deformation $\phi_1=x_1^8x_2x_3x_4$ is in the $Q$--cohomology. The other bulk deformation is obtained by the following deformation of the matrix factorization:
\begin{equation}
\alpha_{(1,0,0)}=x_1^4x_2x_3x_4\bar{\eta}_1
\end{equation}
The brane moduli have the same charges as (\ref{12227mod1}) and (\ref{12227mod3}). The following Massey products are non--zero at order two:
\begin{eqnarray}
y_{(1,0,1)}&=&\{\alpha_{(1,0,0)},\Psi_3\}=\beta_1(x_1,x_2,x_3,x_4)\nonumber\\
y_{(0,2,0)}&=&\Psi_1\cdot\Psi_1=-x_1^1x_2^2x_3^2x_4^2\mathbbm{1}=-\Phi_3\nonumber\\
y_{(0,1,1)}&=&\{\Psi_1,\Psi_3\}=2\Phi_1\nonumber\\
y_{(0,0,2)}&=&\Psi_3\cdot\Psi_3=-x_1^8x_2x_3x_4\mathbbm{1}=\beta_2(x_1,x_2,x_3,x_4)
\end{eqnarray}
The exact terms lead to second order deformations of the matrix factorization, which we call $\alpha_{(1,0,1)}$ and $\alpha_{(0,0,2)}$:
\begin{eqnarray}
\alpha_{(0,0,2)}&=&\alpha_{(1,0,0,0)}\nonumber\\
\alpha_{(1,0,1)}&=&x_1x_2x_3x_4(\eta_5-\bar{\eta}_5)[-\eta_2\eta_3\eta_4\bar{\eta}_1-x_4\eta_2\eta_3\bar{\eta}_1\bar{\eta}_4+x_3\eta_2\eta_4\bar{\eta}_1\bar{\eta}_3-x_3\eta_3\eta_4\bar{\eta}_1\bar{\eta}_2\nonumber\\
&&-x_3x_4\eta_2\bar{\eta}_1\bar{\eta}_3\bar{\eta}_4+x_2x_4\eta_3\bar{\eta}_1\bar{\eta}_2\bar{\eta}_4-x_2x_3\eta_4\bar{\eta}_1\bar{\eta}_2\bar{\eta}_3-x_2x_3x_4\bar{\eta}_1\bar{\eta}_2\bar{\eta}_3\bar{\eta}_4]
\end{eqnarray}
At order $3$, we there are only two non--zero Massey products:
\begin{eqnarray}
y_{(1,0,2)}&=&\{\alpha_{(1,0,1)},\Psi_3\}=-x_1^2x_2^2x_3^2x_4^2\mathbbm{1}=-\Phi_3\nonumber\\
y_{(0,0,3)}&=&\{\alpha_{(0,0,2)},\Psi_3\}=\beta_1(x_1,x_2,x_3,x_4)
\end{eqnarray}
We get another deformation of $Q$:
\begin{eqnarray}
\alpha_{(0,0,3)}&=&\alpha_{(1,0,1)}
\end{eqnarray}
 At order $4$, only one Massey product is allowed:
\begin{eqnarray}
y_{(0,0,4)}&=&\alpha_{(0,0,2)}\cdot\alpha_{(0,0,2)}+\{\alpha_{(0,0,3)},\Psi_3\}=-\Phi_3
\end{eqnarray}
After this step, the algorithm terminates. There are two F--terms:
\begin{eqnarray}
f_1:&\quad&2u_1u_3=0\nonumber \\
f_3:&\quad&-\varphi_2+u_1^2+\varphi_1u_3^2+u_3^4=0 
\end{eqnarray}
There are two solutions to these equations. The one where $u_3=0$ implies the existence of a BPS domain wall, the other one is the solution of a quartic equation. This could be a more general domain wall which separates four vacua. \\
Finally, we also give the deformed matrix factorization:
\begin{eqnarray}
Q_{def}&=&Q+\varphi_1\alpha_{(1,0,0)}+u_1\Psi_1+u_3\Psi_3+\varphi_1u_2\alpha_{(1,0,1)}+u_3^2\alpha_{(0,0,2)}+u_3^4\alpha_{(0,0,3)}
\end{eqnarray}
Up to the F-terms and exact pieces, this squares to the bulk deformed Landau--Ginzburg superpotential. As in the three--parameter case, $Q_{def}^2$ also contains terms which are not in the $Q$--cohomology but all these terms come multiplies with $u_1u_3$ which is the F--term associates to $\Phi_1$.\\
The F--terms can be integrated to give the following effective superpotential:
\begin{equation}
\mathcal{W}_{eff}=u_1^2u_3-\varphi_2u_3+\frac{1}{3}\varphi_1u_3^3+\frac{1}{5}u_3^5
\end{equation}
This result is the same as the one found in the three--moduli case.
\subsubsection*{Correlators}
There are two non--zero correlators which can be computed using the residue formula:
\begin{eqnarray}
\langle\Psi_1\Psi_1\Psi_3 \rangle&=&-1\nonumber\\
\langle \Psi_3\phi_2\rangle&=&\phantom{-}1
\end{eqnarray}
This is in agreement with the deformation theory calculation.
\subsubsection{One Modulus}
The brane with the largest $L$--labels which has one modulus is the one with $L=(4,2,2,1,0)$. The brane modulus is given by (\ref{12227mod1}). In the present case both bulk deformations are $Q$--exact. They are tied to the following deformations of the brane:
\begin{eqnarray}
\alpha_{(1,0,0)}&=&x_1^3x_2x_3x_4\bar{\eta}_1\nonumber\\
\alpha_{(0,1,0)}&=&x_1^2x_2^2x_3^2\bar{\eta}_4
\end{eqnarray}
At order $2$ in deformation theory, there is only one non--vanishing Massey product:
\begin{equation}
y_{(0,0,2)}=\Psi_1\cdot\Psi_1=-x_1^2x_2^2x_3^2x_4^2\mathbbm{1}
\end{equation}
This is $Q$--exact and the corresponding second order deformation is $\alpha_{(0,0,2)}=\alpha_{(0,1,0)}$. There are no more higher products which are non--zero. The deformed matrix factorization is:
\begin{equation}
Q_{def}=Q+\varphi_1\alpha_{(1,0,0)}+\varphi_2\alpha_{(0,1,0)}+u_1\Psi_1+u_1^2\alpha_{(0,0,2)}
\end{equation}
This squares to the deformed Landau--Ginzburg superpotential. All the F--terms are zero. Therefore, the deformation of this brane is unobstructed.
\subsubsection*{Correlators}
In agreement with the Massey products all the two-- and three--point correlators vanish.
\section{The model $\mathbbm{P}(11169)[18]/(\mathbbm{Z}_{18})^2$}
The degree $18$ hypersurface has the the following Landau--Ginzburg superpotential:
\begin{equation}
W=x_1^{18}+x_2^{18}+x_3^{18}+x_4^3+x_5^2
\end{equation}
There is the following $(\mathbbm{Z}_{18})^2$ orbifold action:
\begin{eqnarray}
g_1:&&(1,17,0,0,0)\nonumber\\
g_2:&&(1,0,17,0,0)
\end{eqnarray}
The two bulk moduli are:
\begin{eqnarray}
\phi_1&=&x_1^6x_2^6x_3^6\nonumber\\
\phi_2&=&x_1x_2x_3x_4x_5\nonumber
\end{eqnarray}
Again $\phi_2$ can be replaced by $\phi_2=x_1^2x_2^2x_3^2x_4^2$. In the tables in the appendix we list the branes which have open moduli at the Gepner point.
\subsection{Discussion of Moduli}
Now we discuss the moduli of the brane $L=(8,8,8,0,0)$. The modulus with label $\frac{1}{3}^1\otimes\frac{1}{3}^1\otimes\frac{1}{3}^1\otimes 0^0\otimes 0^0$ looks as follows:
\begin{equation}
\label{11169mod1}
\Psi_1=\left(\begin{array}{cc}
0&x_1^3\\
-x_1^3&0
\end{array}\right)\otimes
\left(\begin{array}{cc}
0&x_2^3\\
-x_2^3&0
\end{array}\right)\otimes
\left(\begin{array}{cc}
0&x_3^3\\
-x_3^3&0
\end{array}\right)\otimes
\left(\begin{array}{cc}
1&0\\
0&1
\end{array}\right)\otimes
\left(\begin{array}{cc}
1&0\\
0&1
\end{array}\right)
\end{equation}
The modulus $\frac{1}{3}^0\otimes\frac{1}{3}^0\otimes\frac{1}{3}^0\otimes 0^0\otimes 0^1$ is expressed by the following matrix:
\begin{equation}
\label{11169mod2}
\Psi_2=\left(\begin{array}{cc}
x_1^3&0\\
0&x_1^3
\end{array}\right)\otimes
\left(\begin{array}{cc}
x_2^3&0\\
0&x_2^3
\end{array}\right)\otimes
\left(\begin{array}{cc}
x_3^3&0\\
0&x_3^3
\end{array}\right)\otimes
\left(\begin{array}{cc}
1&0\\
0&1
\end{array}\right)\otimes
\left(\begin{array}{cc}
0&1\\
-1&0
\end{array}\right)
\end{equation}
The modulus with label $\frac{2}{9}^1\otimes\frac{2}{9}^1\otimes\frac{2}{9}^1\otimes\frac{1}{3}^1\otimes 0^1$ looks like this:
\begin{equation}
\label{11169mod3}
\Psi_3=\left(\begin{array}{cc}
0&x_1^2\\
-x_1^2&0
\end{array}\right)\otimes
\left(\begin{array}{cc}
0&x_2^2\\
-x_2^2&0
\end{array}\right)\otimes
\left(\begin{array}{cc}
0&x_3^2\\
-x_3^2&0
\end{array}\right)\otimes
\left(\begin{array}{cc}
0&1\\
-x_4&0
\end{array}\right)\otimes
\left(\begin{array}{cc}
0&1\\
-1&0
\end{array}\right)
\end{equation}
The last open modulus is has charges $\frac{2}{9}^0\otimes\frac{2}{9}^0\otimes\frac{2}{9}^0\otimes\frac{1}{3}^1\otimes 0^0$:
\begin{equation}
\label{11169mod4}
\Psi_4=\left(\begin{array}{cc}
x_1^2&0\\
0&x_1^2
\end{array}\right)\otimes
\left(\begin{array}{cc}
x_2^2&0\\
0&x_2^2
\end{array}\right)\otimes
\left(\begin{array}{cc}
x_3^2&0\\
0&x_3^2
\end{array}\right)\otimes
\left(\begin{array}{cc}
0&1\\
-x_4&0
\end{array}\right)\otimes
\left(\begin{array}{cc}
1&0\\
0&1
\end{array}\right)
\end{equation}
\subsection{Obstructions}
Let us now discuss the Serre dual obstructions to the deformations (\ref{11169mod1})--(\ref{11169mod4}). The obstruction corresponding to (\ref{11169mod1}) has charges $\frac{5}{9}^0\otimes\frac{5}{9}^0\otimes\frac{5}{9}^0\otimes\frac{1}{3}^1\otimes 0^1$:
\begin{equation}
\label{11169ob1}
\Phi_1=\left(\begin{array}{cc}
x_1^5&0\\
0&x_1^5
\end{array}\right)\otimes
\left(\begin{array}{cc}
x_2^5&0\\
0&x_2^5
\end{array}\right)\otimes
\left(\begin{array}{cc}
x_3^5&0\\
0&x_3^5
\end{array}\right)\otimes
\left(\begin{array}{cc}
0&1\\
-x_4&0
\end{array}\right)\otimes
\left(\begin{array}{cc}
0&1\\
-1&0
\end{array}\right)
\end{equation}
The Serre dual of (\ref{11169mod2}) is $\frac{5}{9}^1\otimes\frac{5}{9}^1\otimes\frac{5}{9}^1\otimes\frac{1}{3}^1\otimes 0^0$
\begin{equation}
\label{11169ob2}
\Phi_2=\left(\begin{array}{cc}
0&x_1^5\\
-x_1^5&0
\end{array}\right)\otimes
\left(\begin{array}{cc}
0&x_2^5\\
-x_2^5&0
\end{array}\right)\otimes
\left(\begin{array}{cc}
0&x_3^5\\
-x_3^5&0
\end{array}\right)\otimes
\left(\begin{array}{cc}
0&1\\
-x_4&0
\end{array}\right)\otimes
\left(\begin{array}{cc}
1&0\\
0&1
\end{array}\right)
\end{equation}
The obstruction corresponding to (\ref{11169mod3}) has charges $\frac{2}{3}^0\otimes\frac{2}{3}^0\otimes\frac{2}{3}^0\otimes0^0\otimes 0^0$ and is the bulk deformation $x_1^6x_2^6x_3^6$:
\begin{equation}
\label{11169ob3}
\Phi_3=\left(\begin{array}{cc}
x_1^6&0\\
0&x_1^6
\end{array}\right)\otimes
\left(\begin{array}{cc}
x_2^6&0\\
0&x_2^6
\end{array}\right)\otimes
\left(\begin{array}{cc}
x_3^6&0\\
0&x_3^6
\end{array}\right)\otimes
\left(\begin{array}{cc}
1&0\\
0&1
\end{array}\right)\otimes
\left(\begin{array}{cc}
1&0\\
0&1
\end{array}\right)
\end{equation}
Finally, the Serre dual of (\ref{11169mod4}) is $\frac{2}{3}^1\otimes\frac{2}{3}^1\otimes\frac{2}{3}^1\otimes0^0\otimes 0^1$:
\begin{equation}
\label{11169ob4}
\Phi_4=\left(\begin{array}{cc}
0&x_1^6\\
-x_1^6&0
\end{array}\right)\otimes
\left(\begin{array}{cc}
0&x_2^6\\
-x_2^6&0
\end{array}\right)\otimes
\left(\begin{array}{cc}
0&x_3^6\\
-x_3^6&0
\end{array}\right)\otimes
\left(\begin{array}{cc}
1&0\\
0&1
\end{array}\right)\otimes
\left(\begin{array}{cc}
0&1\\
-1&0
\end{array}\right)
\end{equation}
\subsection{Massey Products and F--terms}
Finally, we discuss higher products and F--terms for this model. It turns out that the calculation for the three--moduli case is quite involved. We contend ourselves to compute deformation theory only up to order $5$.
\subsubsection{Four Moduli}
\label{sec-11169fourmod}
Our representative example will be the brane $L=(8,8,8,0,0)$. The moduli were given explicitly in (\ref{11169mod1})--(\ref{11169mod4}). For the four--moduli brane, only the bulk deformations $\phi_2=x_1^2x_2^2x_3^2x_4^2$ is $Q$--exact. The associated boundary deformation is:
\begin{equation}
\alpha_{(1,0,0,0,0)}=x_1^2x_2^2x_3^2x_4\bar{\eta}_4
\end{equation}
Below, we list the non--zero Massey products at order $2$:
\begin{eqnarray}
y_{(1,0,0,1,0)}&=&\{\alpha_{(1,0,0,0,0)},\Psi_3\}=\beta_1(x_1,x_2,x_3,x_4)\nonumber\\
y_{(1,0,0,0,1)}&=&\{\alpha_{(1,0,0,0,0)},\Psi_4\}=x_1^4x_2^4x_3^4x_4\mathbbm{1}=\beta_2(x_1,x_2,x_3,x_4)\nonumber\\
y_{(0,2,0,0,0)}&=&\Psi_1\cdot\Psi_1=\Phi_3\nonumber\\
y_{(0,1,0,1,0)}&=&\{\Psi_1,\Psi_3\}=2\Phi_1\nonumber\\
y_{(0,0,2,0,0)}&=&\Psi_2\cdot\Psi_2=-x_1^6x_2^6x_3^6\mathbbm{1}=-\Phi_3\nonumber\\
y_{(0,0,1,1,0)}&=&\{\Psi_2,\Psi_3\}=-2\Phi_2\nonumber\\
y_{(0,0,0,2,0)}&=&\Psi_3\cdot\Psi_3=-x_1^4x_2^4x_3^4x_4\mathbbm{1}=-\beta_2(x_1,x_2,x_3,x_4)\nonumber\\
y_{(0,0,0,1,1)}&=&\{\Psi_3,\Psi_4\}=-2\beta_1(x_1,x_2,x_3,x_4)\nonumber\\
y_{(0,0,0,0,2)}&=&\Psi_4\cdot\Psi_4=-\beta_2(x_1,x_2,x_3,x_4)
\end{eqnarray}
From the $Q$--exact products we get several new $\alpha$'s:
\begin{eqnarray}
-\alpha_{(1,0,0,0,1)}=\alpha_{(0,0,0,2,0)}=\alpha_{(0,0,0,0,2)}&=&x_1^4x_2^4x_3^4\bar{\eta}_4\nonumber\\
\alpha_{(1,0,0,1,0)}=-\frac{1}{2}\alpha_{(0,0,0,1,1)}&=&x_1^4x_2^4x_3^4(\eta_5-\bar{\eta}_5)\bar{\eta}_4(\eta_3-\bar{\eta}_3)(\eta_2-\bar{\eta}_2)(\eta_1-\bar{\eta}_1)\nonumber\\
\end{eqnarray}
At order three, all Massey products are proportional to bosonic open string states:
\begin{eqnarray}
y_{(1,0,0,2,0)}&=&\{\alpha_{(1,0,0,1,0)},\Psi_3\}+\{\alpha_{(0,0,0,2,0)},\alpha_{(1,0,0,0,0)}\}=-\Phi_3\nonumber\\
y_{(1,0,0,1,1)}&=&\{\alpha_{(1,0,0,1,0)},\Psi_4\}+\{\alpha_{(1,0,0,0,1)},\Psi_3\}+\{\alpha_{(0,0,0,1,1)},\alpha_{(1,0,0,0,0)}\}=2\Phi_4\nonumber\\
y_{(1,0,0,0,2)}&=&\{\alpha_{(1,0,0,0,1)},\Psi_4\}+\{\alpha_{(1,0,0,0,0)},\alpha_{(0,0,0,0,2)}\}=-\Phi_3\nonumber\\
y_{(0,0,0,3,0)}&=&\{\alpha_{(0,0,0,2,0)},\Psi_3\}=-\Phi_4\nonumber\\
y_{(0,0,0,2,1)}&=&\{\alpha_{(0,0,0,2,0)},\Psi_4\}+\{\alpha_{(0,0,0,1,1)},\Psi_3\}=3\Phi_3\nonumber\\
y_{(0,0,0,1,2)}&=&\{\alpha_{(0,0,0,1,1)},\Psi_4\}+\{\alpha_{(0,0,0,0,2)},\Psi_3\}=-3\Phi_4\nonumber\\
y_{(0,0,0,0,3)}&=&\{\alpha_{(0,0,0,0,2)},\Psi_4\}=\Phi_3
\end{eqnarray}
Since we get no new deformations at order three the Massey products at order four must be products of the deformations of order $2$. One can show that these all anticommute, which is why the algorithm terminates at order four. Furthermore, the additional terms which arise through the F--terms at lower order all give zero.\\
Collecting all the products which yield obstructions, we obtain the following F--terms:
\begin{eqnarray}
f_1:&\quad&2u_1u_3=0 \nonumber\\
f_2:&\quad&-2u_2u_3=0 \nonumber\\
f_3:&\quad&\varphi_1+u_1^2-u_2^2-\varphi_2u_3^2 -\varphi_2u_4^2+3u_3^2u_4+u_4^3=0\nonumber\\
f_4:&\quad&2\varphi_2u_3u_4 -u_3^3-3u_3u_4^2=0
\end{eqnarray}
This can be integrated to the following effective superpotential:
\begin{equation}
\mathcal{W}_{eff}=u_1^2u_3-u_2^2u_3+\varphi_1u_3-\frac{1}{3}\varphi_2u_3^3-\varphi_2u_3u_4^2+u_3^3u_4+u_3u_4^3
\end{equation}
The deformed matrix factorization looks as follows:
\begin{eqnarray}
Q_{def}&=&Q+\varphi_2\alpha_{(1,0,0,0,0)}+u_1\Psi_1+u_2\Psi_2+u_3\Psi_3+u_4\Psi_4\nonumber\\
&&+\varphi_2u_3\alpha_{(1,0,0,1,0)}+\varphi_2u_4\alpha_{(1,0,0,0,1)}+u_3u_4\alpha_{(0,0,0,1,1)}+u_3^2\alpha_{(0,0,0,2,0)}+u_4^2\alpha_{(0,0,0,0,2)}\nonumber\\
\end{eqnarray}
Up to the F--terms and exact expressions, this squares to the deformed Landau--Ginzburg superpotential. As encountered in the $12227$--model there are additional terms in $Q_{def}^2$ which are not in the cohomology of the matrix factorization. This is not inconsistent because these terms are proportional to the F--terms associated to $\Phi_1$ and $\Phi_2$.
\subsubsection*{Correlators}
The following correlators are non--zero:
\begin{eqnarray}
\langle\Psi_1\Psi_1\Psi_3\rangle&=&\phantom{-}1\nonumber\\
\langle\Psi_2\Psi_2\Psi_3\rangle&=&-1\nonumber\\
\langle\Psi_1\Psi_2\Psi_4\rangle=-\langle\Psi_1\Psi_4\Psi_2\rangle&=&-1\nonumber\\
\langle\Psi_3\phi_1\rangle&=&\phantom{-}1
\end{eqnarray}
This is in agreement with the deformation theory results.
\subsubsection{Three Moduli}
Let us now discuss a brane which has three moduli. We choose the brane with label $L=(8,8,5,0,0)$. The moduli have the same charge decomposition as (\ref{11169mod1}), (\ref{11169mod2}) and (\ref{11169mod4}). Both bulk deformations are $Q$--exact. The corresponding linear deformations of the matrix factorization are:
\begin{eqnarray}
\alpha_{(1,0,0,0,0)}&=&x_1^6x_2^6\bar{\eta}_3\nonumber\\
\alpha_{(0,1,0,0,0)}&=&x_1^2x_2^2x_3^2x_4\bar{\eta}_4
\end{eqnarray}
In the following, we list the non--zero Massey products at order $2$:
\begin{eqnarray}
y_{(1,0,1,0,0)}&=&\{\alpha_{(1,0,0,0,0)},\Psi_1\}=\beta_1(x_1,x_2)\nonumber\\
y_{(0,1,0,0,1)}&=&\{\alpha_{(0,1,0,0,0)},\Psi_4\}=x_1^4x_2^4x_3^4x_4\mathbbm{1}=\beta_2(x_1,x_2,x_3,x_4)\nonumber\\
y_{(0,0,2,0,0)}&=&\Psi_1\cdot\Psi_1=x_1^6x_2^6x_3^6\mathbbm{1}=\beta_3(x_1,x_2,x_3)\nonumber\\
y_{(0,0,0,2,0)}&=&\Psi_2\cdot\Psi_2=-x_1^6x_2^6x_3^6\mathbbm{1}=-\beta_3(x_1,x_2,x_3)\nonumber\\
y_{(0,0,0,0,2)}&=&-x_1^4x_2^4x_3^4x_4\mathbbm{1}=-\beta_2(x_1,x_2,x_3,x_4)
\end{eqnarray}
All these Massey products give contributions to the deformations of the matrix factorizations. These are:
\begin{eqnarray}
\alpha_{(0,0,0,0,2)}=-\alpha_{(0,1,0,0,1)}&=&x_1^4x_2^4x_3^4\bar{\eta}_4\nonumber\\
\alpha_{(0,0,0,2,0)}=-\alpha_{(0,0,2,0,0)}&=&\alpha_{(1,0,0,0,0)}\nonumber\\
\alpha_{(1,0,1,0,0)}&=&x_2^9(\eta_2-\bar{\eta}_2)\eta_1\bar{\eta}_1
\end{eqnarray}
At order $3$, we again have five non--zero Massey products:
\begin{eqnarray}
y_{(1,0,2,0,0)}&=&\{\alpha_{(1,0,1,0,0)},\Psi_1\}+\{\alpha_{(1,0,0,0,0)},\alpha_{(0,0,2,0,0)}\}=\beta_4(x_1,x_2,x_3)\nonumber\\
y_{(0,1,0,0,2)}&=&\{\alpha_{(1,0,0,0,1)},\Psi_4\}=-x_1^6x_2^6x_3^6\mathbbm{1}=-\beta_3(x_1,x_2,x_3)\nonumber\\
y_{(0,0,3,0,0)}&=&\{\alpha_{(0,0,2,0,0)},\Psi_1\}=\beta_5(x_1,x_2)\nonumber\\
y_{(0,0,1,2,0)}&=&\{\alpha_{(0,0,0,2,0)},\Psi_1\}=-\beta_5(x_1,x_2)\nonumber\\
y_{(0,0,0,0,3)}&=&\{\alpha_{(0,0,0,0,2)},\Psi_4\}=x_1^6x_2^6x_3^6\mathbbm{1}=\beta_3(x_1,x_2,x_3)
\end{eqnarray}
The new deformations are:
\begin{eqnarray}
\alpha_{(0,1,0,0,2)}=-\alpha_{(0,0,0,0,3)}&=&-\alpha_{(1,0,0,0,0)}\nonumber\\
\alpha_{(0,0,1,2,0)}=-\alpha_{(0,0,3,0,0)}&=&\alpha_{(1,0,1,0,0)}\nonumber\\
\alpha_{(1,0,2,0,0)}&=&x_1^3x_2^3\eta_2\eta_3(\eta_1-\bar{\eta}_1)-x_1^3x_2^3x_3^6\eta_2\bar{\eta}_3(\eta_1-\bar{\eta}_1) 
\end{eqnarray}
and two more which are more complicated. There are no obstructions and we have to go to order four in deformation theory:
\begin{eqnarray}
y_{(2,0,2,0,0)}&=&\alpha_{(1,0,1,0,0)}\cdot\alpha_{(1,0,1,0,0)}+\{\alpha_{(1,0,2,0,0)},\alpha_{(1,0,0,0,0)}\}=\beta_6(x_1,x_2)\nonumber\\
y_{(1,0,3,0,0)}&=&\{\alpha_{(1,0,1,0,0)},\alpha_{(0,0,2,0,0)}\}+\{\alpha_{(1,0,2,0,0)},\Psi_1\}+\{\alpha_{(0,0,3,0,0)},\alpha_{(1,0,0,0,0)}\}=\beta_3(x_1,x_2,x_3)\nonumber\\
y_{(0,0,4,0,0)}&=&\alpha_{(0,0,2,0,0)}\cdot\alpha_{(0,0,2,0,0)}+\{\alpha_{(0,0,3,0,0)},\Psi_1\}=-\beta_1(x_1,x_2,x_3)\nonumber\\
y_{(0,0,2,2,0)}&=&\{\alpha_{(0,0,2,0,0)},\alpha_{(0,0,0,2,0)}\}+\{\alpha_{(0,0,1,2,0)},\Psi_1\}=\beta_1(x_1,x_2,x_3)\nonumber\\
y_{(0,0,1,0,3)}&=&\{\alpha_{(0,0,0,0,3)},\Psi_1\}=\beta_2(x_1,x_2)
\end{eqnarray}
We get six new exact states:
\begin{eqnarray}
\alpha_{(0,0,1,0,3)}=-\alpha_{(0,1,1,0,3)}&=&\alpha_{(0,0,3,0,0)}\nonumber\\
\alpha_{(0,0,2,2,0)}=-\alpha_{(0,0,4,0,0)}&=&\alpha_{(1,0,2,0,0)}\nonumber\\
\alpha_{(1,0,3,0,0)}&=&-\alpha_{(1,0,0,0,0)}\nonumber\\
\alpha_{(2,0,2,0,0)}&=&x_2^9\eta_2\eta_1\bar{\eta}_1
\end{eqnarray}
At order five there are six non--zero Massey products:
\begin{eqnarray}
y_{(2,0,3,0,0)}&=&\{\alpha_{(1,0,1,0,0)},\alpha_{(1,0,2,0,0)}\}+\{\alpha_{(1,0,3,0,0)},\alpha_{(1,0,0,0,0)}\}+\{\alpha_{(2,0,2,0,0)},\Psi_2\}=\beta_4(x_1,x_2,x_3)\nonumber\\
y_{(1,0,4,0,0)}&=&\{\alpha_{(1,0,1,0,0)},\alpha_{(0,0,3,0,0)}\}+\{\alpha_{(0,0,2,0,0)},\alpha_{(1,0,2,0,0)}\}+\{\alpha_{(0,0,4,0,0)},\alpha_{(1,0,0,0,0)}\}\nonumber\\
&&+\{\alpha_{(1,0,3,0,0)},\Psi_1\}=\beta_7(x_1,x_2)\nonumber\\
y_{(1,0,2,2,0)}&=&\{\alpha_{(1,0,1,0,0)},\alpha_{(0,0,1,2,0)}\}+\{\alpha_{(0,0,0,2,0)},\alpha_{(1,0,2,0,0)}\}+\{\alpha_{(0,0,2,2,0)},\alpha_{(1,0,0,0,0)}\}=\beta_8(x_1,x_2)\nonumber\\
y_{(0,1,2,0,2)}&=&\{\alpha_{(0,0,2,0,0)},\alpha_{(0,1,0,0,2)}\}+\{\alpha_{(0,1,1,0,2)},\Psi_1\}=\beta_4(x_1,x_2,x_3)\nonumber\\
y_{(0,0,5,0,0)}&=&\{\alpha_{(0,0,2,0,0)},\alpha_{(0,0,3,0,0)}\}+\{\alpha_{(0,0,4,0,0)},\Psi_1\}=-x_1^6x_2^6x_3^6\mathbbm{1}=-\beta_3(x_1,x_2,x_3)\nonumber\\
y_{(0,0,2,0,3)}&=&\{\alpha_{(0,0,2,0,0)},\alpha_{(0,0,0,0,3)}\}+\{\alpha_{(0,0,1,0,3)},\Psi_1\}=-\beta_4(x_1,x_2,x_3)
\end{eqnarray}   
We observe recurring patterns and seem to get more and more deformations at every order. This suggests that the deformation theory algorithm may not terminate. Furthermore we cannot exclude that obstructions appear at higher order because structure arguments that imply that the Massey products can never yield obstructions are not obvious.\\
We therefore conclude our discussion with the statement that up to order five there are no obstructions to the deformations of this brane.
\subsubsection*{Correlators}
All the correlators which do not contain integrated insertions are zero. This confirms the results above.
\subsubsection{Two Moduli}
As an example for a brane with two moduli we discuss the brane with labels $L=(8,8,4,0,0)$. The moduli of this brane, and all the other branes with two moduli, are (\ref{11169mod2}) and (\ref{11169mod4}). Furthermore we have two first order boundary deformations since both bulk deformations are exact on the boundary:
\begin{eqnarray}
\alpha_{(1,0,0,0)}&=&x_1^6x_2^6x_3\bar{\eta}_3\nonumber\\
\alpha_{(0,1,0,0)}&=&x_1^2x_2^2x_3^2x_4\bar{\eta}_4
\end{eqnarray} 
The non--zero Massey products at the boundary are:
\begin{eqnarray}
y_{(0,1,0,1)}&=&\{\alpha_{(0,1,0,0)},\Psi_4\}=x_1^4x_2^4x_3^4x_4\mathbbm{1}=\beta_1(x_1,x_2,x_3,x_4)\nonumber\\
y_{(0,0,2,0)}&=&\Psi_2\cdot\Psi_2=-x_1^6x_2^6x_3^6\mathbbm{1}=\beta_2(x_1,x_2,x_3)\nonumber\\
y_{(0,0,0,2)}&=&\Psi_4\cdot\Psi_4=-x_1^4x_2^4x_3^4x_4\mathbbm{1}=-\beta_1(x_1,x_2,x_3,x_4)
\end{eqnarray}
All these products are $Q$--exact and the corresponding second order deformations are the following simple expressions:
\begin{eqnarray}
\alpha_{(0,1,0,1)}=-\alpha_{(0,0,0,2)}&=&-x_1^4x_2^4x_3^4\bar{\eta}_4\nonumber\\
\alpha_{(0,0,2,0)}&=&\alpha_{(1,0,0,0)}
\end{eqnarray}
We continue with computing the higher products at order $3$:
\begin{eqnarray}
y_{(0,1,0,2)}&=&\{\alpha_{(0,1,0,1)},\Psi_4\}+\{\alpha_{(0,0,0,2)},\alpha_{(0,1,0,0)}\}=-x_1^6x_2^6x_3^6\mathbbm{1}=\beta_2(x_1,x_2,x_3)\nonumber\\
y_{(0,0,0,3)}&=&\{\alpha_{(0,0,0,2)},\Psi_4\}=x_1^6x_2^6x_3^6\mathbbm{1}=-\beta_2(x_1,x_2,x_3)
\end{eqnarray}
Again, we encounter no obstructions and the third order deformations are simply:
\begin{eqnarray}
\alpha_{(0,1,0,2)}=-\alpha_{(0,0,0,3)}=\alpha_{(1,0,0,0)}
\end{eqnarray}
Now we are done. Since $\alpha_{(1,0,0,0)}$ commutes with all moduli and all second order deformations the are no non--vanishing Massey products at higher order. All the F--terms remain zero and therefore both brane moduli are unobstructed by bulk deformations. The deformed matrix factorization looks as follows:
\begin{eqnarray}
Q_{def}&=&Q+\varphi_1\alpha_{(1,0,0,0)}+\varphi_2\alpha_{(0,1,0,0)}+u_2\Psi_2+u_4\Psi_4+u_2^2\alpha_{(0,0,2,0)}+u_4^2\alpha_{(0,0,0,2)}\nonumber\\
&&+\varphi_2u_4^2\alpha_{(0,1,0,2)}+u_4^3\alpha_{(0,0,0,3)}
\end{eqnarray}
One can check easily that this squares to the deformed Landau--Ginzburg superpotential.
\subsubsection*{Correlators}
As expected from the deformation theory calculation, all the correlators which are computable by the Kapustin--Li residue formula are zero. This is in agreement with the result that both boundary moduli are unobstructed.
\subsubsection{One Modulus}
The one--modulus brane with the maximal $L$--label has $L=(8,8,2,0,0)$. Its only boundary deformation is (\ref{11169mod4}). The two bulk deformations are $Q$--exact and we introduce the following first order deformations of the matrix factorization:
\begin{eqnarray}
\alpha_{(1,0,0)}&=&x_1^6x_2^6x_3^3\bar{\eta}_3\nonumber\\
\alpha_{(0,1,0)}&=&x_1^2x_2^2x_3^2x_4\bar{\eta}_4
\end{eqnarray}
Computing the first set of Massey products, we find that only two are non--vanishing:
\begin{eqnarray}
y_{(0,1,1)}&=&\{\alpha_{(0,1,0)},\Psi_4\}=x_1^4x_2^4x_3^4x_4\mathbbm{1}=\beta_1(x_1,x_2,x_3,x_4)\nonumber\\
y_{(0,0,2)}&=&\Psi_4\cdot\Psi_4=-x_1^4x_2^4x_3^4x_4\mathbbm{1}=-\beta_1(x_1,x_2,x_3,x_4)
\end{eqnarray}
From this, we obtain two new deformation $\alpha_{(0,1,1)}$ and $\alpha_{(0,0,2)}$ which have a particularly simple representation:
\begin{equation}
\alpha_{(0,1,1)}=-\alpha_{(0,0,2)}=-x_1^4x_2^4x_3^4\bar{\eta}_4
\end{equation}
At the next order, there are two non--zero Massey products:
\begin{eqnarray}
y_{(0,1,2)}&=&\{\alpha_{(0,1,1)},\Psi_4\}+\{\alpha_{(0,0,2)},\alpha_{(0,1,0)}\}=-x_1^6x_2^6x_3^6\mathbbm{1}=\beta_2(x_1,x_2,x_3)\nonumber\\
y_{(0,0,3)}&=&\{\alpha_{(0,0,2)},\Psi_4\}=-x_1^6x_2^6x_3^6\mathbbm{1}=-\beta_2(x_1,x_2,x_3)
\end{eqnarray}
We find that:
\begin{equation}
\alpha_{(0,1,2)}=-\alpha_{(0,0,3)}=\alpha_{(1,0,0)}
\end{equation}
The algorithm terminates here. One can see this as follows. The $\alpha$'s at order three anticommute with the modulus and all the other deformations. Therefore all higher Massey products are $0$. Since we have not encountered any obstructions, the only possible F--term remains $0$. The deformed matrix factorization looks as follows:
\begin{eqnarray}
Q_{def}&=&Q+\varphi_1\alpha_{(1,0,0)}+\varphi_2\alpha_{(0,1,0)}+u_4\Psi_4\nonumber\\
&&+\varphi_2u_4\alpha_{(0,1,1)}+u_4^2\alpha_{(0,0,2)}+\varphi_2u_4^2\alpha_{(0,1,2)}+u_4^3\alpha_{(0,0,3)}
\end{eqnarray}
\subsubsection*{Correlators}
The three--point disk amplitude and the bulk--boundary two--point functions vanish which confirms that the brane modulus is unobstructed.
\section{Conclusions}
\label{sec-conclusions}
In this paper we have discussed tensor product branes and their moduli for the mirrors of two--parameter Calabi--Yau hypersurfaces. Making use of the deformation theory of matrix factorizations we determined which of the brane moduli are obstructed by computing F--terms. Let us now discuss some open problems and further directions of research.\\
There have been some unexpected problems with the deformation theory algorithm itself. In a few cases the F--terms could only be integrated to an effective superpotential for a particular choice of deformations. Furthermore some deformed matrix factorizations exhibited an unusual factorization behavior. Despite the self--consistency of the results, these phenomena might also indicate that the deformation theory algorithm has to be modified in some way. There may also be a deeper reason for these issues which may be related to some special properties of the D--brane under consideration. It would be very useful to have an independent method to determine the structure of the brane moduli spaces in order to verify or falsify these results. \\
Note also that the deformation theory algorithm is only partially useful to probe the combined open/closed moduli space near the Gepner point. The Massey product algorithm is best suited for situations where all boundary moduli are obstructed. Although this should be the case generically for Calabi--Yau threefolds, it depends very much on the model whether this really happens. Only if all moduli are obstructed it is guaranteed that the algorithm really terminates. For unobstructed moduli it should be expected that the algorithm never stops. It is actually quite remarkable that we have found several examples where the algorithm terminates even though the moduli are unobstructed. \\
As was demonstrated in great detail, the Massey product algorithm is a rather tedious and technically challenging way to answer the questions we posed. It may be necessary to find more elegant methods to find out whether a brane has obstructed moduli or not. One immediately thinks of some geometric input which could help. In most cases it is easy to relate a matrix factorization to a boundary state in conformal field theory but the geometric interpretation of these is often very involved. Of course matrix factorizations can be related to geometry via the techniques of \cite{Herbst:2008jq} but simple matrix factorizations do usually not lead to simple geometric boundary conditions. \\
A further possible line of investigation concerns the "fake F--terms" we have found whenever an open modulus squares to a closed string deformation which is exact on the boundary. These conditions lead to a pair of linearly deformed matrix factorizations at a special point in brane moduli space but not to an effective superpotential. Nevertheless, a non--trivial BPS domain wall tension has been found in \cite{Knapp:2008uw} for such a configuration. The effective superpotential only seems to come out correctly when one considers the full non--linear bulk-- and boundary deformations of a D--brane and not just linear boundary deformations. These issues raise the question under which conditions it makes sense to turn on just a subset of the moduli. \\
We have seen that our Massey product algorithm yields an explicit description of the deformation theory of B--type D--branes at the Gepner point. From a different point of view, it describes the $A_\infty$--structure on the category of matrix factorizations. By the open string version of the Calabi--Yau/Landau--Ginzburg correspondence~\cite{Herbst:2008jq}, we have an explicit map to the category of coherent sheaves at large volume. However, no explicit description of the deformation theory of coherent sheaves, or equivalently, complexes $E$ of holomorphic vector bundles is known. While we have explicit representatives of $H^{odd}(Q)$ and $H^{even}(Q)$, it is in general difficult to obtain explicit representatives of $H^1(\mathrm{End} E)$ and $H^2(\mathrm{End} E)$, respectively. One possible way has been presented in \cite{Aspinwall:2004bs} where the Massey products have been computed through the $A_\infty$--products, albeit in the simpler context of non--compact Calabi--Yau threefolds. It would be very interesting to compare these two approaches of computing Massey products. In particular, the open string Calabi--Yau/Landau--Ginzburg correspondence should be extended to include the $A_\infty$--structure in both categories.\\
While we have focused on the region in the closed string moduli space near the Gepner point, there are further methods of computing the effective superpotential near the large volume point or even everywhere in the complex structure moduli space. For an example of the latter see~\cite{Baumgartl:2007an}. For the former, a new method has been proposed recently in~\cite{Grimm:2008dq}. Applying this method to the branes we have discussed here could shed light on some of the issues mentioned above such as the "fake" F--terms. More generally, the discrimination between open and closed moduli at different points in the full brane moduli space needs to be properly understood.\\
The main application of our results is in the context open string mirror symmetry calculations. We can now take the deformed matrix factorizations and F--terms and relate them to geometric boundary conditions which are necessary for deriving Picard--Fuchs equations can to compute domain wall tensions and disk instanton numbers or effective superpotentials in flat coordinates. As compared to the one--parameter models the combined bulk/boundary moduli space for branes in two--parameter models has a much richer structure and we expect to find interesting new phenomena. This will be discussed elsewhere.
\begin{appendix}
\section{Moduli of Tensor Product Branes in Two--Parameter Hypersurfaces}
In this appendix we list all tensor product boundary states with moduli for two--parameter hypersurfaces. Furthermore we give a decomposition of the moduli in terms of their minimal model components.
\subsection{The model $\mathbbm{P}(11222)[8]/\mathbbm{Z}_8\times (\mathbbm{Z}_4)^2$}
\begin{longtable}{|c||c|c|}
\caption{Tensor product branes with moduli for $\mathbbm{P}(11222)[8]/\mathbbm{Z}_8\times (\mathbbm{Z}_4)^2$}\label{tab11222mod}\\
\hline
Boundary state & Number of Moduli & Structure of Moduli \\
\hline\hline
\endfirsthead
\caption{(continued)}\\
\hline
Boundary state & Number of Moduli & Structure of Moduli \\
\hline\hline
\endhead
$(3,3,1,1,1)$&$2$&$\begin{array}{c}\frac{1}{2}^1\otimes\frac{1}{2}^1\otimes 0^1\otimes 0^1\otimes 0^1\\ \frac{1}{2}^0\otimes\frac{1}{2}^0\otimes 0^1\otimes 0^1\otimes 0^1\end{array}$\\
\hline
$(3,2,1,1,1)$&$2$&$\begin{array}{c}\frac{1}{2}^1\otimes\frac{1}{2}^1\otimes 0^1\otimes 0^1\otimes 0^1\\ \frac{1}{2}^0\otimes\frac{1}{2}^0\otimes 0^1\otimes 0^1\otimes 0^1\end{array}$\\
\hline
$(2,2,1,1,1)$&$2$&$\begin{array}{c}\frac{1}{2}^1\otimes\frac{1}{2}^1\otimes 0^1\otimes 0^1\otimes 0^1\\ \frac{1}{2}^0\otimes\frac{1}{2}^0\otimes 0^1\otimes 0^1\otimes 0^1\end{array}$\\
\hline
$(3,1,1,1,1)$&$1$&$\frac{1}{2}^1\otimes\frac{1}{2}^1\otimes 0^1\otimes 0^1\otimes 0^1$\\
\hline
$(2,1,1,1,1)$&$1$&$ \frac{1}{2}^1\otimes\frac{1}{2}^1\otimes 0^1\otimes 0^1\otimes 0^1$\\
\hline
$(1,1,1,1,1)$&$1$&$ \frac{1}{2}^1\otimes\frac{1}{2}^1\otimes 0^1\otimes 0^1\otimes 0^1$\\
\hline\hline
\end{longtable}
\newpage
\subsection{The model $\mathbbm{P}(11226)[12]/\mathbbm{Z}_{12}\times (\mathbbm{Z}_6)^2$}
\begin{longtable}{|c||c|c|}
\caption{Tensor product branes with moduli for $\mathbbm{P}(11226)[12]/\mathbbm{Z}_{12}\times (\mathbbm{Z}_6)^2$}\label{tab11226mod}\\
\hline
Boundary state & Number of Moduli & Structure of Moduli \\
\hline\hline
\endfirsthead
\caption{(continued)}\\
\hline
Boundary state & Number of Moduli & Structure of Moduli \\
\hline\hline
\endhead
$(5,5,2,2,0)$&$4$&$\begin{array}{c}\frac{1}{6}^1\otimes\frac{1}{6}^1\otimes\frac{1}{3}^0\otimes\frac{1}{3}^0\otimes 0^1\\ \frac{1}{6}^0\otimes\frac{1}{6}^0\otimes\frac{1}{3}^0\otimes\frac{1}{3}^0\otimes 0^1\\\frac{1}{2}^1\otimes\frac{1}{2}^1\otimes 0^1\otimes 0^1\otimes 0^1\\ \frac{1}{2}^0\otimes\frac{1}{2}^0\otimes 0^1\otimes 0^1\otimes 0^1\end{array}$\\
\hline
$(5,4,2,2,0)$&$4$&$\begin{array}{c}\frac{1}{6}^1\otimes\frac{1}{6}^1\otimes\frac{1}{3}^0\otimes\frac{1}{3}^0\otimes 0^1\\ \frac{1}{6}^0\otimes\frac{1}{6}^0\otimes\frac{1}{3}^0\otimes\frac{1}{3}^0\otimes 0^1\\\frac{1}{2}^1\otimes\frac{1}{2}^1\otimes 0^1\otimes 0^1\otimes 0^1\\ \frac{1}{2}^0\otimes\frac{1}{2}^0\otimes 0^1\otimes 0^1\otimes 0^1\end{array}$\\
\hline
$(4,4,2,2,0)$&$4$&$\begin{array}{c}\frac{1}{6}^1\otimes\frac{1}{6}^1\otimes\frac{1}{3}^0\otimes\frac{1}{3}^0\otimes 0^1\\ \frac{1}{6}^0\otimes\frac{1}{6}^0\otimes\frac{1}{3}^0\otimes\frac{1}{3}^0\otimes 0^1\\\frac{1}{2}^1\otimes\frac{1}{2}^1\otimes 0^1\otimes 0^1\otimes 0^1\\ \frac{1}{2}^0\otimes\frac{1}{2}^0\otimes 0^1\otimes 0^1\otimes 0^1\end{array}$\\
\hline
$(5,3,2,2,0)$&$3$&$\begin{array}{c} \frac{1}{6}^0\otimes\frac{1}{6}^0\otimes\frac{1}{3}^0\otimes\frac{1}{3}^0\otimes 0^1\\\frac{1}{2}^1\otimes\frac{1}{2}^1\otimes 0^1\otimes 0^1\otimes 0^1\\ \frac{1}{2}^0\otimes\frac{1}{2}^0\otimes 0^1\otimes 0^1\otimes 0^1\end{array}$\\
\hline
$(4,3,2,2,0)$&$3$&$\begin{array}{c} \frac{1}{6}^0\otimes\frac{1}{6}^0\otimes\frac{1}{3}^0\otimes\frac{1}{3}^0\otimes 0^1\\\frac{1}{2}^1\otimes\frac{1}{2}^1\otimes 0^1\otimes 0^1\otimes 0^1\\ \frac{1}{2}^0\otimes\frac{1}{2}^0\otimes 0^1\otimes 0^1\otimes 0^1\end{array}$\\
\hline
$(3,3,2,2,0)$&$3$&$\begin{array}{c} \frac{1}{6}^0\otimes\frac{1}{6}^0\otimes\frac{1}{3}^0\otimes\frac{1}{3}^0\otimes 0^1\\\frac{1}{2}^1\otimes\frac{1}{2}^1\otimes 0^1\otimes 0^1\otimes 0^1\\ \frac{1}{2}^0\otimes\frac{1}{2}^0\otimes 0^1\otimes 0^1\otimes 0^1\end{array}$\\
\hline
$(5,5,2,1,0)$&$2$&$\begin{array}{c}\frac{1}{6}^1\otimes\frac{1}{6}^1\otimes\frac{1}{3}^0\otimes\frac{1}{3}^0\otimes 0^1\\ \frac{1}{6}^0\otimes\frac{1}{6}^0\otimes\frac{1}{3}^0\otimes\frac{1}{3}^0\otimes 0^1\end{array}$\\
\hline
$(5,5,1,1,0)$&$2$&$\begin{array}{c}\frac{1}{6}^1\otimes\frac{1}{6}^1\otimes\frac{1}{3}^0\otimes\frac{1}{3}^0\otimes 0^1\\ \frac{1}{6}^0\otimes\frac{1}{6}^0\otimes\frac{1}{3}^0\otimes\frac{1}{3}^0\otimes 0^1\end{array}$\\
\hline
$(5,4,2,1,0)$&$2$&$\begin{array}{c}\frac{1}{6}^1\otimes\frac{1}{6}^1\otimes\frac{1}{3}^0\otimes\frac{1}{3}^0\otimes 0^1\\ \frac{1}{6}^0\otimes\frac{1}{6}^0\otimes\frac{1}{3}^0\otimes\frac{1}{3}^0\otimes 0^1\end{array}$\\
\hline
$(5,4,1,1,0)$&$2$&$\begin{array}{c}\frac{1}{6}^1\otimes\frac{1}{6}^1\otimes\frac{1}{3}^0\otimes\frac{1}{3}^0\otimes 0^1\\ \frac{1}{6}^0\otimes\frac{1}{6}^0\otimes\frac{1}{3}^0\otimes\frac{1}{3}^0\otimes 0^1\end{array}$\\
\hline
$(5,2,2,2,0)$&$2$&$\begin{array}{c}\frac{1}{6}^0\otimes\frac{1}{6}^0\otimes\frac{1}{3}^0\otimes\frac{1}{3}^0\otimes 0^1\\\frac{1}{2}^1\otimes\frac{1}{2}^1\otimes 0^1\otimes 0^1\otimes 0^1 \end{array}$\\
\hline
$(4,4,2,1,0)$&$2$&$\begin{array}{c}\frac{1}{6}^1\otimes\frac{1}{6}^1\otimes\frac{1}{3}^0\otimes\frac{1}{3}^0\otimes 0^1\\ \frac{1}{6}^0\otimes\frac{1}{6}^0\otimes\frac{1}{3}^0\otimes\frac{1}{3}^0\otimes 0^1\end{array}$\\
\hline
$(4,4,1,1,0)$&$2$&$\begin{array}{c}\frac{1}{6}^1\otimes\frac{1}{6}^1\otimes\frac{1}{3}^0\otimes\frac{1}{3}^0\otimes 0^1\\ \frac{1}{6}^0\otimes\frac{1}{6}^0\otimes\frac{1}{3}^0\otimes\frac{1}{3}^0\otimes 0^1\end{array}$\\
\hline
$(4,2,2,2,0)$&$2$&$\begin{array}{c}\frac{1}{6}^0\otimes\frac{1}{6}^0\otimes\frac{1}{3}^0\otimes\frac{1}{3}^0\otimes 0^1\\\frac{1}{2}^1\otimes\frac{1}{2}^1\otimes 0^1\otimes 0^1\otimes 0^1 \end{array}$\\
\hline
$(3,2,2,2,0)$&$2$&$\begin{array}{c}\frac{1}{6}^0\otimes\frac{1}{6}^0\otimes\frac{1}{3}^0\otimes\frac{1}{3}^0\otimes 0^1\\\frac{1}{2}^1\otimes\frac{1}{2}^1\otimes 0^1\otimes 0^1\otimes 0^1 \end{array}$\\
\hline
$(2,2,2,2,0)$&$2$&$\begin{array}{c}\frac{1}{6}^0\otimes\frac{1}{6}^0\otimes\frac{1}{3}^0\otimes\frac{1}{3}^0\otimes 0^1\\\frac{1}{2}^1\otimes\frac{1}{2}^1\otimes 0^1\otimes 0^1\otimes 0^1 \end{array}$\\
\hline
$(5,3,2,1,0)$&$1$&$\begin{array}{c}\frac{1}{6}^0\otimes\frac{1}{6}^0\otimes\frac{1}{3}^0\otimes\frac{1}{3}^0\otimes 0^1\end{array}$\\
\hline
$(5,3,1,1,0)$&$1$&$\begin{array}{c}\frac{1}{6}^0\otimes\frac{1}{6}^0\otimes\frac{1}{3}^0\otimes\frac{1}{3}^0\otimes 0^1\end{array}$\\
\hline
$(5,2,2,1,0)$&$1$&$\begin{array}{c}\frac{1}{6}^0\otimes\frac{1}{6}^0\otimes\frac{1}{3}^0\otimes\frac{1}{3}^0\otimes 0^1\end{array}$\\
\hline
$(5,2,1,1,0)$&$1$&$\begin{array}{c}\frac{1}{6}^0\otimes\frac{1}{6}^0\otimes\frac{1}{3}^0\otimes\frac{1}{3}^0\otimes 0^1\end{array}$\\
\hline
$(5,1,2,2,0)$&$1$&$\begin{array}{c}\frac{1}{6}^0\otimes\frac{1}{6}^0\otimes\frac{1}{3}^0\otimes\frac{1}{3}^0\otimes 0^1\end{array}$\\
\hline
$(5,1,2,1,0)$&$1$&$\begin{array}{c}\frac{1}{6}^0\otimes\frac{1}{6}^0\otimes\frac{1}{3}^0\otimes\frac{1}{3}^0\otimes 0^1\end{array}$\\
\hline
$(5,1,1,1,0)$&$1$&$\begin{array}{c}\frac{1}{6}^0\otimes\frac{1}{6}^0\otimes\frac{1}{3}^0\otimes\frac{1}{3}^0\otimes 0^1\end{array}$\\
\hline
$(4,3,2,1,0)$&$1$&$\begin{array}{c}\frac{1}{6}^0\otimes\frac{1}{6}^0\otimes\frac{1}{3}^0\otimes\frac{1}{3}^0\otimes 0^1\end{array}$\\
\hline
$(4,3,1,1,0)$&$1$&$\begin{array}{c}\frac{1}{6}^0\otimes\frac{1}{6}^0\otimes\frac{1}{3}^0\otimes\frac{1}{3}^0\otimes 0^1\end{array}$\\
\hline
$(4,2,2,1,0)$&$1$&$\begin{array}{c}\frac{1}{6}^0\otimes\frac{1}{6}^0\otimes\frac{1}{3}^0\otimes\frac{1}{3}^0\otimes 0^1\end{array}$\\
\hline
$(4,2,1,1,0)$&$1$&$\begin{array}{c}\frac{1}{6}^0\otimes\frac{1}{6}^0\otimes\frac{1}{3}^0\otimes\frac{1}{3}^0\otimes 0^1\end{array}$\\
\hline
$(4,1,2,2,0)$&$1$&$\begin{array}{c}\frac{1}{6}^0\otimes\frac{1}{6}^0\otimes\frac{1}{3}^0\otimes\frac{1}{3}^0\otimes 0^1\end{array}$\\
\hline
$(4,1,2,1,0)$&$1$&$\begin{array}{c}\frac{1}{6}^0\otimes\frac{1}{6}^0\otimes\frac{1}{3}^0\otimes\frac{1}{3}^0\otimes 0^1\end{array}$\\
\hline
$(4,1,1,1,0)$&$1$&$\begin{array}{c}\frac{1}{6}^0\otimes\frac{1}{6}^0\otimes\frac{1}{3}^0\otimes\frac{1}{3}^0\otimes 0^1\end{array}$\\
\hline
$(3,3,2,1,0)$&$1$&$\begin{array}{c}\frac{1}{6}^0\otimes\frac{1}{6}^0\otimes\frac{1}{3}^0\otimes\frac{1}{3}^0\otimes 0^1\end{array}$\\
\hline
$(3,3,1,1,0)$&$1$&$\begin{array}{c}\frac{1}{6}^0\otimes\frac{1}{6}^0\otimes\frac{1}{3}^0\otimes\frac{1}{3}^0\otimes 0^1\end{array}$\\
\hline
$(3,2,2,1,0)$&$1$&$\begin{array}{c}\frac{1}{6}^0\otimes\frac{1}{6}^0\otimes\frac{1}{3}^0\otimes\frac{1}{3}^0\otimes 0^1\end{array}$\\
\hline
$(3,2,1,1,0)$&$1$&$\begin{array}{c}\frac{1}{6}^0\otimes\frac{1}{6}^0\otimes\frac{1}{3}^0\otimes\frac{1}{3}^0\otimes 0^1\end{array}$\\
\hline
$(3,1,2,2,0)$&$1$&$\begin{array}{c}\frac{1}{6}^0\otimes\frac{1}{6}^0\otimes\frac{1}{3}^0\otimes\frac{1}{3}^0\otimes 0^1\end{array}$\\
\hline
$(3,1,2,1,0)$&$1$&$\begin{array}{c}\frac{1}{6}^0\otimes\frac{1}{6}^0\otimes\frac{1}{3}^0\otimes\frac{1}{3}^0\otimes 0^1\end{array}$\\
\hline
$(3,1,1,1,0)$&$1$&$\begin{array}{c}\frac{1}{6}^0\otimes\frac{1}{6}^0\otimes\frac{1}{3}^0\otimes\frac{1}{3}^0\otimes 0^1\end{array}$\\
\hline
$(2,2,2,1,0)$&$1$&$\begin{array}{c}\frac{1}{6}^0\otimes\frac{1}{6}^0\otimes\frac{1}{3}^0\otimes\frac{1}{3}^0\otimes 0^1\end{array}$\\
\hline
$(2,2,1,1,0)$&$1$&$\begin{array}{c}\frac{1}{6}^0\otimes\frac{1}{6}^0\otimes\frac{1}{3}^0\otimes\frac{1}{3}^0\otimes 0^1\end{array}$\\
\hline
$(2,1,2,2,0)$&$1$&$\begin{array}{c}\frac{1}{6}^0\otimes\frac{1}{6}^0\otimes\frac{1}{3}^0\otimes\frac{1}{3}^0\otimes 0^1\end{array}$\\
\hline
$(2,1,2,1,0)$&$1$&$\begin{array}{c}\frac{1}{6}^0\otimes\frac{1}{6}^0\otimes\frac{1}{3}^0\otimes\frac{1}{3}^0\otimes 0^1\end{array}$\\
\hline
$(2,1,1,1,0)$&$1$&$\begin{array}{c}\frac{1}{6}^0\otimes\frac{1}{6}^0\otimes\frac{1}{3}^0\otimes\frac{1}{3}^0\otimes 0^1\end{array}$\\
\hline
$(1,1,2,2,0)$&$1$&$\begin{array}{c}\frac{1}{6}^0\otimes\frac{1}{6}^0\otimes\frac{1}{3}^0\otimes\frac{1}{3}^0\otimes 0^1\end{array}$\\
\hline
$(1,1,2,1,0)$&$1$&$\begin{array}{c}\frac{1}{6}^0\otimes\frac{1}{6}^0\otimes\frac{1}{3}^0\otimes\frac{1}{3}^0\otimes 0^1\end{array}$\\
\hline
$(1,1,1,1,0)$&$1$&$\begin{array}{c}\frac{1}{6}^0\otimes\frac{1}{6}^0\otimes\frac{1}{3}^0\otimes\frac{1}{3}^0\otimes 0^1\end{array}$\\
\hline\hline
\end{longtable}
\newpage
\subsection{The Hypersurface in $\mathbbm{P}(12234)[12]/(\mathbbm{Z}_6)^2$}
\begin{longtable}{|c||c|c|}
\caption{Tensor product branes with moduli for $\mathbbm{P}(12234)[12]/(\mathbbm{Z}_6)^2$}\label{tab12234mod}\\
\hline
Boundary state & Number of Moduli & Structure of Moduli \\
\hline\hline
\endfirsthead
\caption{continued}\\
\hline
Boundary state & Number of Moduli & Structure of Moduli \\
\hline\hline
\endhead
$(5,2,2,1,0)$&$2$&$\begin{array}{c}\frac{1}{2}^1\otimes0^1\otimes 0^1\otimes\frac{1}{2}^0\otimes 0^0\\ \frac{1}{2}^0\otimes 0^1\otimes 0^1\otimes\frac{1}{2}^1\otimes 0^0\end{array}$\\
\hline
$(4,2,2,1,0)$&$2$&$\begin{array}{c}\frac{1}{2}^1\otimes0^1\otimes 0^1\otimes\frac{1}{2}^0\otimes 0^0\\ \frac{1}{2}^0\otimes 0^1\otimes 0^1\otimes\frac{1}{2}^1\otimes 0^0\end{array}$\\
\hline
$(3,2,2,1,0)$&$2$&$\begin{array}{c}\frac{1}{2}^1\otimes0^1\otimes 0^1\otimes\frac{1}{2}^0\otimes 0^0\\ \frac{1}{2}^0\otimes 0^1\otimes 0^1\otimes\frac{1}{2}^1\otimes 0^0\end{array}$\\
\hline
$(5,2,2,0,0)$&$1$&$\begin{array}{c} \frac{1}{2}^0\otimes 0^1\otimes 0^1\otimes\frac{1}{2}^1\otimes 0^0\end{array}$\\
\hline
$(4,2,2,0,0)$&$1$&$\begin{array}{c} \frac{1}{2}^0\otimes 0^1\otimes 0^1\otimes\frac{1}{2}^1\otimes 0^0\end{array}$\\
\hline
$(3,2,2,0,0)$&$1$&$\begin{array}{c} \frac{1}{2}^0\otimes 0^1\otimes 0^1\otimes\frac{1}{2}^1\otimes 0^0\end{array}$\\
\hline
$(2,2,2,1,0)$&$1$&$\begin{array}{c} \frac{1}{2}^1\otimes 0^1\otimes 0^1\otimes\frac{1}{2}^0\otimes 0^0\end{array}$\\
\hline\hline
\end{longtable}
\subsection{The model $\mathbbm{P}(12227)[14]/(\mathbbm{Z}_7)^2$}
\begin{longtable}{|c||c|c|}
\caption{Tensor product branes with moduli for $\mathbbm{P}(12227)[14]/(\mathbbm{Z}_7)^2$}\label{tab12227mod}\\
\hline
Boundary state & Number of Moduli & Structure of Moduli \\
\hline\hline
\endfirsthead
\caption{(continued)}\\
\hline
Boundary state & Number of Moduli & Structure of Moduli \\
\hline\hline
\endhead
$(6,2,2,2,0)$&$4$&$\begin{array}{c}\frac{1}{7}^0\otimes\frac{2}{7}^0\otimes\frac{2}{7}^0\otimes \frac{2}{7}^0\otimes 0^1 \\ \frac{1}{7}^1\otimes\frac{2}{7}^0\otimes\frac{2}{7}^0\otimes \frac{2}{7}^0\otimes 0^0\\\frac{4}{7}^1\otimes\frac{1}{7}^1\otimes\frac{1}{7}^1\otimes\frac{1}{7}^1\otimes 0^1\\\frac{4}{7}^0\otimes\frac{1}{7}^1\otimes\frac{1}{7}^1\otimes\frac{1}{7}^1\otimes 0^0 \end{array}$\\
\hline
$(5,2,2,2,0)$&$4$&$\begin{array}{c}\frac{1}{7}^0\otimes\frac{2}{7}^0\otimes\frac{2}{7}^0\otimes \frac{2}{7}^0\otimes 0^1 \\ \frac{1}{7}^1\otimes\frac{2}{7}^0\otimes\frac{2}{7}^0\otimes \frac{2}{7}^0\otimes 0^0\\\frac{4}{7}^1\otimes\frac{1}{7}^1\otimes\frac{1}{7}^1\otimes\frac{1}{7}^1\otimes 0^1\\\frac{4}{7}^0\otimes\frac{1}{7}^1\otimes\frac{1}{7}^1\otimes\frac{1}{7}^1\otimes 0^0 \end{array}$\\
\hline
$(4,2,2,2,0)$&$3$&$\begin{array}{c}\frac{1}{7}^0\otimes\frac{2}{7}^0\otimes\frac{2}{7}^0\otimes \frac{2}{7}^0\otimes 0^1 \\ \frac{4}{7}^1\otimes\frac{1}{7}^1\otimes\frac{1}{7}^1\otimes\frac{1}{7}^1\otimes 0^1\\\frac{4}{7}^0\otimes\frac{1}{7}^1\otimes\frac{1}{7}^1\otimes\frac{1}{7}^1\otimes 0^0 \end{array}$\\
\hline
$(6,2,2,1,0)$&$2$&$\begin{array}{c}\frac{1}{7}^0\otimes\frac{2}{7}^0\otimes\frac{2}{7}^0\otimes \frac{2}{7}^0\otimes 0^1 \\ \frac{1}{7}^1\otimes\frac{2}{7}^0\otimes\frac{2}{7}^0\otimes \frac{2}{7}^0\otimes 0^0 \end{array}$\\
\hline
$(6,2,1,1,0)$&$2$&$\begin{array}{c}\frac{1}{7}^0\otimes\frac{2}{7}^0\otimes\frac{2}{7}^0\otimes \frac{2}{7}^0\otimes 0^1 \\ \frac{1}{7}^1\otimes\frac{2}{7}^0\otimes\frac{2}{7}^0\otimes \frac{2}{7}^0\otimes 0^0 \end{array}$\\
\hline
$(6,1,1,1,0)$&$2$&$\begin{array}{c}\frac{1}{7}^0\otimes\frac{2}{7}^0\otimes\frac{2}{7}^0\otimes \frac{2}{7}^0\otimes 0^1 \\ \frac{1}{7}^1\otimes\frac{2}{7}^0\otimes\frac{2}{7}^0\otimes \frac{2}{7}^0\otimes 0^0 \end{array}$\\
\hline
$(5,2,2,1,0)$&$2$&$\begin{array}{c}\frac{1}{7}^0\otimes\frac{2}{7}^0\otimes\frac{2}{7}^0\otimes \frac{2}{7}^0\otimes 0^1 \\ \frac{1}{7}^1\otimes\frac{2}{7}^0\otimes\frac{2}{7}^0\otimes \frac{2}{7}^0\otimes 0^0 \end{array}$\\
\hline
$(5,2,1,1,0)$&$2$&$\begin{array}{c}\frac{1}{7}^0\otimes\frac{2}{7}^0\otimes\frac{2}{7}^0\otimes \frac{2}{7}^0\otimes 0^1 \\ \frac{1}{7}^1\otimes\frac{2}{7}^0\otimes\frac{2}{7}^0\otimes \frac{2}{7}^0\otimes 0^0 \end{array}$\\
\hline
$(5,1,1,1,0)$&$2$&$\begin{array}{c}\frac{1}{7}^0\otimes\frac{2}{7}^0\otimes\frac{2}{7}^0\otimes \frac{2}{7}^0\otimes 0^1 \\ \frac{1}{7}^1\otimes\frac{2}{7}^0\otimes\frac{2}{7}^0\otimes \frac{2}{7}^0\otimes 0^0 \end{array}$\\
\hline
$(3,2,2,2,0)$&$2$&$\begin{array}{c}\frac{1}{7}^0\otimes\frac{2}{7}^0\otimes\frac{2}{7}^0\otimes \frac{2}{7}^0\otimes 0^1 \\ \frac{4}{7}^1\otimes\frac{1}{7}^1\otimes\frac{1}{7}^1\otimes\frac{1}{7}^1\otimes 0^1\\ \end{array}$\\
\hline
$(2,2,2,2,0)$&$2$&$\begin{array}{c}\frac{1}{7}^0\otimes\frac{2}{7}^0\otimes\frac{2}{7}^0\otimes \frac{2}{7}^0\otimes 0^1 \\ \frac{4}{7}^1\otimes\frac{1}{7}^1\otimes\frac{1}{7}^1\otimes\frac{1}{7}^1\otimes 0^1\\ \end{array}$\\
\hline
$(4,2,2,1,0)$&$1$&$\begin{array}{c}\frac{1}{7}^0\otimes\frac{2}{7}^0\otimes\frac{2}{7}^0\otimes \frac{2}{7}^0\otimes 0^1 \end{array}$\\
\hline
$(4,2,1,1,0)$&$1$&$\begin{array}{c}\frac{1}{7}^0\otimes\frac{2}{7}^0\otimes\frac{2}{7}^0\otimes \frac{2}{7}^0\otimes 0^1 \end{array}$\\
\hline
$(4,1,1,1,0)$&$1$&$\begin{array}{c}\frac{1}{7}^0\otimes\frac{2}{7}^0\otimes\frac{2}{7}^0\otimes \frac{2}{7}^0\otimes 0^1 \end{array}$\\
\hline
$(3,2,2,1,0)$&$1$&$\begin{array}{c}\frac{1}{7}^0\otimes\frac{2}{7}^0\otimes\frac{2}{7}^0\otimes \frac{2}{7}^0\otimes 0^1 \end{array}$\\
\hline
$(3,2,1,1,0)$&$1$&$\begin{array}{c}\frac{1}{7}^0\otimes\frac{2}{7}^0\otimes\frac{2}{7}^0\otimes \frac{2}{7}^0\otimes 0^1 \end{array}$\\
\hline
$(3,1,1,1,0)$&$1$&$\begin{array}{c}\frac{1}{7}^0\otimes\frac{2}{7}^0\otimes\frac{2}{7}^0\otimes \frac{2}{7}^0\otimes 0^1 \end{array}$\\
\hline
$(2,2,2,1,0)$&$1$&$\begin{array}{c}\frac{1}{7}^0\otimes\frac{2}{7}^0\otimes\frac{2}{7}^0\otimes \frac{2}{7}^0\otimes 0^1 \end{array}$\\
\hline
$(2,2,1,1,0)$&$1$&$\begin{array}{c}\frac{1}{7}^0\otimes\frac{2}{7}^0\otimes\frac{2}{7}^0\otimes \frac{2}{7}^0\otimes 0^1 \end{array}$\\
\hline
$(2,1,1,1,0)$&$1$&$\begin{array}{c}\frac{1}{7}^0\otimes\frac{2}{7}^0\otimes\frac{2}{7}^0\otimes \frac{2}{7}^0\otimes 0^1 \end{array}$\\
\hline
$(1,2,2,2,0)$&$1$&$\begin{array}{c}\frac{1}{7}^0\otimes\frac{2}{7}^0\otimes\frac{2}{7}^0\otimes \frac{2}{7}^0\otimes 0^1 \end{array}$\\
\hline
$(1,2,2,1,0)$&$1$&$\begin{array}{c}\frac{1}{7}^0\otimes\frac{2}{7}^0\otimes\frac{2}{7}^0\otimes \frac{2}{7}^0\otimes 0^1 \end{array}$\\
\hline
$(1,2,1,1,0)$&$1$&$\begin{array}{c}\frac{1}{7}^0\otimes\frac{2}{7}^0\otimes\frac{2}{7}^0\otimes \frac{2}{7}^0\otimes 0^1 \end{array}$\\
\hline
$(1,1,1,1,0)$&$1$&$\begin{array}{c}\frac{1}{7}^0\otimes\frac{2}{7}^0\otimes\frac{2}{7}^0\otimes \frac{2}{7}^0\otimes 0^1 \end{array}$\\
\hline\hline
\end{longtable}
\subsection{The model $\mathbbm{P}(11169)[18]/(\mathbbm{Z}_{18})^2$}
\begin{longtable}{|c||c|c|}
\caption{Tensor product branes with moduli for $\mathbbm{P}(11169)[18]/(\mathbbm{Z}_{18})^2$}\label{tab11169mod}\\
\hline
Boundary state & Number of Moduli & Structure of Moduli \\
\hline\hline
\endfirsthead
\caption{(continued)}\\
\hline
Boundary state & Number of Moduli & Structure of Moduli \\
\hline\hline
\endhead
$(8,8,8,0,0)$&$4$&$\begin{array}{c}\frac{1}{3}^1\otimes\frac{1}{3}^1\otimes\frac{1}{3}^1\otimes 0^0\otimes 0^0\\ \frac{1}{3}^0\otimes\frac{1}{3}^0\otimes\frac{1}{3}^0\otimes 0^0\otimes 0^1 \\ \frac{2}{9}^1\otimes\frac{2}{9}^1\otimes\frac{2}{9}^1\otimes\frac{1}{3}^1\otimes 0^1 \\ \ \frac{2}{9}^0\otimes\frac{2}{9}^0\otimes\frac{2}{9}^0\otimes\frac{1}{3}^1\otimes 0^0\end{array}$\\
\hline
$(8,8,7,0,0)$&$4$&$\begin{array}{c}\frac{1}{3}^1\otimes\frac{1}{3}^1\otimes\frac{1}{3}^1\otimes 0^0\otimes 0^0\\ \frac{1}{3}^0\otimes\frac{1}{3}^0\otimes\frac{1}{3}^0\otimes 0^0\otimes 0^1 \\ \frac{2}{9}^1\otimes\frac{2}{9}^1\otimes\frac{2}{9}^1\otimes\frac{1}{3}^1\otimes 0^1 \\ \ \frac{2}{9}^0\otimes\frac{2}{9}^0\otimes\frac{2}{9}^0\otimes\frac{1}{3}^1\otimes 0^0\end{array}$\\
\hline
$(8,8,6,0,0)$&$4$&$\begin{array}{c}\frac{1}{3}^1\otimes\frac{1}{3}^1\otimes\frac{1}{3}^1\otimes 0^0\otimes 0^0\\ \frac{1}{3}^0\otimes\frac{1}{3}^0\otimes\frac{1}{3}^0\otimes 0^0\otimes 0^1 \\ \frac{2}{9}^1\otimes\frac{2}{9}^1\otimes\frac{2}{9}^1\otimes\frac{1}{3}^1\otimes 0^1 \\ \ \frac{2}{9}^0\otimes\frac{2}{9}^0\otimes\frac{2}{9}^0\otimes\frac{1}{3}^1\otimes 0^0\end{array}$\\
\hline
$(8,7,7,0,0)$&$4$&$\begin{array}{c}\frac{1}{3}^1\otimes\frac{1}{3}^1\otimes\frac{1}{3}^1\otimes 0^0\otimes 0^0\\ \frac{1}{3}^0\otimes\frac{1}{3}^0\otimes\frac{1}{3}^0\otimes 0^0\otimes 0^1 \\ \frac{2}{9}^1\otimes\frac{2}{9}^1\otimes\frac{2}{9}^1\otimes\frac{1}{3}^1\otimes 0^1 \\ \ \frac{2}{9}^0\otimes\frac{2}{9}^0\otimes\frac{2}{9}^0\otimes\frac{1}{3}^1\otimes 0^0\end{array}$\\
\hline
$(8,7,6,0,0)$&$4$&$\begin{array}{c}\frac{1}{3}^1\otimes\frac{1}{3}^1\otimes\frac{1}{3}^1\otimes 0^0\otimes 0^0\\ \frac{1}{3}^0\otimes\frac{1}{3}^0\otimes\frac{1}{3}^0\otimes 0^0\otimes 0^1 \\ \frac{2}{9}^1\otimes\frac{2}{9}^1\otimes\frac{2}{9}^1\otimes\frac{1}{3}^1\otimes 0^1 \\ \ \frac{2}{9}^0\otimes\frac{2}{9}^0\otimes\frac{2}{9}^0\otimes\frac{1}{3}^1\otimes 0^0\end{array}$\\
\hline
$(8,6,6,0,0)$&$4$&$\begin{array}{c}\frac{1}{3}^1\otimes\frac{1}{3}^1\otimes\frac{1}{3}^1\otimes 0^0\otimes 0^0\\ \frac{1}{3}^0\otimes\frac{1}{3}^0\otimes\frac{1}{3}^0\otimes 0^0\otimes 0^1 \\ \frac{2}{9}^1\otimes\frac{2}{9}^1\otimes\frac{2}{9}^1\otimes\frac{1}{3}^1\otimes 0^1 \\ \ \frac{2}{9}^0\otimes\frac{2}{9}^0\otimes\frac{2}{9}^0\otimes\frac{1}{3}^1\otimes 0^0\end{array}$\\
\hline
$(7,7,7,0,0)$&$4$&$\begin{array}{c}\frac{1}{3}^1\otimes\frac{1}{3}^1\otimes\frac{1}{3}^1\otimes 0^0\otimes 0^0\\ \frac{1}{3}^0\otimes\frac{1}{3}^0\otimes\frac{1}{3}^0\otimes 0^0\otimes 0^1 \\ \frac{2}{9}^1\otimes\frac{2}{9}^1\otimes\frac{2}{9}^1\otimes\frac{1}{3}^1\otimes 0^1 \\ \ \frac{2}{9}^0\otimes\frac{2}{9}^0\otimes\frac{2}{9}^0\otimes\frac{1}{3}^1\otimes 0^0\end{array}$\\
\hline
$(7,7,6,0,0)$&$4$&$\begin{array}{c}\frac{1}{3}^1\otimes\frac{1}{3}^1\otimes\frac{1}{3}^1\otimes 0^0\otimes 0^0\\ \frac{1}{3}^0\otimes\frac{1}{3}^0\otimes\frac{1}{3}^0\otimes 0^0\otimes 0^1 \\ \frac{2}{9}^1\otimes\frac{2}{9}^1\otimes\frac{2}{9}^1\otimes\frac{1}{3}^1\otimes 0^1 \\ \ \frac{2}{9}^0\otimes\frac{2}{9}^0\otimes\frac{2}{9}^0\otimes\frac{1}{3}^1\otimes 0^0\end{array}$\\
\hline
$(7,6,6,0,0)$&$4$&$\begin{array}{c}\frac{1}{3}^1\otimes\frac{1}{3}^1\otimes\frac{1}{3}^1\otimes 0^0\otimes 0^0\\ \frac{1}{3}^0\otimes\frac{1}{3}^0\otimes\frac{1}{3}^0\otimes 0^0\otimes 0^1 \\ \frac{2}{9}^1\otimes\frac{2}{9}^1\otimes\frac{2}{9}^1\otimes\frac{1}{3}^1\otimes 0^1 \\ \ \frac{2}{9}^0\otimes\frac{2}{9}^0\otimes\frac{2}{9}^0\otimes\frac{1}{3}^1\otimes 0^0\end{array}$\\
\hline
$(6,6,6,0,0)$&$4$&$\begin{array}{c}\frac{1}{3}^1\otimes\frac{1}{3}^1\otimes\frac{1}{3}^1\otimes 0^0\otimes 0^0\\ \frac{1}{3}^0\otimes\frac{1}{3}^0\otimes\frac{1}{3}^0\otimes 0^0\otimes 0^1 \\ \frac{2}{9}^1\otimes\frac{2}{9}^1\otimes\frac{2}{9}^1\otimes\frac{1}{3}^1\otimes 0^1 \\ \ \frac{2}{9}^0\otimes\frac{2}{9}^0\otimes\frac{2}{9}^0\otimes\frac{1}{3}^1\otimes 0^0\end{array}$\\
\hline
$(8,8,5,0,0)$&$3$&$\begin{array}{c}\frac{1}{3}^1\otimes\frac{1}{3}^1\otimes\frac{1}{3}^1\otimes 0^0\otimes 0^0\\ \frac{1}{3}^0\otimes\frac{1}{3}^0\otimes\frac{1}{3}^0\otimes 0^0\otimes 0^1 \\  \frac{2}{9}^0\otimes\frac{2}{9}^0\otimes\frac{2}{9}^0\otimes\frac{1}{3}^1\otimes 0^0\end{array}$\\
\hline
$(8,7,5,0,0)$&$3$&$\begin{array}{c}\frac{1}{3}^1\otimes\frac{1}{3}^1\otimes\frac{1}{3}^1\otimes 0^0\otimes 0^0\\ \frac{1}{3}^0\otimes\frac{1}{3}^0\otimes\frac{1}{3}^0\otimes 0^0\otimes 0^1 \\  \frac{2}{9}^0\otimes\frac{2}{9}^0\otimes\frac{2}{9}^0\otimes\frac{1}{3}^1\otimes 0^0\end{array}$\\
\hline
$(8,6,5,0,0)$&$3$&$\begin{array}{c}\frac{1}{3}^1\otimes\frac{1}{3}^1\otimes\frac{1}{3}^1\otimes 0^0\otimes 0^0\\ \frac{1}{3}^0\otimes\frac{1}{3}^0\otimes\frac{1}{3}^0\otimes 0^0\otimes 0^1 \\  \frac{2}{9}^0\otimes\frac{2}{9}^0\otimes\frac{2}{9}^0\otimes\frac{1}{3}^1\otimes 0^0\end{array}$\\
\hline
$(8,5,5,0,0)$&$3$&$\begin{array}{c}\frac{1}{3}^1\otimes\frac{1}{3}^1\otimes\frac{1}{3}^1\otimes 0^0\otimes 0^0\\ \frac{1}{3}^0\otimes\frac{1}{3}^0\otimes\frac{1}{3}^0\otimes 0^0\otimes 0^1 \\  \frac{2}{9}^0\otimes\frac{2}{9}^0\otimes\frac{2}{9}^0\otimes\frac{1}{3}^1\otimes 0^0\end{array}$\\
\hline
$(7,7,5,0,0)$&$3$&$\begin{array}{c}\frac{1}{3}^1\otimes\frac{1}{3}^1\otimes\frac{1}{3}^1\otimes 0^0\otimes 0^0\\ \frac{1}{3}^0\otimes\frac{1}{3}^0\otimes\frac{1}{3}^0\otimes 0^0\otimes 0^1 \\  \frac{2}{9}^0\otimes\frac{2}{9}^0\otimes\frac{2}{9}^0\otimes\frac{1}{3}^1\otimes 0^0\end{array}$\\
\hline
$(7,6,5,0,0)$&$3$&$\begin{array}{c}\frac{1}{3}^1\otimes\frac{1}{3}^1\otimes\frac{1}{3}^1\otimes 0^0\otimes 0^0\\ \frac{1}{3}^0\otimes\frac{1}{3}^0\otimes\frac{1}{3}^0\otimes 0^0\otimes 0^1 \\  \frac{2}{9}^0\otimes\frac{2}{9}^0\otimes\frac{2}{9}^0\otimes\frac{1}{3}^1\otimes 0^0\end{array}$\\
\hline
$(7,5,5,0,0)$&$3$&$\begin{array}{c}\frac{1}{3}^1\otimes\frac{1}{3}^1\otimes\frac{1}{3}^1\otimes 0^0\otimes 0^0\\ \frac{1}{3}^0\otimes\frac{1}{3}^0\otimes\frac{1}{3}^0\otimes 0^0\otimes 0^1 \\  \frac{2}{9}^0\otimes\frac{2}{9}^0\otimes\frac{2}{9}^0\otimes\frac{1}{3}^1\otimes 0^0\end{array}$\\
\hline
$(6,6,5,0,0)$&$3$&$\begin{array}{c}\frac{1}{3}^1\otimes\frac{1}{3}^1\otimes\frac{1}{3}^1\otimes 0^0\otimes 0^0\\ \frac{1}{3}^0\otimes\frac{1}{3}^0\otimes\frac{1}{3}^0\otimes 0^0\otimes 0^1 \\  \frac{2}{9}^0\otimes\frac{2}{9}^0\otimes\frac{2}{9}^0\otimes\frac{1}{3}^1\otimes 0^0\end{array}$\\
\hline
$(6,5,5,0,0)$&$3$&$\begin{array}{c}\frac{1}{3}^1\otimes\frac{1}{3}^1\otimes\frac{1}{3}^1\otimes 0^0\otimes 0^0\\ \frac{1}{3}^0\otimes\frac{1}{3}^0\otimes\frac{1}{3}^0\otimes 0^0\otimes 0^1 \\  \frac{2}{9}^0\otimes\frac{2}{9}^0\otimes\frac{2}{9}^0\otimes\frac{1}{3}^1\otimes 0^0\end{array}$\\
\hline
$(5,5,5,0,0)$&$3$&$\begin{array}{c}\frac{1}{3}^1\otimes\frac{1}{3}^1\otimes\frac{1}{3}^1\otimes 0^0\otimes 0^0\\ \frac{1}{3}^0\otimes\frac{1}{3}^0\otimes\frac{1}{3}^0\otimes 0^0\otimes 0^1 \\  \frac{2}{9}^0\otimes\frac{2}{9}^0\otimes\frac{2}{9}^0\otimes\frac{1}{3}^1\otimes 0^0\end{array}$\\
\hline
$(8,8,4,0,0)$&$2$&$\begin{array}{c} \frac{1}{3}^0\otimes\frac{1}{3}^0\otimes\frac{1}{3}^0\otimes 0^0\otimes 0^1 \\  \frac{2}{9}^0\otimes\frac{2}{9}^0\otimes\frac{2}{9}^0\otimes\frac{1}{3}^1\otimes 0^0\end{array}$\\
\hline
$(8,8,3,0,0)$&$2$&$\begin{array}{c} \frac{1}{3}^0\otimes\frac{1}{3}^0\otimes\frac{1}{3}^0\otimes 0^0\otimes 0^1 \\  \frac{2}{9}^0\otimes\frac{2}{9}^0\otimes\frac{2}{9}^0\otimes\frac{1}{3}^1\otimes 0^0\end{array}$\\
\hline
$(8,7,4,0,0)$&$2$&$\begin{array}{c} \frac{1}{3}^0\otimes\frac{1}{3}^0\otimes\frac{1}{3}^0\otimes 0^0\otimes 0^1 \\  \frac{2}{9}^0\otimes\frac{2}{9}^0\otimes\frac{2}{9}^0\otimes\frac{1}{3}^1\otimes 0^0\end{array}$\\
\hline
$(8,7,3,0,0)$&$2$&$\begin{array}{c} \frac{1}{3}^0\otimes\frac{1}{3}^0\otimes\frac{1}{3}^0\otimes 0^0\otimes 0^1 \\  \frac{2}{9}^0\otimes\frac{2}{9}^0\otimes\frac{2}{9}^0\otimes\frac{1}{3}^1\otimes 0^0\end{array}$\\
\hline
$(8,6,4,0,0)$&$2$&$\begin{array}{c} \frac{1}{3}^0\otimes\frac{1}{3}^0\otimes\frac{1}{3}^0\otimes 0^0\otimes 0^1 \\  \frac{2}{9}^0\otimes\frac{2}{9}^0\otimes\frac{2}{9}^0\otimes\frac{1}{3}^1\otimes 0^0\end{array}$\\
\hline
$(8,6,3,0,0)$&$2$&$\begin{array}{c} \frac{1}{3}^0\otimes\frac{1}{3}^0\otimes\frac{1}{3}^0\otimes 0^0\otimes 0^1 \\  \frac{2}{9}^0\otimes\frac{2}{9}^0\otimes\frac{2}{9}^0\otimes\frac{1}{3}^1\otimes 0^0\end{array}$\\
\hline
$(8,5,4,0,0)$&$2$&$\begin{array}{c} \frac{1}{3}^0\otimes\frac{1}{3}^0\otimes\frac{1}{3}^0\otimes 0^0\otimes 0^1 \\  \frac{2}{9}^0\otimes\frac{2}{9}^0\otimes\frac{2}{9}^0\otimes\frac{1}{3}^1\otimes 0^0\end{array}$\\
\hline
$(8,5,3,0,0)$&$2$&$\begin{array}{c} \frac{1}{3}^0\otimes\frac{1}{3}^0\otimes\frac{1}{3}^0\otimes 0^0\otimes 0^1 \\  \frac{2}{9}^0\otimes\frac{2}{9}^0\otimes\frac{2}{9}^0\otimes\frac{1}{3}^1\otimes 0^0\end{array}$\\
\hline
$(8,4,4,0,0)$&$2$&$\begin{array}{c} \frac{1}{3}^0\otimes\frac{1}{3}^0\otimes\frac{1}{3}^0\otimes 0^0\otimes 0^1 \\  \frac{2}{9}^0\otimes\frac{2}{9}^0\otimes\frac{2}{9}^0\otimes\frac{1}{3}^1\otimes 0^0\end{array}$\\
\hline
$(8,4,3,0,0)$&$2$&$\begin{array}{c} \frac{1}{3}^0\otimes\frac{1}{3}^0\otimes\frac{1}{3}^0\otimes 0^0\otimes 0^1 \\  \frac{2}{9}^0\otimes\frac{2}{9}^0\otimes\frac{2}{9}^0\otimes\frac{1}{3}^1\otimes 0^0\end{array}$\\
\hline
$(8,3,3,0,0)$&$2$&$\begin{array}{c} \frac{1}{3}^0\otimes\frac{1}{3}^0\otimes\frac{1}{3}^0\otimes 0^0\otimes 0^1 \\  \frac{2}{9}^0\otimes\frac{2}{9}^0\otimes\frac{2}{9}^0\otimes\frac{1}{3}^1\otimes 0^0\end{array}$\\
\hline
$(7,7,4,0,0)$&$2$&$\begin{array}{c} \frac{1}{3}^0\otimes\frac{1}{3}^0\otimes\frac{1}{3}^0\otimes 0^0\otimes 0^1 \\  \frac{2}{9}^0\otimes\frac{2}{9}^0\otimes\frac{2}{9}^0\otimes\frac{1}{3}^1\otimes 0^0\end{array}$\\
\hline
$(7,7,3,0,0)$&$2$&$\begin{array}{c} \frac{1}{3}^0\otimes\frac{1}{3}^0\otimes\frac{1}{3}^0\otimes 0^0\otimes 0^1 \\  \frac{2}{9}^0\otimes\frac{2}{9}^0\otimes\frac{2}{9}^0\otimes\frac{1}{3}^1\otimes 0^0\end{array}$\\
\hline
$(7,6,4,0,0)$&$2$&$\begin{array}{c} \frac{1}{3}^0\otimes\frac{1}{3}^0\otimes\frac{1}{3}^0\otimes 0^0\otimes 0^1 \\  \frac{2}{9}^0\otimes\frac{2}{9}^0\otimes\frac{2}{9}^0\otimes\frac{1}{3}^1\otimes 0^0\end{array}$\\
\hline
$(7,6,3,0,0)$&$2$&$\begin{array}{c} \frac{1}{3}^0\otimes\frac{1}{3}^0\otimes\frac{1}{3}^0\otimes 0^0\otimes 0^1 \\  \frac{2}{9}^0\otimes\frac{2}{9}^0\otimes\frac{2}{9}^0\otimes\frac{1}{3}^1\otimes 0^0\end{array}$\\
\hline
$(7,5,4,0,0)$&$2$&$\begin{array}{c} \frac{1}{3}^0\otimes\frac{1}{3}^0\otimes\frac{1}{3}^0\otimes 0^0\otimes 0^1 \\  \frac{2}{9}^0\otimes\frac{2}{9}^0\otimes\frac{2}{9}^0\otimes\frac{1}{3}^1\otimes 0^0\end{array}$\\
\hline
$(7,5,3,0,0)$&$2$&$\begin{array}{c} \frac{1}{3}^0\otimes\frac{1}{3}^0\otimes\frac{1}{3}^0\otimes 0^0\otimes 0^1 \\  \frac{2}{9}^0\otimes\frac{2}{9}^0\otimes\frac{2}{9}^0\otimes\frac{1}{3}^1\otimes 0^0\end{array}$\\
\hline
$(7,4,4,0,0)$&$2$&$\begin{array}{c} \frac{1}{3}^0\otimes\frac{1}{3}^0\otimes\frac{1}{3}^0\otimes 0^0\otimes 0^1 \\  \frac{2}{9}^0\otimes\frac{2}{9}^0\otimes\frac{2}{9}^0\otimes\frac{1}{3}^1\otimes 0^0\end{array}$\\
\hline
$(7,4,3,0,0)$&$2$&$\begin{array}{c} \frac{1}{3}^0\otimes\frac{1}{3}^0\otimes\frac{1}{3}^0\otimes 0^0\otimes 0^1 \\  \frac{2}{9}^0\otimes\frac{2}{9}^0\otimes\frac{2}{9}^0\otimes\frac{1}{3}^1\otimes 0^0\end{array}$\\
\hline
$(7,3,3,0,0)$&$2$&$\begin{array}{c} \frac{1}{3}^0\otimes\frac{1}{3}^0\otimes\frac{1}{3}^0\otimes 0^0\otimes 0^1 \\  \frac{2}{9}^0\otimes\frac{2}{9}^0\otimes\frac{2}{9}^0\otimes\frac{1}{3}^1\otimes 0^0\end{array}$\\
\hline
$(6,6,4,0,0)$&$2$&$\begin{array}{c} \frac{1}{3}^0\otimes\frac{1}{3}^0\otimes\frac{1}{3}^0\otimes 0^0\otimes 0^1 \\  \frac{2}{9}^0\otimes\frac{2}{9}^0\otimes\frac{2}{9}^0\otimes\frac{1}{3}^1\otimes 0^0\end{array}$\\
\hline
$(6,6,3,0,0)$&$2$&$\begin{array}{c} \frac{1}{3}^0\otimes\frac{1}{3}^0\otimes\frac{1}{3}^0\otimes 0^0\otimes 0^1 \\  \frac{2}{9}^0\otimes\frac{2}{9}^0\otimes\frac{2}{9}^0\otimes\frac{1}{3}^1\otimes 0^0\end{array}$\\
\hline
$(6,5,4,0,0)$&$2$&$\begin{array}{c} \frac{1}{3}^0\otimes\frac{1}{3}^0\otimes\frac{1}{3}^0\otimes 0^0\otimes 0^1 \\  \frac{2}{9}^0\otimes\frac{2}{9}^0\otimes\frac{2}{9}^0\otimes\frac{1}{3}^1\otimes 0^0\end{array}$\\
\hline
$(6,5,3,0,0)$&$2$&$\begin{array}{c} \frac{1}{3}^0\otimes\frac{1}{3}^0\otimes\frac{1}{3}^0\otimes 0^0\otimes 0^1 \\  \frac{2}{9}^0\otimes\frac{2}{9}^0\otimes\frac{2}{9}^0\otimes\frac{1}{3}^1\otimes 0^0\end{array}$\\
\hline
$(6,4,4,0,0)$&$2$&$\begin{array}{c} \frac{1}{3}^0\otimes\frac{1}{3}^0\otimes\frac{1}{3}^0\otimes 0^0\otimes 0^1 \\  \frac{2}{9}^0\otimes\frac{2}{9}^0\otimes\frac{2}{9}^0\otimes\frac{1}{3}^1\otimes 0^0\end{array}$\\
\hline
$(6,4,3,0,0)$&$2$&$\begin{array}{c} \frac{1}{3}^0\otimes\frac{1}{3}^0\otimes\frac{1}{3}^0\otimes 0^0\otimes 0^1 \\  \frac{2}{9}^0\otimes\frac{2}{9}^0\otimes\frac{2}{9}^0\otimes\frac{1}{3}^1\otimes 0^0\end{array}$\\
\hline
$(6,3,3,0,0)$&$2$&$\begin{array}{c} \frac{1}{3}^0\otimes\frac{1}{3}^0\otimes\frac{1}{3}^0\otimes 0^0\otimes 0^1 \\  \frac{2}{9}^0\otimes\frac{2}{9}^0\otimes\frac{2}{9}^0\otimes\frac{1}{3}^1\otimes 0^0\end{array}$\\
\hline
$(5,5,4,0,0)$&$2$&$\begin{array}{c} \frac{1}{3}^0\otimes\frac{1}{3}^0\otimes\frac{1}{3}^0\otimes 0^0\otimes 0^1 \\  \frac{2}{9}^0\otimes\frac{2}{9}^0\otimes\frac{2}{9}^0\otimes\frac{1}{3}^1\otimes 0^0\end{array}$\\
\hline
$(5,5,3,0,0)$&$2$&$\begin{array}{c} \frac{1}{3}^0\otimes\frac{1}{3}^0\otimes\frac{1}{3}^0\otimes 0^0\otimes 0^1 \\  \frac{2}{9}^0\otimes\frac{2}{9}^0\otimes\frac{2}{9}^0\otimes\frac{1}{3}^1\otimes 0^0\end{array}$\\
\hline
$(5,4,4,0,0)$&$2$&$\begin{array}{c} \frac{1}{3}^0\otimes\frac{1}{3}^0\otimes\frac{1}{3}^0\otimes 0^0\otimes 0^1 \\  \frac{2}{9}^0\otimes\frac{2}{9}^0\otimes\frac{2}{9}^0\otimes\frac{1}{3}^1\otimes 0^0\end{array}$\\
\hline
$(5,4,3,0,0)$&$2$&$\begin{array}{c} \frac{1}{3}^0\otimes\frac{1}{3}^0\otimes\frac{1}{3}^0\otimes 0^0\otimes 0^1 \\  \frac{2}{9}^0\otimes\frac{2}{9}^0\otimes\frac{2}{9}^0\otimes\frac{1}{3}^1\otimes 0^0\end{array}$\\
\hline
$(5,3,3,0,0)$&$2$&$\begin{array}{c} \frac{1}{3}^0\otimes\frac{1}{3}^0\otimes\frac{1}{3}^0\otimes 0^0\otimes 0^1 \\  \frac{2}{9}^0\otimes\frac{2}{9}^0\otimes\frac{2}{9}^0\otimes\frac{1}{3}^1\otimes 0^0\end{array}$\\
\hline
$(4,4,4,0,0)$&$2$&$\begin{array}{c} \frac{1}{3}^0\otimes\frac{1}{3}^0\otimes\frac{1}{3}^0\otimes 0^0\otimes 0^1 \\  \frac{2}{9}^0\otimes\frac{2}{9}^0\otimes\frac{2}{9}^0\otimes\frac{1}{3}^1\otimes 0^0\end{array}$\\
\hline
$(4,4,3,0,0)$&$2$&$\begin{array}{c} \frac{1}{3}^0\otimes\frac{1}{3}^0\otimes\frac{1}{3}^0\otimes 0^0\otimes 0^1 \\  \frac{2}{9}^0\otimes\frac{2}{9}^0\otimes\frac{2}{9}^0\otimes\frac{1}{3}^1\otimes 0^0\end{array}$\\
\hline
$(4,3,3,0,0)$&$2$&$\begin{array}{c} \frac{1}{3}^0\otimes\frac{1}{3}^0\otimes\frac{1}{3}^0\otimes 0^0\otimes 0^1 \\  \frac{2}{9}^0\otimes\frac{2}{9}^0\otimes\frac{2}{9}^0\otimes\frac{1}{3}^1\otimes 0^0\end{array}$\\
\hline
$(3,3,3,0,0)$&$2$&$\begin{array}{c} \frac{1}{3}^0\otimes\frac{1}{3}^0\otimes\frac{1}{3}^0\otimes 0^0\otimes 0^1 \\  \frac{2}{9}^0\otimes\frac{2}{9}^0\otimes\frac{2}{9}^0\otimes\frac{1}{3}^1\otimes 0^0\end{array}$\\
\hline
$(8,7,2,0,0)$&$1$&$ \frac{2}{9}^0\otimes\frac{2}{9}^0\otimes\frac{2}{9}^0\otimes\frac{1}{3}^1\otimes 0^0$\\
\hline
$(8,6,2,0,0)$&$1$&$ \frac{2}{9}^0\otimes\frac{2}{9}^0\otimes\frac{2}{9}^0\otimes\frac{1}{3}^1\otimes 0^0$\\
\hline
$(8,5,2,0,0)$&$1$&$ \frac{2}{9}^0\otimes\frac{2}{9}^0\otimes\frac{2}{9}^0\otimes\frac{1}{3}^1\otimes 0^0$\\
\hline
$(8,4,2,0,0)$&$1$&$ \frac{2}{9}^0\otimes\frac{2}{9}^0\otimes\frac{2}{9}^0\otimes\frac{1}{3}^1\otimes 0^0$\\
\hline
$(8,3,2,0,0)$&$1$&$ \frac{2}{9}^0\otimes\frac{2}{9}^0\otimes\frac{2}{9}^0\otimes\frac{1}{3}^1\otimes 0^0$\\
\hline
$(7,7,2,0,0)$&$1$&$ \frac{2}{9}^0\otimes\frac{2}{9}^0\otimes\frac{2}{9}^0\otimes\frac{1}{3}^1\otimes 0^0$\\
\hline
$(7,6,2,0,0)$&$1$&$ \frac{2}{9}^0\otimes\frac{2}{9}^0\otimes\frac{2}{9}^0\otimes\frac{1}{3}^1\otimes 0^0$\\
\hline
$(7,5,2,0,0)$&$1$&$ \frac{2}{9}^0\otimes\frac{2}{9}^0\otimes\frac{2}{9}^0\otimes\frac{1}{3}^1\otimes 0^0$\\
\hline
$(7,4,2,0,0)$&$1$&$ \frac{2}{9}^0\otimes\frac{2}{9}^0\otimes\frac{2}{9}^0\otimes\frac{1}{3}^1\otimes 0^0$\\
\hline
$(7,3,2,0,0)$&$1$&$ \frac{2}{9}^0\otimes\frac{2}{9}^0\otimes\frac{2}{9}^0\otimes\frac{1}{3}^1\otimes 0^0$\\
\hline
$(7,2,2,0,0)$&$1$&$ \frac{2}{9}^0\otimes\frac{2}{9}^0\otimes\frac{2}{9}^0\otimes\frac{1}{3}^1\otimes 0^0$\\
\hline
$(6,6,2,0,0)$&$1$&$ \frac{2}{9}^0\otimes\frac{2}{9}^0\otimes\frac{2}{9}^0\otimes\frac{1}{3}^1\otimes 0^0$\\
\hline
$(6,5,2,0,0)$&$1$&$ \frac{2}{9}^0\otimes\frac{2}{9}^0\otimes\frac{2}{9}^0\otimes\frac{1}{3}^1\otimes 0^0$\\
\hline
$(6,4,2,0,0)$&$1$&$ \frac{2}{9}^0\otimes\frac{2}{9}^0\otimes\frac{2}{9}^0\otimes\frac{1}{3}^1\otimes 0^0$\\
\hline
$(6,3,2,0,0)$&$1$&$ \frac{2}{9}^0\otimes\frac{2}{9}^0\otimes\frac{2}{9}^0\otimes\frac{1}{3}^1\otimes 0^0$\\
\hline
$(6,2,2,0,0)$&$1$&$ \frac{2}{9}^0\otimes\frac{2}{9}^0\otimes\frac{2}{9}^0\otimes\frac{1}{3}^1\otimes 0^0$\\
\hline
$(5,5,2,0,0)$&$1$&$ \frac{2}{9}^0\otimes\frac{2}{9}^0\otimes\frac{2}{9}^0\otimes\frac{1}{3}^1\otimes 0^0$\\
\hline
$(5,4,2,0,0)$&$1$&$ \frac{2}{9}^0\otimes\frac{2}{9}^0\otimes\frac{2}{9}^0\otimes\frac{1}{3}^1\otimes 0^0$\\
\hline
$(5,3,2,0,0)$&$1$&$ \frac{2}{9}^0\otimes\frac{2}{9}^0\otimes\frac{2}{9}^0\otimes\frac{1}{3}^1\otimes 0^0$\\
\hline
$(5,2,2,0,0)$&$1$&$ \frac{2}{9}^0\otimes\frac{2}{9}^0\otimes\frac{2}{9}^0\otimes\frac{1}{3}^1\otimes 0^0$\\
\hline
$(4,4,2,0,0)$&$1$&$ \frac{2}{9}^0\otimes\frac{2}{9}^0\otimes\frac{2}{9}^0\otimes\frac{1}{3}^1\otimes 0^0$\\
\hline
$(4,3,2,0,0)$&$1$&$ \frac{2}{9}^0\otimes\frac{2}{9}^0\otimes\frac{2}{9}^0\otimes\frac{1}{3}^1\otimes 0^0$\\
\hline
$(4,2,2,0,0)$&$1$&$ \frac{2}{9}^0\otimes\frac{2}{9}^0\otimes\frac{2}{9}^0\otimes\frac{1}{3}^1\otimes 0^0$\\
\hline
$(3,3,2,0,0)$&$1$&$ \frac{2}{9}^0\otimes\frac{2}{9}^0\otimes\frac{2}{9}^0\otimes\frac{1}{3}^1\otimes 0^0$\\
\hline
$(3,2,2,0,0)$&$1$&$ \frac{2}{9}^0\otimes\frac{2}{9}^0\otimes\frac{2}{9}^0\otimes\frac{1}{3}^1\otimes 0^0$\\
\hline
$(2,2,2,0,0)$&$1$&$ \frac{2}{9}^0\otimes\frac{2}{9}^0\otimes\frac{2}{9}^0\otimes\frac{1}{3}^1\otimes 0^0$\\
\hline\hline
\end{longtable}
\end{appendix}
\bibliographystyle{fullsort}
\bibliography{bibliography}

\providecommand{\href}[2]{#2}\begingroup\raggedright\begin{thebibliography}{10}

\bibitem{Walcher:2006rs}
J.~Walcher, ``{O}pening {M}irror {S}ymmetry on the {Q}uintic,'' {\em Commun.
  Math. Phys.} {\bf 276} (2007) 671--689,
\href{http://www.arXiv.org/abs/hep-th/0605162}{{\tt hep-th/0605162}}.

\bibitem{Morrison:2007bm}
D.~R. Morrison and J.~Walcher, ``{D}-branes and {N}ormal {F}unctions,''
\href{http://www.arXiv.org/abs/arXiv:0709.4028 [hep-th]}{{\tt arXiv:0709.4028
  [hep-th]}}.

\bibitem{Walcher:2007tp}
J.~Walcher, ``{Extended Holomorphic Anomaly and Loop Amplitudes in Open
  Topological String},''
\href{http://www.arXiv.org/abs/arXiv:0705.4098[hep-th]}{{\tt
  arXiv:0705.4098[hep-th]}}.

\bibitem{Walcher:2007qp}
J.~Walcher, ``{Evidence for Tadpole Cancellation in the Topological String},''
\href{http://www.arXiv.org/abs/arXiv:0712.2775[hep-th]}{{\tt
  arXiv:0712.2775[hep-th]}}.

\bibitem{Krefl:2008sj}
D.~Krefl and J.~Walcher, ``{Real Mirror Symmetry for One-parameter
  Hypersurfaces},'' {\em JHEP} {\bf 09} (2008) 031,
\href{http://www.arXiv.org/abs/arXiv:0805.0792[hep-th]}{{\tt
  arXiv:0805.0792[hep-th]}}.

\bibitem{Knapp:2008uw}
J.~Knapp and E.~Scheidegger, ``{Towards Open String Mirror Symmetry for
  One-Parameter Calabi-Yau Hypersurfaces},''
\href{http://www.arXiv.org/abs/arXiv:0805.1013[hep-th]}{{\tt
  arXiv:0805.1013[hep-th]}}.

\bibitem{Jockers:2008pe}
H.~Jockers and M.~Soroush, ``{Effective superpotentials for compact D5-brane
  Calabi-Yau geometries},''
\href{http://www.arXiv.org/abs/arXiv:0808.0761[hep-th]}{{\tt
  arXiv:0808.0761[hep-th]}}.

\bibitem{Mayr:2001xk}
P.~Mayr, ``{N = 1 mirror symmetry and open/closed string duality},'' {\em Adv.
  Theor. Math. Phys.} {\bf 5} (2002) 213--242,
\href{http://www.arXiv.org/abs/hep-th/0108229}{{\tt hep-th/0108229}}.

\bibitem{Lerche:2001cw}
W.~Lerche and P.~Mayr, ``{On N = 1 mirror symmetry for open type II strings},''
\href{http://www.arXiv.org/abs/hep-th/0111113}{{\tt hep-th/0111113}}.

\bibitem{Candelas:1993dm}
P.~Candelas, X.~De~La~Ossa, A.~Font, S.~H. Katz, and D.~R. Morrison, ``{Mirror
  symmetry for two parameter models. I},'' {\em Nucl. Phys.} {\bf B416} (1994)
  481--538,
\href{http://www.arXiv.org/abs/hep-th/9308083}{{\tt hep-th/9308083}}.

\bibitem{Hosono:1993qy}
S.~Hosono, A.~Klemm, S.~Theisen, and S.-T. Yau, ``{Mirror symmetry, mirror map
  and applications to Calabi-Yau hypersurfaces},'' {\em Commun. Math. Phys.}
  {\bf 167} (1995) 301--350,
\href{http://www.arXiv.org/abs/hep-th/9308122}{{\tt hep-th/9308122}}.

\bibitem{Candelas:1994hw}
P.~Candelas, A.~Font, S.~H. Katz, and D.~R. Morrison, ``{Mirror symmetry for
  two parameter models. 2},'' {\em Nucl. Phys.} {\bf B429} (1994) 626--674,
\href{http://www.arXiv.org/abs/hep-th/9403187}{{\tt hep-th/9403187}}.

\bibitem{Hosono:1994ax}
S.~Hosono, A.~Klemm, S.~Theisen, and S.-T. Yau, ``{Mirror symmetry, mirror map
  and applications to complete intersection Calabi-Yau spaces},'' {\em Nucl.
  Phys.} {\bf B433} (1995) 501--554,
\href{http://www.arXiv.org/abs/hep-th/9406055}{{\tt hep-th/9406055}}.

\bibitem{Siqveland1}
A.~Siqveland, ``{T}he {M}ethod of {C}omputing {F}ormal {M}oduli,'' {\em J.
  Alg.} {\bf 241} (2001) 292--327.

\bibitem{siqvelandPHD}
A.~Siqveland, ``{M}atric {M}assey {P}roducts and {F}ormal {M}oduli {L}ocal and
  {G}lobal.'' Phd thesis, University of Oslo, Dept. of Mathematics, 1995.

\bibitem{Knapp:2006rd}
J.~Knapp and H.~Omer, ``{M}atrix {F}actorizations, {M}inimal {M}odels and
  {M}assey {P}roducts,'' {\em JHEP} {\bf 05} (2006) 064,
\href{http://www.arXiv.org/abs/hep-th/0604189}{{\tt hep-th/0604189}}.

\bibitem{Hori:2004ja}
K.~Hori and J.~Walcher, ``{F-term equations near Gepner points},'' {\em JHEP}
  {\bf 01} (2005) 008,
\href{http://www.arXiv.org/abs/hep-th/0404196}{{\tt hep-th/0404196}}.

\bibitem{Ashok:2004zb}
S.~K. Ashok, E.~Dell'Aquila, and D.-E. Diaconescu, ``{Fractional branes in
  Landau-Ginzburg orbifolds},'' {\em Adv. Theor. Math. Phys.} {\bf 8} (2004)
  461--513,
\href{http://www.arXiv.org/abs/hep-th/0401135}{{\tt hep-th/0401135}}.

\bibitem{Kapustin:2003ga}
A.~Kapustin and Y.~Li, ``{T}opological {C}orrelators in {L}andau-{G}inzburg
  {M}odels with {B}oundaries,'' {\em Adv. Theor. Math. Phys.} {\bf 7} (2004)
  727--749,
\href{http://www.arXiv.org/abs/hep-th/0305136}{{\tt hep-th/0305136}}.

\bibitem{Knapp:2007vc}
J.~Knapp, ``{D-Branes in Topological String Theory},''
\href{http://www.arXiv.org/abs/arXiv:0709.2045[hep-th]}{{\tt
  arXiv:0709.2045[hep-th]}}.

\bibitem{Herbst:2004jp}
M.~Herbst, C.-I. Lazaroiu, and W.~Lerche, ``{S}uperpotentials, {A}(infinity)
  {R}elations and {WDVV} {E}quations for open topological {S}trings,'' {\em
  JHEP} {\bf 02} (2005) 071,
\href{http://www.arXiv.org/abs/hep-th/0402110}{{\tt hep-th/0402110}}.

\bibitem{Fuchs:2000gv}
J.~Fuchs, C.~Schweigert, and J.~Walcher, ``{Projections in string theory and
  boundary states for Gepner models},'' {\em Nucl. Phys.} {\bf B588} (2000)
  110--148,
\href{http://www.arXiv.org/abs/hep-th/0003298}{{\tt hep-th/0003298}}.

\bibitem{Brunner:2004zd}
I.~Brunner, K.~Hori, K.~Hosomichi, and J.~Walcher, ``{Orientifolds of Gepner
  models},'' {\em JHEP} {\bf 02} (2007) 001,
\href{http://www.arXiv.org/abs/hep-th/0401137}{{\tt hep-th/0401137}}.

\bibitem{Recknagel:2002qq}
A.~Recknagel, ``{Permutation branes},'' {\em JHEP} {\bf 04} (2003) 041,
\href{http://www.arXiv.org/abs/hep-th/0208119}{{\tt hep-th/0208119}}.

\bibitem{Brunner:2005fv}
I.~Brunner and M.~R. Gaberdiel, ``{M}atrix {F}actorisations and {P}ermutation
  {B}ranes,'' {\em JHEP} {\bf 07} (2005) 012,
\href{http://www.arXiv.org/abs/hep-th/0503207}{{\tt hep-th/0503207}}.

\bibitem{Enger:2005jk}
H.~Enger, A.~Recknagel, and D.~Roggenkamp, ``{Permutation branes and linear
  matrix factorisations},'' {\em JHEP} {\bf 01} (2006) 087,
\href{http://www.arXiv.org/abs/hep-th/0508053}{{\tt hep-th/0508053}}.

\bibitem{Fredenhagen:2005an}
S.~Fredenhagen and T.~Quella, ``{Generalised permutation branes},'' {\em JHEP}
  {\bf 11} (2005) 004,
\href{http://www.arXiv.org/abs/hep-th/0509153}{{\tt hep-th/0509153}}.

\bibitem{Caviezel:2005th}
C.~Caviezel, S.~Fredenhagen, and M.~R. Gaberdiel, ``{The RR charges of A-type
  Gepner models},'' {\em JHEP} {\bf 01} (2006) 111,
\href{http://www.arXiv.org/abs/hep-th/0511078}{{\tt hep-th/0511078}}.

\bibitem{GPS05}
G.-M. Greuel, G.~Pfister, and H.~Sch\"onemann, ``{\sc Singular} 3.0,'' {A
  Computer Algebra System for Polynomial Computations}, Centre for Computer
  Algebra, University of Kaiserslautern, 2005.
\newblock {\tt http://www.singular.uni-kl.de}.

\bibitem{Herbst:2008jq}
M.~Herbst, K.~Hori, and D.~Page, ``{Phases Of N=2 Theories In 1+1 Dimensions
  With Boundary},''
\href{http://www.arXiv.org/abs/arXiv:0803.2045 [hep-th]}{{\tt arXiv:0803.2045
  [hep-th]}}.

\bibitem{Aspinwall:2004bs}
P.~S. Aspinwall and S.~H. Katz, ``{Computation of superpotentials for
  D-Branes},'' {\em Commun. Math. Phys.} {\bf 264} (2006) 227--253,
\href{http://www.arXiv.org/abs/hep-th/0412209}{{\tt hep-th/0412209}}.

\bibitem{Baumgartl:2007an}
M.~Baumgartl, I.~Brunner, and M.~R. Gaberdiel, ``{D-brane superpotentials and
  RG flows on the quintic},'' {\em JHEP} {\bf 07} (2007) 061,
\href{http://www.arXiv.org/abs/arXiv:0704.2666[hep-th]}{{\tt
  arXiv:0704.2666[hep-th]}}.

\bibitem{Grimm:2008dq}
T.~W. Grimm, T.-W. Ha, A.~Klemm, and D.~Klevers, ``{The D5-brane effective
  action and superpotential in N=1 compactifications},''
\href{http://www.arXiv.org/abs/arXiv:0811.2996[hep-th]}{{\tt
  arXiv:0811.2996[hep-th]}}.

\end{thebibliography}\endgroup
\end{document}